\documentclass[twocolumn,iop]{emulateapj}

\usepackage{hyperref}
\usepackage{amsmath,amssymb}
\usepackage{rotating}
\usepackage{natbib}
\usepackage{morefloats}
\usepackage{subfigure}
\hypersetup{colorlinks, citecolor=blue}
\slugcomment{Accepted for publication on ApJ}

\shorttitle{Gas kinematics in the nucleus of NGC 1386}
\shortauthors{Lena et al.}

\begin{document}


\title{The complex gas kinematics in the nucleus of the Seyfert 2 galaxy NGC 1386: rotation, outflows and inflows}


\author{
D. Lena\altaffilmark{1}, 
A. Robinson\altaffilmark{1},
T. Storchi-Bergman\altaffilmark{2,7},
A. Schnorr-M\"{u}ller\altaffilmark{3,4},
T. Seelig\altaffilmark{1},\\
R. A. Riffel\altaffilmark{5},
N. M. Nagar\altaffilmark{6},
G. S. Couto\altaffilmark{2},
L. Shadler\altaffilmark{1}
}
\email{dxl1840@g.rit.edu}


\altaffiltext{1}{School of Physics and Astronomy, Rochester Institute of Technology, 84 Lomb Memorial Drive, Rochester, NY 14623-5603, USA}
\altaffiltext{2}{Instituto de Fisica, Universidade Federal do Rio Grande do Sul, 91501-970, Porto Alegre, Brazil}
\altaffiltext{3}{Max-Planck-Institut f\"{u}r extraterrestrische Physik, Giessenbachstr. 1, D-85741, Garching, Germany}
\altaffiltext{4}{CAPES Foundation, Ministry of Education of Brazil, 70040-020, Bras'lia, Brazil}
\altaffiltext{5}{Universidade Federal de Santa Maria, Departamento de Fisica, 97105-900, Santa Maria, RS, Brazil}
\altaffiltext{6}{Astronomy Department, Universidad de Concepci\'{o}n, Casilla 160-C, Concepci\'{o}n, Chile}
\altaffiltext{7}{Harvard-Smithsonian Center for Astrophysics, 60 Garden Street, Cambridge, MA 02138, USA}


\begin{abstract}
We present optical integral field spectroscopy of the circum-nuclear gas of the Seyfert 2 galaxy NGC 1386. The data cover the central 7$^{\prime\prime} \times 9^{\prime\prime}$ (530 $\times$ 680 pc) at a spatial resolution of 0\farcs9 (68 pc), and the spectral range 5700-7000 \AA\ at a resolution of 66 km s$^{-1}$. The line emission is dominated by a bright central component, with two lobes extending $\approx$ 3$^{\prime\prime}$ north and south of the nucleus. We identify three main kinematic components. The first has low velocity dispersion ($\bar \sigma \approx $ 90 km s$^{-1}$), extends over the whole field-of-view, and has a velocity field consistent with gas rotating in the galaxy disk. We interpret the lobes as resulting from photoionization of disk gas in regions where the AGN radiation cones intercept the disk. The second has higher velocity dispersion ($\bar \sigma \approx$ 200 km s$^{-1}$) and is observed in the inner 150 pc around the continuum peak. This component is double peaked, with redshifted and blueshifted components separated by $\approx$ 500 km s$^{-1}$. Together with previous HST imaging, these features suggest the presence of a bipolar outflow for which we estimate a mass outflow rate of $\mathrm{\dot M} \gtrsim $ 0.1 M$_{\odot}$ yr$^{-1}$. The third component is revealed by velocity residuals associated with enhanced velocity dispersion and suggests that outflow and/or rotation is occurring approximately in the equatorial plane of the torus. A second system of velocity residuals may indicate the presence of streaming motions along dusty spirals in the disk.
\end{abstract}


%\keywords{galaxies: evolution, galaxies: formation, galaxies:interactions, galaxies: kinematics and dynamics, galaxies: general}



\section{Introduction} \label{sec: intro}
Gas accretion onto a supermassive black hole (SBH) is believed to be the mechanism which powers active galactic nuclei \citep[AGNs, e.g.][]{Lynden-Bell69,BegelmanBR84}.
However, understanding how mass is transferred from galactic scales (kiloparsecs) down to nuclear scales (parsec and sub-parsecs) to create a reservoir of gas to be accreted has been a long-standing problem in the study of nuclear activity in galaxies.

Theoretical studies and simulations have shown that non-axisymmetric potentials efficiently promote gas inflow towards the inner regions of galaxies: close encounters and galactic mergers have been identified as mechanisms capable of driving gas from large scales (10 kpc) down to the inner kiloparsec \citep[e.g.][]{Hernquist89,DiMatteoSH05,SpringelMH05,SpringelDMH05,HopkinsSH06}. Stellar bars (and bars-within-bars) are also important non-axysimmetric features deemed able to funnel gas from the inner kpc down to few hundreds of pc from the SBH \citep[e.g.][]{ShlosmanFB89,FriedliM93,EnglmaierS04}. Simulations presented in \citet{HopkinsQ10} suggest that, in gas-rich systems, on scales of $\sim$ 10 - 100 pc, inflows are achieved through a system of nested gravitational instabilities, similar to the bars-within-bars scenario, but with a range of morphologies, such as nuclear spirals (including one- and three-armed spirals), bars, rings, barred rings, clumpy disks and streams.

Observations support the hypothesis that large-scale bars funnel gas toward the center of galaxies \citep[e.g.][]{CrenshawKG03}. Imaging has revealed that structures such as small-scale disks or nuclear bars and associated spiral arms are frequently observed in the inner kiloparsec of active galaxies \citep[e.g][]{ErwinS99,PoggeM02,LaineEtAl03}. The most common nuclear structures are dusty spirals, estimated to reside in more than half of active galaxies \citep{MartiniRM03}. Strong correlation between the presence of nuclear dust structures (filaments, spirals and disks) and activity in galaxies has also been reported \citep{LopesSFM07}. This correlation between dust structures and activity, along with the enhanced frequency of dusty spirals, supports the hypothesis that nuclear spirals are a mechanism for fueling the SBH, transporting gas from kiloparsec scales down to within a few tens of parsecs of the active nucleus.

In recent years integral field spectroscopy of nearby AGNs has provided a clearer view of the processes of AGN fueling on scales in the range $\sim$10 - 100 pc, revealing kinematical features consistent with gas inflowing along bars and nuclear spirals \citep[e.g.][]{KnapenSHRBR00,FathiSBREtAl06,SBergmannEtAl07,RiffelEtAl08,MullerSanchezDG09,RiffelSB11,SMullerEtAl11,RiffellRASBW13,CombesGBC14,DaviesMH14,SchnorrMSBNF14, SMSBN14}.

Removal of angular momentum can be achieved not only through gravitational torques, but also via the outflows, or winds, originating from the interaction between ionized gas and the magnetic field \citep[e.g.][]{Rees87}. Outflows are prevalent among powerful AGNs (e.g. \citealt{CrenshawKG03MassLoss, BarbosaSBCFWS09}), however studies of nearby Seyferts suggest that compact outflows ($\sim$ 100 pc in extent) with velocities of $\sim$ 100 km s$^{-1}$ and mass outflow rates of a few solar masses per year are common even in low-luminosity AGNs \citep[e.g.][]{SBergmannEtAl07,DaviesMHT09,SMullerEtAl11,MuellerSanchezPH11,DaviesMH14}.

Outflows transfer mechanical energy to the host interstellar medium, therefore quantifying mass (and kinetic energy) outflow rates is fundamental to understanding the process and pace of galaxy evolution (e.g. \citealt{DiMatteoSH05}). Key questions are: how are the inflow and outflow rates related? Does the presence of one preclude the other? In order to answer such questions we have assembled a hard X-ray selected sample of 20 nearby (z $<$0.01) Seyferts covering a range of accretion power. With this sample we aim to (1) determine if there is a transition in the behavior of the circumnuclear gas kinematics from low-luminosity to high-luminosity AGNs and investigate the effect of other factors such as host galaxy properties; (2) measure the mass inflow/outflow rate and relate it to AGN accretion power. 

Here we present a quantitative analysis of the gas and stellar kinematics in the inner 250 parsec of NGC 1386 as derived from integral field spectroscopic (IFS) observations carried out with the GEMINI South telescope. 

NGC 1386 is a Sb/c spiral galaxy \citep[][]{MalkanGT98}, located in the Fornax cluster \citep{Sandage78}, which hosts a Seyfert 2 nucleus.
At a distance of 15.6 Mpc \citep[determined from surface brightness fluctuations,][]{JensenEtAl2003} the linear scale is 76 pc arcsec$^{-1}$. The inclination with respect to the line of sight is $\theta \approx$ 65$^{\circ}$, as derived from the ratio of the apparent axes using values from the 2MASS catalog \citep{JCCS03} available in NED\footnote{NED is the NASA/IPAC Extragalactic Database (NED) which is operated by the Jet Propulsion Laboratory, California Institute of Technology, under contract with the National Aeronautics and Space Administration.}. 

Being one of the nearest Seyferts, NGC 1386 has been extensively studied. \citet{Sandage78} classified NGC 1386 as a high excitation galaxy on the basis of its high [OIII]$\lambda$5007/H$\beta$ line intensity ratio. High resolution observations at 8.4 GHz show the presence of a non-variable compact source confined within the central arcsecond and elongated southward in position angle PA $\approx$ 170 - 175$^{\circ}$ \citep{nagarWMG99,MundellEtAl09}. HST imaging of the [OIII] and H$\alpha$ + [NII] emission reveals a jet-like distribution of ionized gas organized in knots along the north-south axis and extending over a distance of about 2$^{\prime\prime}$ from the nucleus \citep{ferruit2000}. Despite the close alignment between ionized gas and radio emission, direct comparison suggests that there is no detailed association amongst the two \citep{MundellEtAl09}.
Early 2D spectroscopic observations \citep{WeaverWilson91,rossa2000}, supported by more recent studies \citep[e.g.][]{SchHnk03,bennert06}, have already revealed the presence of a complex nuclear spectrum showing the presence of a strong narrow component, following rotational motion, superposed on a broader component suggestive of high-velocity outflows or inflows.    
\vskip10pt

The paper is organized as follows: observations and data reduction are presented in \textsection \ref{sec: analysis}, the data analysis is described in \textsection \ref{sec: line_fitting}. Our results are presented in \textsection \ref{sec: results} and discussed in \textsection \ref{sec: disc}. In \textsection \ref{sec: nuclear_conf} we propose an interpretation of the results summarizing our main findings and the proposed model in \textsection \ref{sec: conclusion}.

% ==========================================================================================

% \begin{minipage}[position]{width}
\begin{figure*}[t]
\begin{center}$
\begin{array}{ccc}
\includegraphics[trim=4cm 1.2cm 5cm 1.1cm, clip=true, scale=0.625]{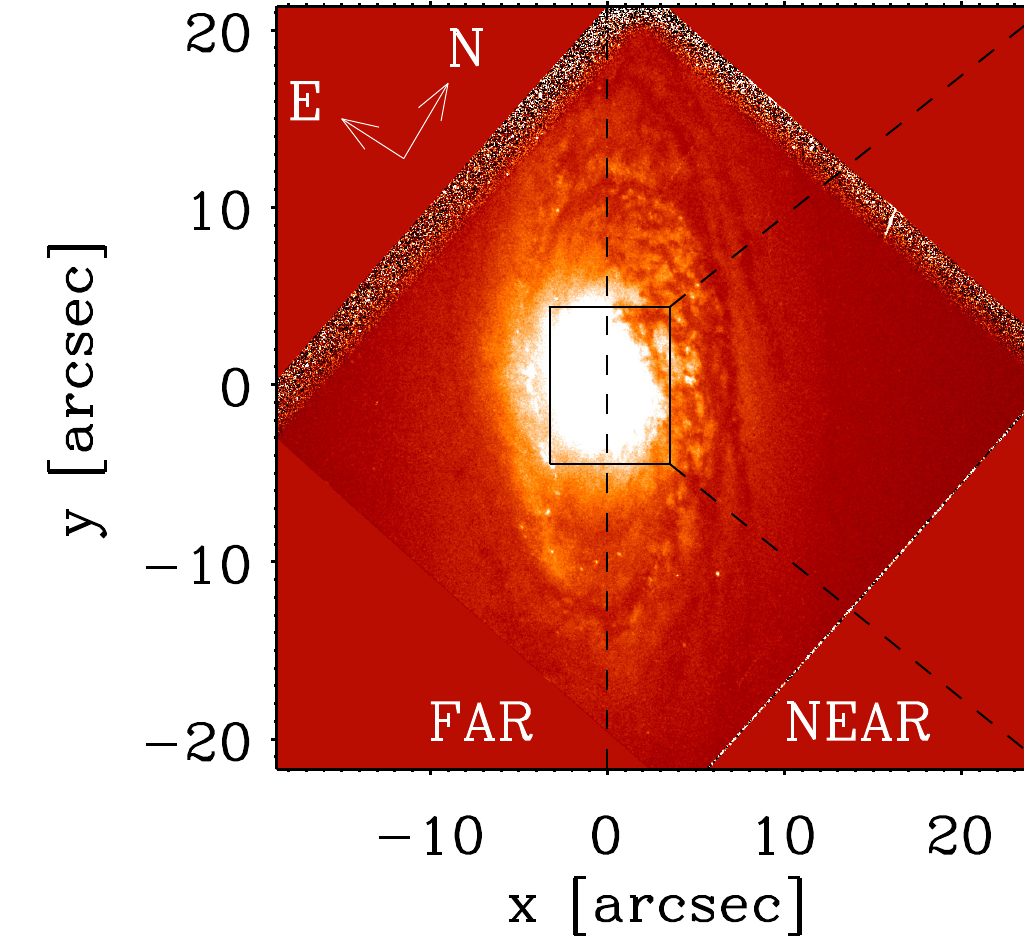}	& \includegraphics[trim=5.5cm 1.2cm 4.5cm 1.1cm, clip=true, scale=0.625]{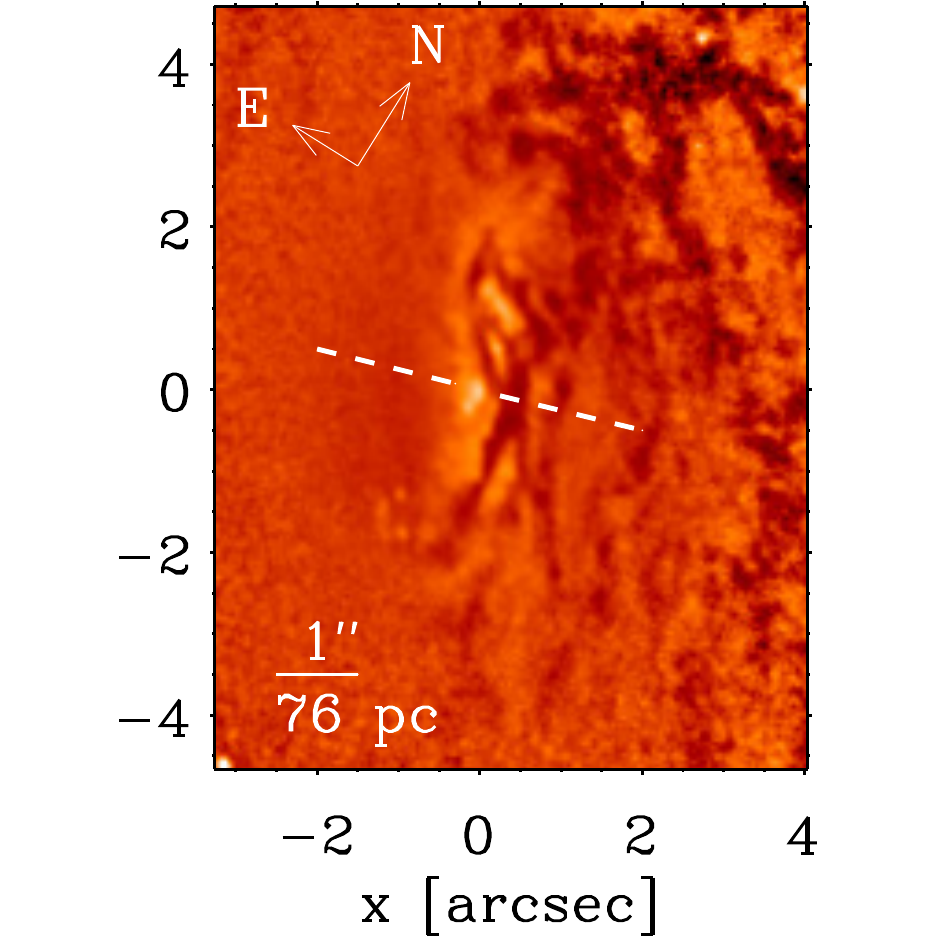} & \includegraphics[trim=6.5cm 1.2cm 3cm 1.1cm, clip=true, scale=0.625]{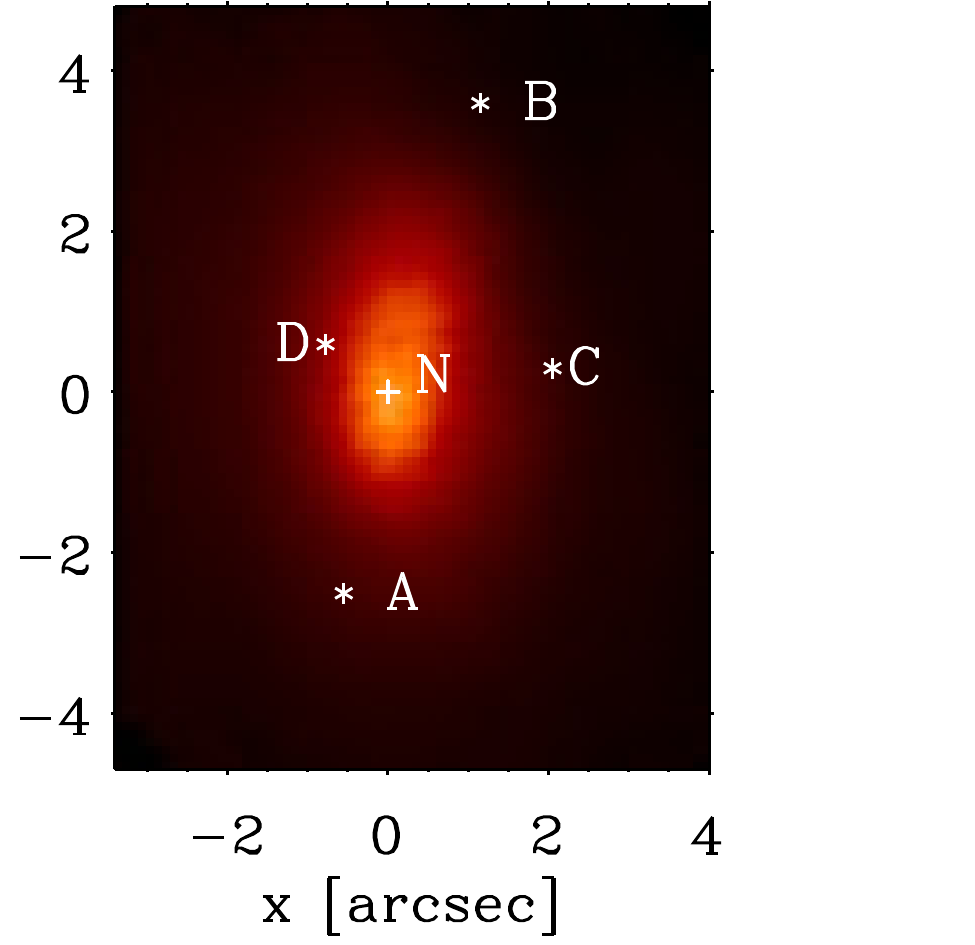}\\
\end{array}$

        \subfigure{
        % \label{fig:first}
        \includegraphics[trim=0cm 0.25cm 0cm 0cm, clip=true, width=1\textwidth]{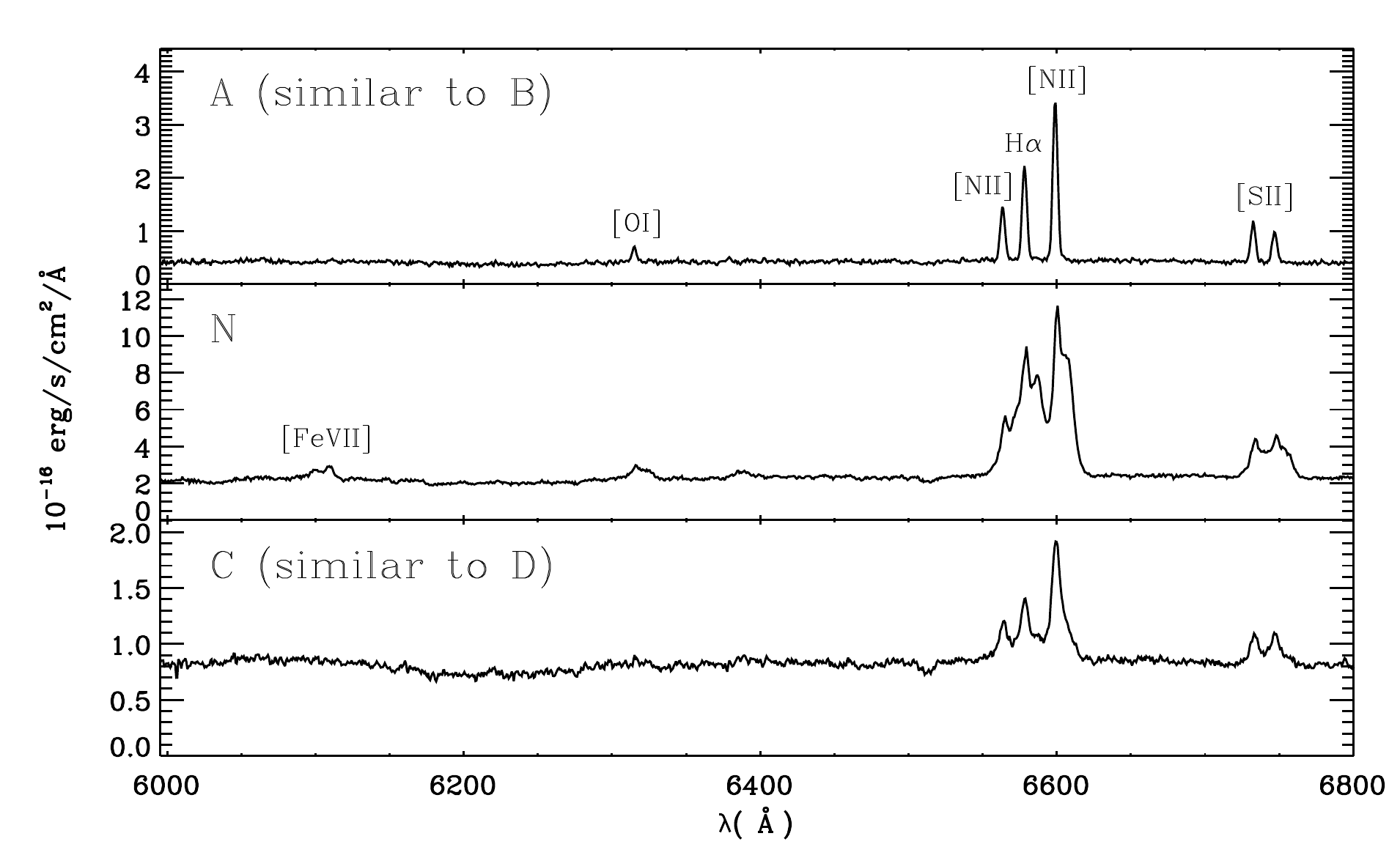}
         }

\end{center}
\caption{\textit{Top left:} HST image of NGC 1386 (WFPC2/F606W-PC, proposal ID: 5446). The vertical dashed line marks the semi-major axis (PA $= 25^{\circ}$ east of north). \textit{Top center:} zoom-in on the box in the left panel showing the ``structure map" \citep{PoggeM02} corresponding to the observed field-of-view. The map was derived from the image in the left panel. The dashed line indicates the approximate position angle (PA $= 76^{\circ}$ east of north) and extension of a faint bar-like [OIII] emission observed by \citet{ferruit2000}. \textit{Top right:} continuum map as obtained from our data; labels mark spaxels from which the representative spectra plotted below have been extracted. \textit{Bottom:} representative spectra from individual spaxels A, C and N, as marked in the top right panel.}
\label{fig: NGC1386}
\end{figure*} 
% \end{minipage}

\begin{figure}[h]
$
\begin{array}{cc}
\includegraphics[trim=0cm 0.65cm 0cm 0cm, clip=true, scale=0.64]{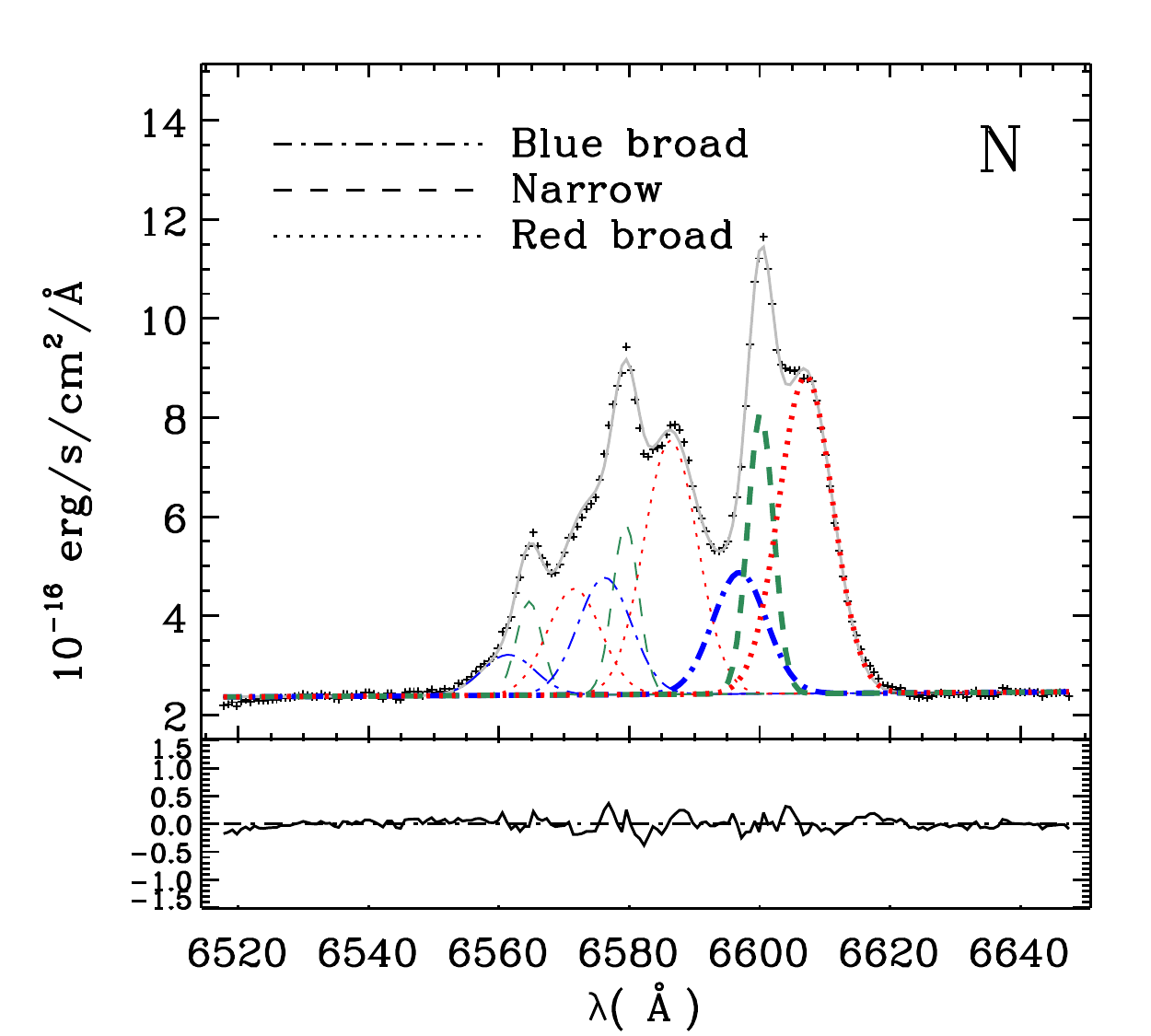}\\
\includegraphics[trim=0cm 0.65cm 0cm 0cm, clip=true, scale=0.64]{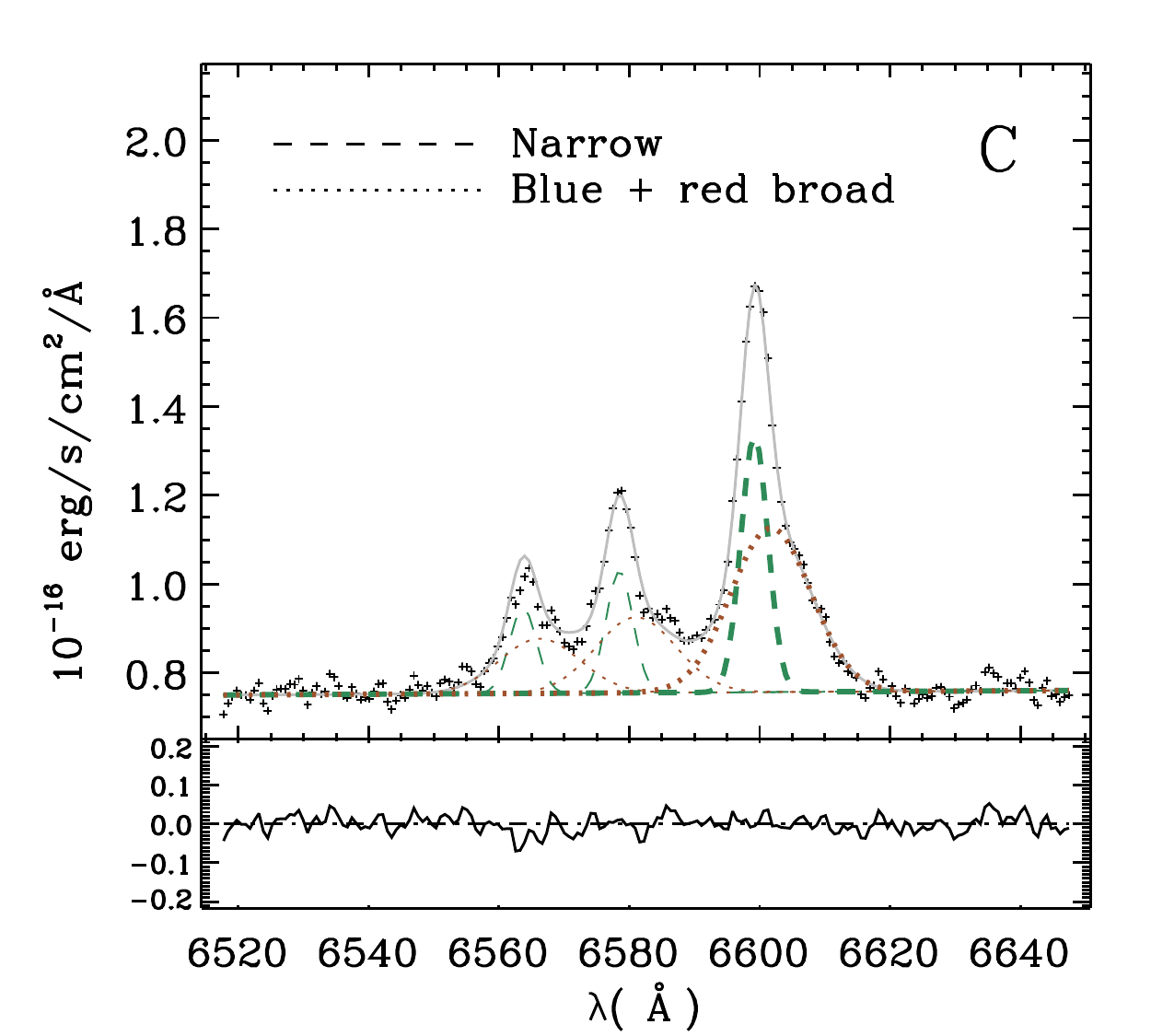}\\
\includegraphics[trim=0cm 0cm 0cm 0cm, clip=true, scale=0.64]{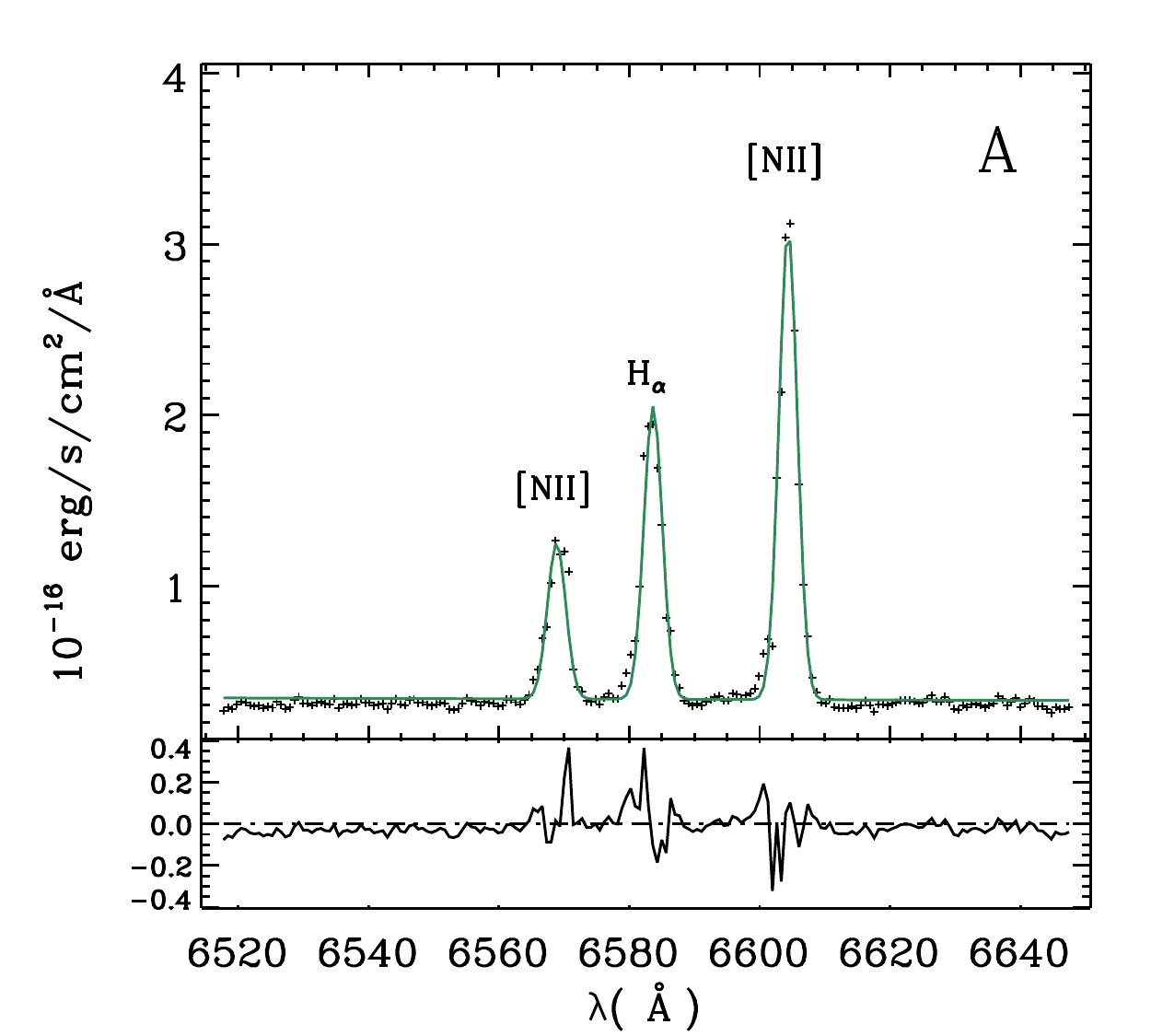} \\
\end{array}$
\caption{Gaussian fits to the representative spectra shown in Fig.\ref{fig: NGC1386}. \textit{Top:} three component fit. \textit{Middle:} two component fit. \textit{Bottom:} single component fit. Crosses show the data points and the grey solid lines show the summed fit, with dashed and dashed-dotted lines showing the individual components as labelled. The lower sub-panel in each figure shows the residuals of the fit.}
\label{fig: fits}
\end{figure}

% ===============
% ===============
\section{Observations} \label{sec: analysis}

NGC 1386 was observed on September 23, 2011 with the Integral Field Unit (IFU) of the Gemini Multi-Object Spectrograph \citep[GMOS;][]{AllingtonSEtAl2002, HookEtAl2004} mounted on the Gemini South Telescope (program ID: GS-2011B-Q23). Two adjacent fields were observed, each covering 7 $\times$ 5 arcsec$^{2}$, resulting in a total angular coverage of 7.5 $\times$ 9.5 arcsec$^{2}$ with 0\farcs2 sampling, centered on the nucleus and extended along the galaxy semi-major axis (position angle PA = 25$^{\circ}$). In the top panels of Fig.\ref{fig: NGC1386} we show the observed field-of-view (FOV) on an HST image, the corresponding ``structure map" created via the contrast-enhancement technique defined by \citet{PoggeM02}, and a continuum image obtained from our observation.

In order to account for dead fibers in the IFU, $\pm$ 0\farcs3 spatial dithering was applied along both axes while a 5 nm spectral dithering was applied to account for CCD gaps. This resulted in 12 exposures of 375 seconds each.
From measurements on the standard star, observed immediately after the galaxy, we determined that the seeing limited the spatial resolution to 0\farcs9 $\pm$ 0\farcs1, corresponding to 68 $\pm$ 8 pc at the galaxy distance.

In order to cover the wavelength range 5600-7000 \AA, which includes the emission lines [FeVII]$\lambda$6087, [OI]$\lambda$6302, H$\alpha$+[NII]$\lambda\lambda$6548,6583, [SII]$\lambda\lambda$6716,6731 and several stellar absorption features, we used the IFU in two-slits mode with the grating GMOS R400 in combination with the r(630 nm) filter, this yields a spectral resolution R = 1918 corresponding to a velocity dispersion $\sigma \approx $ 66 km s$^{-1}$.
A portion of the spectrum including the most prominent emission lines is shown in the bottom panel of Fig.\ref{fig: NGC1386} for some representative regions of the FOV.

Data reduction was performed using the IRAF\footnote{IRAF is the Image Reduction and Analysis Facility, a general purpose software system for the reduction and analysis of astronomical data. IRAF is written and supported by the National Optical Astronomy Observatories (NOAO) in Tucson, Arizona. NOAO is operated by the Association of Universities for Research in Astronomy (AURA), Inc. under cooperative agreement with the National Science Foundation.} packages provided by the GEMINI Observatory and specifically developed for the GMOS instrument\footnote{\url{http://www.gemini.edu/sciops/data-and-results/processing-software?q=node/11822}}. The data reduction process includes bias, overscan and sky subtraction, flat-fielding, trimming, wavelength and flux calibration, building of the data cubes at a sampling of 0.1 $\times$ 0.1 arcsec/pixel, final alignment and combination of the 12 cubes. The resulting cube contains 7030 spectra.

The typical accuracy in the wavelength calibration is 0.14 \AA\ (or 6 km s$^{-1}$ at $\lambda$ = 6583 \AA). Flux calibration is expected to be accurate to $\approx$ 10\%.

% ==========================================================================================
% ==========================================================================================
\section{Emission line fitting} 
\label{sec: line_fitting}

We used customized IDL\footnote{IDL, or Interactive Data Language, is a programming language used for data analysis and visualization.} routines to model the continuum and the profiles of the most prominent emission lines (H$\alpha$, [NII]$\lambda\lambda$6548,6583, [SII]$\lambda\lambda$6716,6731, [OI]$\lambda$6302 and [FeVII]$\lambda$6087).
Gaussian profiles were fitted to the lines in order to derive centroid velocities, velocity dispersions and fluxes.
Four separate runs were performed to fit the H$\alpha$ + [NII] lines, the [SII] doublet, the [OI] line and [FeVII].
Given the limited range in wavelength of the fitted spectral regions, the local continuum adjacent to the lines was modeled with a first order polynomial. 

Visual inspection of the spectra suggests that, besides [FeVII], emission line profiles within $\approx1^{\prime\prime}$ of the nucleus mostly result from the superposition of one relatively narrow component ($\bar \sigma \approx 90$ km s$^{-1}$) and two broader components ($\bar \sigma \approx 200$ km s$^{-1}$), see for example the [NII] and [SII] profile for spectrum N in Fig.\ref{fig: NGC1386}). Hereafter we will refer to the components as ``narrow" and ``broad" (note, however, that the broad components are not from the ``classical" broad line region of AGN where line widths exceed 1000 km s$^{-1}$). Beyond the central 2$^{\prime\prime}$ the lines are clearly unblended and single-peaked requiring only one Gaussian to be satisfactorily modeled. Between 1$^{\prime\prime}$ and 2$^{\prime\prime}$ the line profiles mostly present a narrow core and a broad base. We interpret the latter as due to the blending of the two broad components and we fit these lines with two Gaussians: one for the narrow core and one for the broad base. Note, however, that for a fraction of the spaxels in this region the spectra can still be successfully modeled with three Gaussians.

In order to fit the H$\alpha$+[NII] lines we assumed, for each component of the fit (i.e. the narrow and the two broad components), that H$\alpha$ and [NII]$\lambda$6548 have the same redshift and width as [NII]$\lambda$6583 (the strongest line). 
Therefore, the velocity and velocity dispersion of each component of [NII]$\lambda$6548 and H$\alpha$ are the same as those of the corresponding components of [NII]$\lambda$6583. The amplitude of each [NII]$\lambda$6548 component was fixed at the theoretical value of 1/2.96 of that of the corresponding [NII]$\lambda$6583 amplitude. A similar approach was used to fit the [SII] doublet.

A simpler double peaked profile gives a better representation of the [FeVII] line which, as it lacks a detectable narrow component, was fitted with two Gaussians of the same width. Because of its high ionization potential ($E = 125$ eV, \citealt{CorlissSugar82}), this line is believed to be produced by a combination of hard UV continuum and hot collisionally ionized plasma \citep{ContiniPV98}. Given the high energies required for its production, the [Fe VII] line is believed to originate close to the AGN (a few tens of pc, \citealt{FergusonKBF97}). 
\vskip10pt

\textit{Three Gaussians fit}: 
in order to determine the region over which the three Gaussian fit is necessary, we require that for each component the line amplitude is at least twice the standard deviation in the continuum adjacent to the line. We also assume that the two broad components have the same width (without this condition, the fit would be be underconstrained in many spaxels, where the profiles are heavily blended).
With this approach we fit three Gaussian components to each of H$\alpha$, [NII] and [SII] mostly within a radius of $\approx1^{\prime\prime}$ from the nucleus, but also at larger distances, between 1$^{\prime\prime}$ and 2$^{\prime\prime}$ (the nucleus, denoted N, is identified with the continuum brightness peak). Due to the lower S/N, three Gaussians were fitted to [OI] only for a handful of pixels within $\approx$ 0\farcs7 from the center. 
An example of the three components fit is shown in the top panel of Fig.\ref{fig: fits} for [NII] and H$\alpha$.

\textit{Two Gaussians fit}: broad components are still visible as a broad base mainly between 1$^{\prime\prime}$ - 2$^{\prime\prime}$ from the nucleus. However, for many pixels the two components are blended and the S/N is insufficient to justify fitting three components. Therefore in this region we mostly fit two, representing the broad base and the narrow core, respectively. 
An example of a two-components fit is shown in the middle panel of Fig.\ref{fig: fits}. This fit was performed with the goal of extracting the narrow component.

\begin{figure}[t]
\center
\includegraphics[trim=6cm 1.5cm 4cm 0cm, clip=true, scale=0.67]{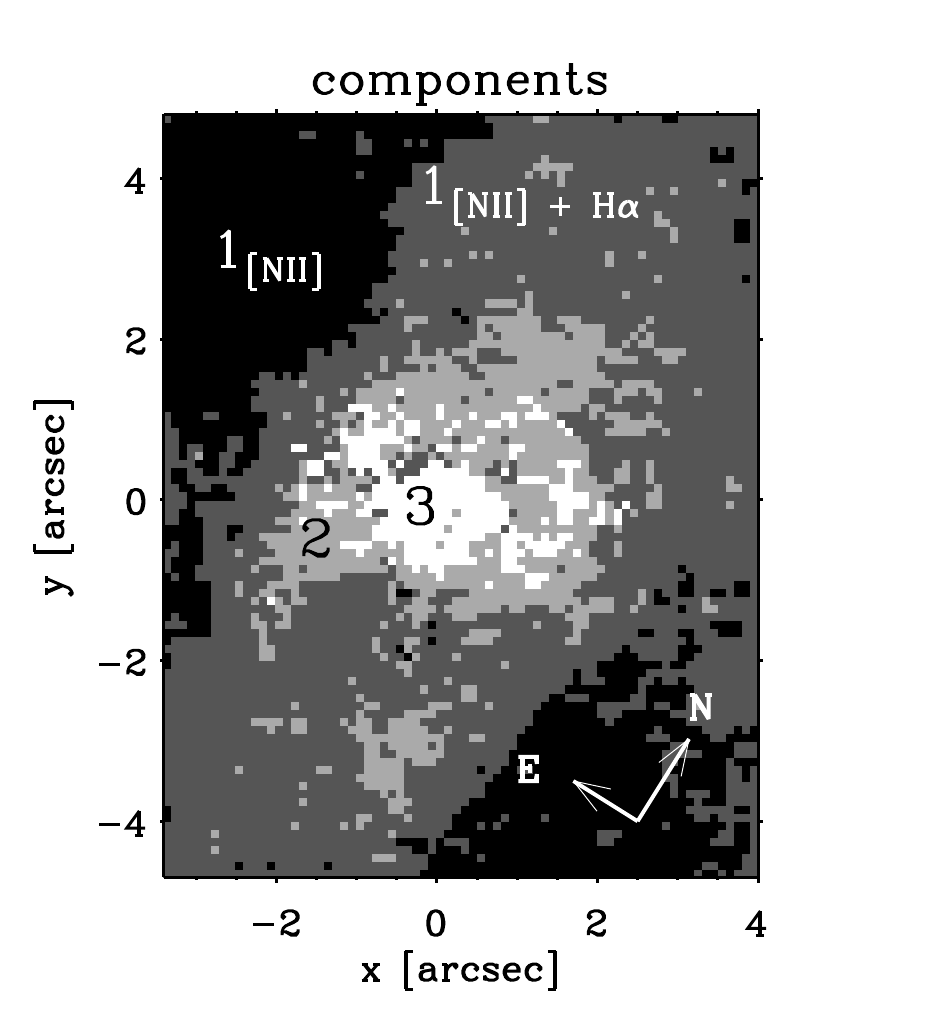}
\caption{Map showing the regions where one, two or three Gaussians were fitted to the emission lines. No significant H$_{\alpha}$ emission was detected in the darker region labelled as ``1$_{[NII]}$".}
\label{fig: components}
\end{figure}

\textit{Single Gaussian fit}: a single Gaussian fit was used beyond the central 2$^{\prime\prime}$ where the broad components become negligible and the two-components fit failed.
An example of single component fit is shown in the bottom panel of Fig.\ref{fig: fits}. 
A map showing the number of components fitted to the spectra in a given spaxel is shown in Fig.\ref{fig: components}.

\vskip10pt

\begin{figure}[p]  
\begin{center}$
\begin{array}{c}
\includegraphics[trim= 1cm 0cm 0cm 0cm, clip=true, scale=0.66]{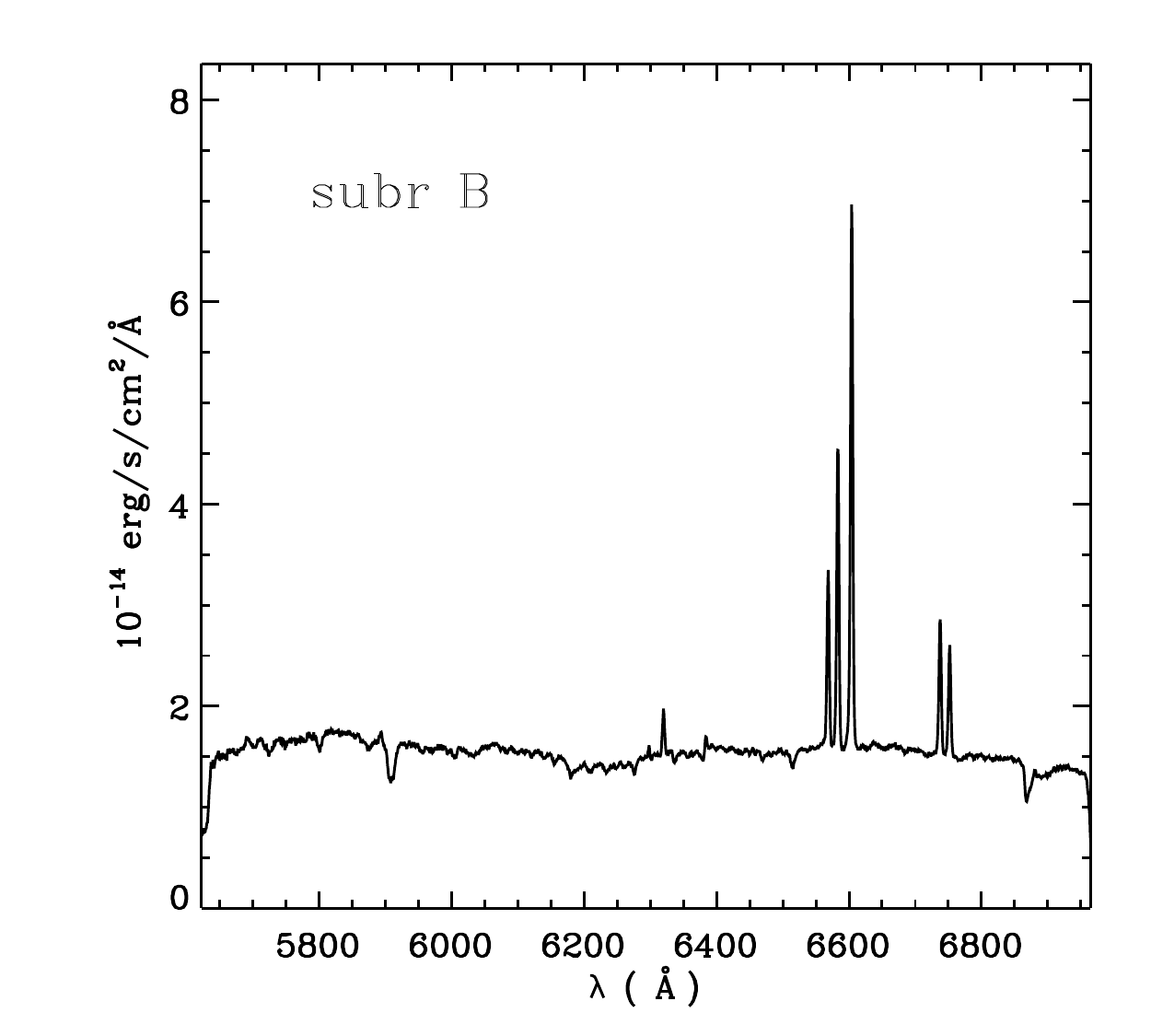} \\
\includegraphics[trim= 0.85cm 0cm 0cm 0.5cm, clip=true, scale=0.66]{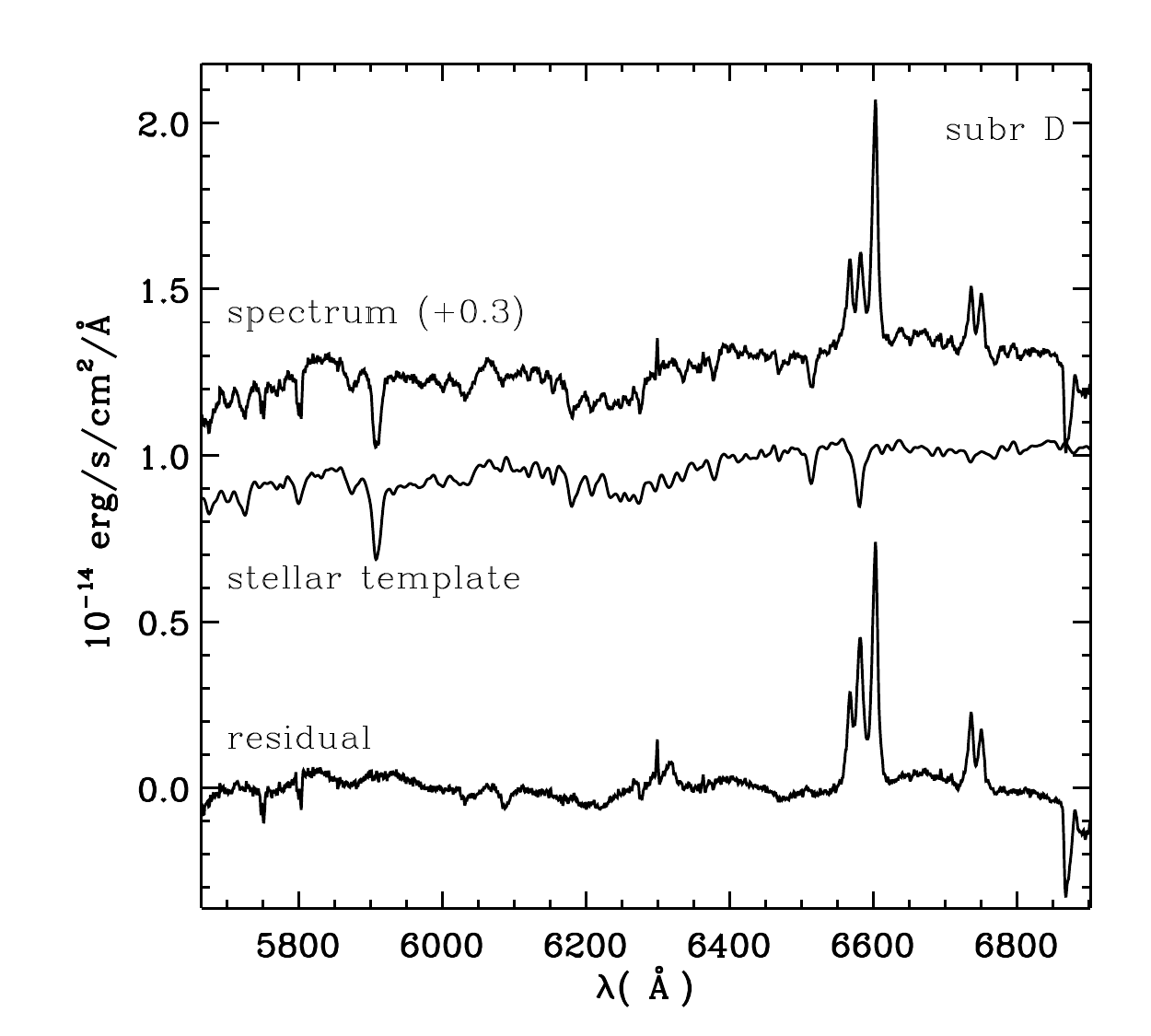} \\
\includegraphics[trim= 1cm 0.3cm 0cm 0.5cm, clip=true, scale=0.66]{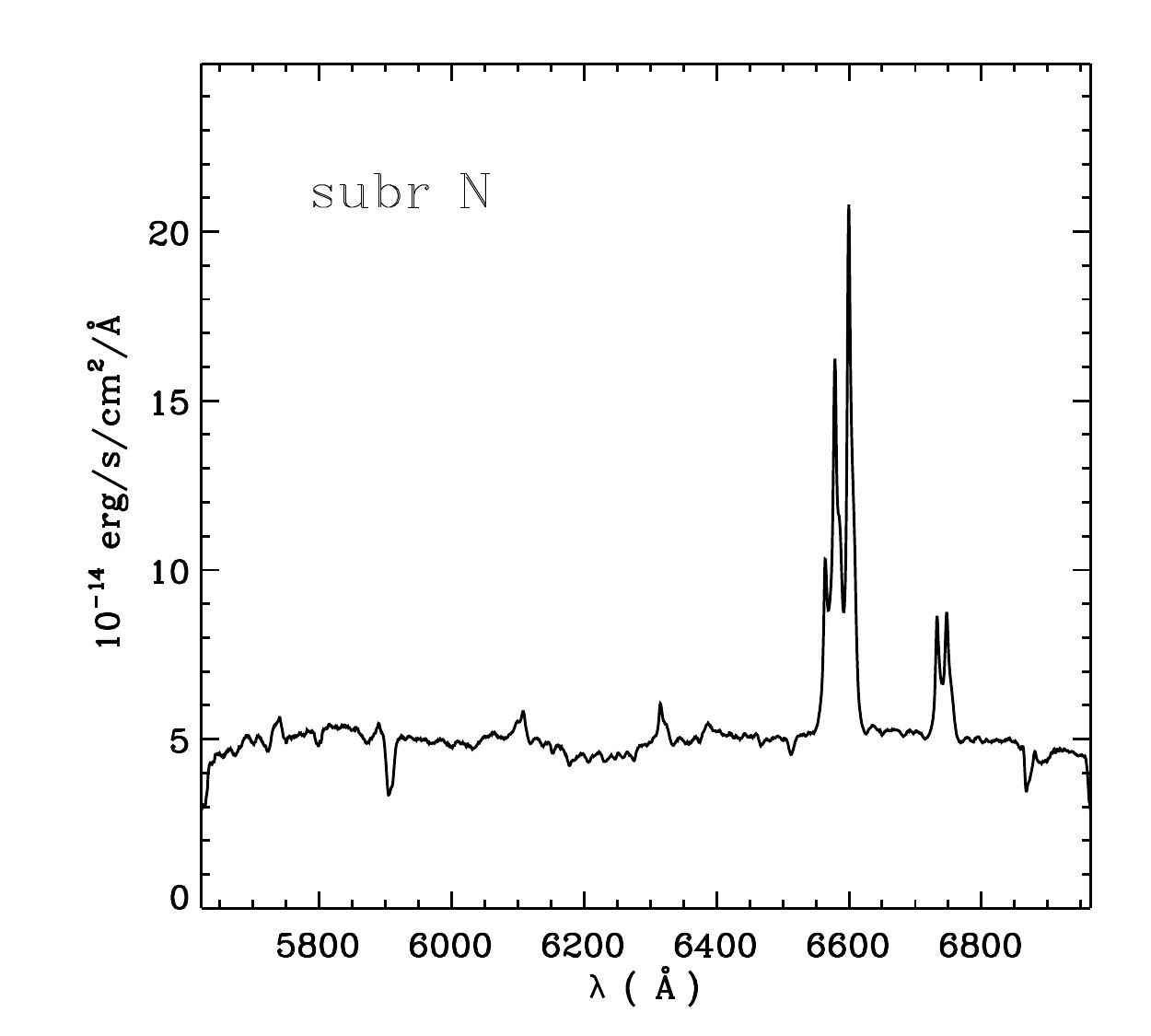} \\
\end{array}$
\end{center}
\caption{Spectra extracted from the subregions specified in Fig.\ref{fig: [NII]maps}. In order to correct for H$\alpha$ absorption, a stellar template has been fitted and subtracted from subregion D.}
\label{fig: subr_spec}
\end{figure}

To determine accurate line intensity ratios representative of the various kinematic components, five subregions of the FOV were identified (see velocity map for the narrow and broad blue component in Fig.\ref{fig: [NII]maps}). The subregions represent, respectively, the northern and southern ``lobes" (subregions A and B), regions associated with velocity residuals approximately perpendicular to the axis of the lobes (see \textsection \ref{subsec: gasv_model} for details), east and west of the nucleus (C and D), and the nucleus itself (N). The spectra obtained for sub-regions B, D and N are presented in Fig.\ref{fig: subr_spec}; the spectra for subregions A and C are very similar to those of subregion B and D respectively. 

The continuum and all the most prominent emission lines ([OI], H$\alpha$, [NII], [SII]) were fitted simultaneously with the constraint that all the lines are characterized by the same redshift and the same width as the [NII]$\lambda$6583 line. With these assumptions we fitted single Gaussians to the emission lines in the subregion B spectrum and two Gaussians to each line in subregion A. This was done to take into account the non-Gaussian base of the lines. In this case, the total flux of each line was considered to be the sum of the fluxes of the two components.
Line emission is weak relative to the stellar continuum in subregions C and D, therefore stellar templates were fitted and subtracted. The templates, a subset of the models presented in \cite{Vazdekis99}, represent an old stellar population of spectral type O-M and metallicity -0.7 $\leq$ [Fe/H] $+0.2$. 
After this procedure, line profiles in subregion C and D were fitted with the same approach described for subregion A.  
The profiles of subregion N, which includes the nucleus, were fitted with three components: a narrow core and two broader Gaussians characterized by the same width. In order to minimize contamination from H$\alpha$, the blueshifted broad component was fitted to the blue side of the [NII]$\lambda$6548. 

The stellar continuum was not removed prior to fitting the subregion A, B and N spectra as this did not produce any significant difference in the results.

% Subregions 1 - 6: ..............................................

% [NII] single components fit: ==================
\begin{figure*}[t]
\begin{center}$
\begin{array}{ccc}
\includegraphics[trim=6cm 1.7cm 4cm 0.5cm, clip=true, scale=0.67]{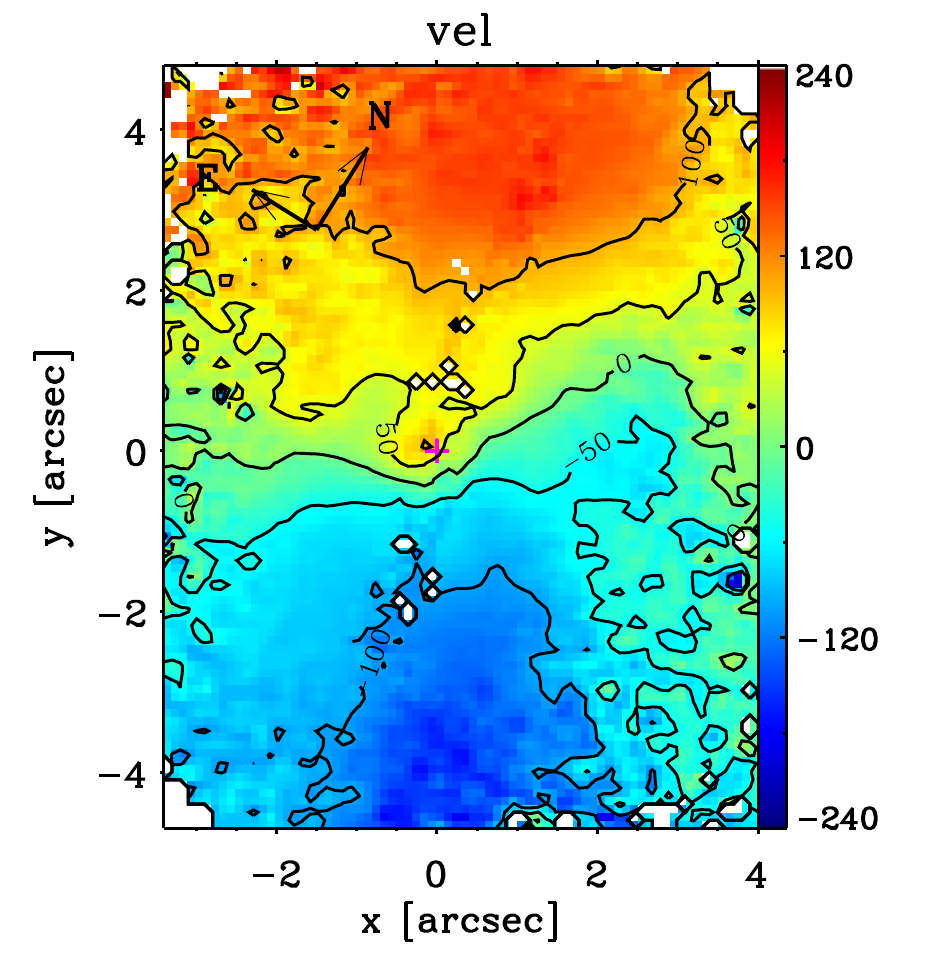} 	& \includegraphics[trim=7cm 1.7cm 4cm 0.5cm, clip=true, scale=0.67]{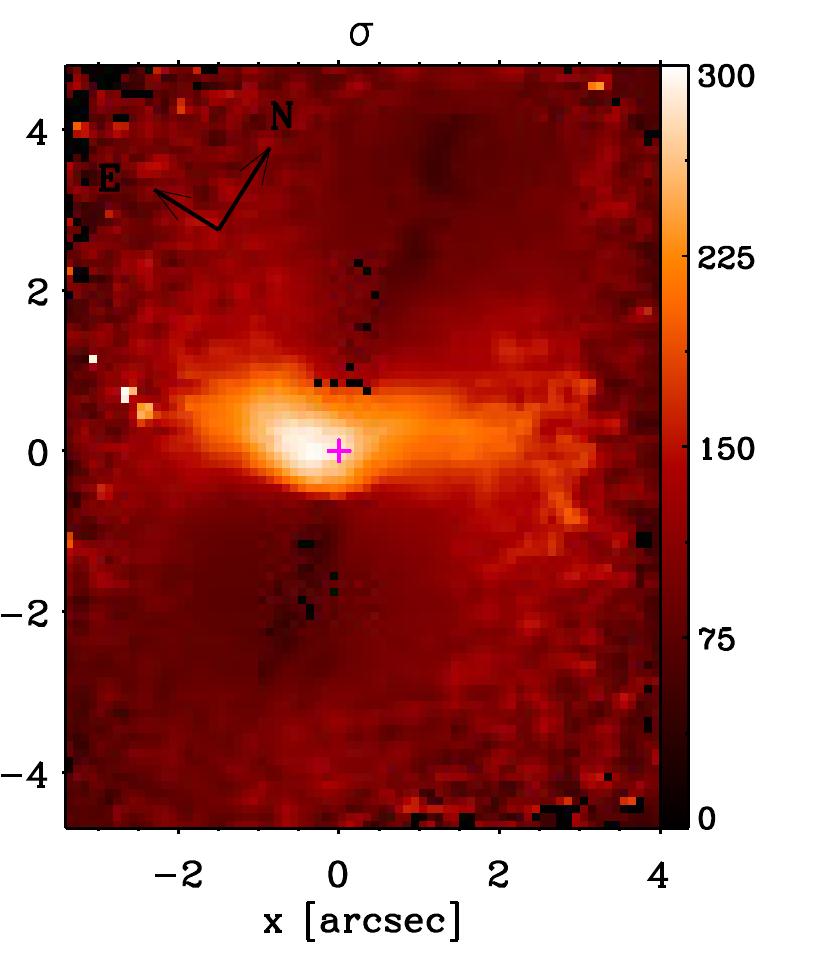} & \includegraphics[trim=7cm 1.7cm 4cm 0.5cm, clip=true, scale=0.67]{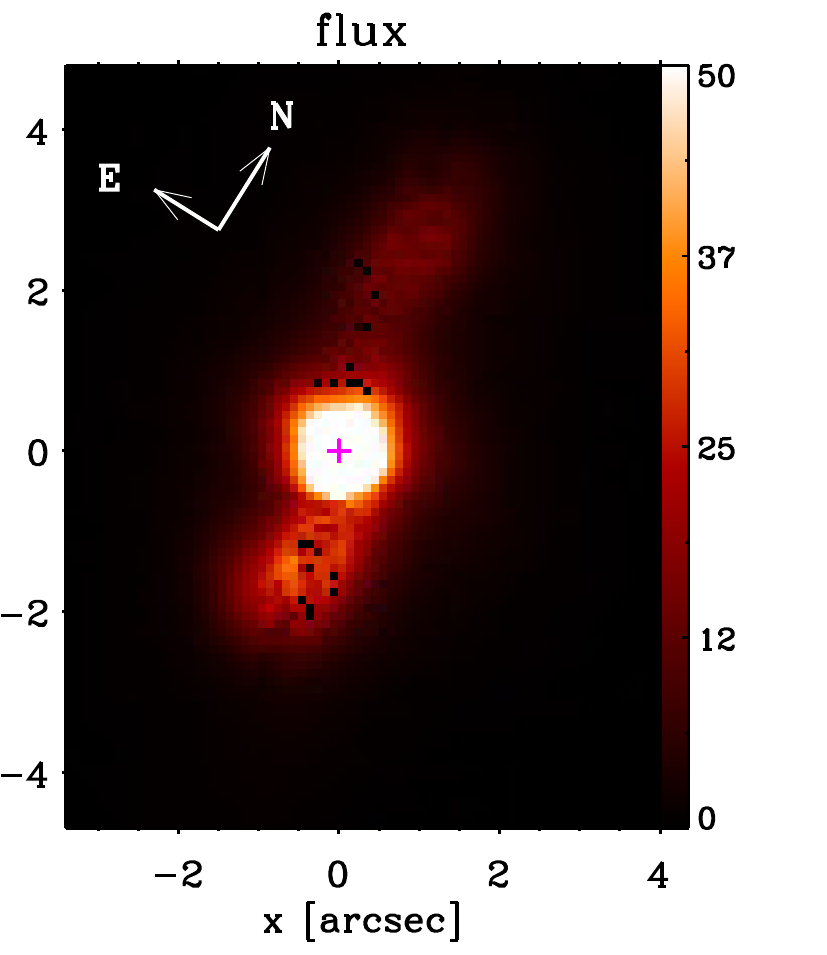}
\end{array}$
\end{center}
\caption{Velocity [km s$^{-1}$], velocity dispersion [km s$^{-1}$] and flux [10$^{-16}$ erg/s/cm$^{2}$/spaxel] as derived by fitting a single Gaussian to the [NII]$\lambda$6583 emission line. From the flux map it is evident that the strongest emission is mainly concentrated in an unresolved region within 1$^{\prime\prime}$ from the nucleus and along the north-south direction. Low-level emission is nevertheless present over the entire FOV. A systemic velocity $v_\mathrm{sys}$ = 786 km s$^{-1}$ was derived from the modeling of the [NII] kinematics and was subtracted from the velocity map. The magenta cross marks the continuum peak.}
\label{fig: [NII]1c}
\end{figure*}

% ============================================================
% ============================================================
\subsection{Uncertainties} \label{sec: uncertainties}

Errors on the measured quantities were derived from Monte Carlo simulations. For each spaxel, we constructed one hundred realizations of the spectrum by adding Gaussian noise with amplitude comparable to the noise measured in the original spectrum. Mean values and standard deviations for the centroid velocities, velocity dispersions and fluxes were derived for each spaxel, with the standard deviation of the distribution in each parameter being taken as the uncertainty.

% ===============
% ===============
\section{Results} 
\label{sec: results}
\subsection{Gas kinematics}
\label{subset: gask}
\subsubsection{Velocity and velocity dispersion}

As [NII]$\lambda$6583 has the highest S/N, we use results obtained from the fits to this line to represent and model the gas velocity field. The velocities recovered from other emission lines are consistent within the errors. 

We first show maps obtained by fitting a single Gaussian to the [NII] line throughout the entire FOV. Centroid velocity and velocity dispersion maps obtained with this approach are presented in Fig.\ref{fig: [NII]1c} after the subtraction of a systemic velocity $v_\mathrm{sys}$ = 786 km s$^{-1}$ determined from the modeling of the gas kinematics (see \textsection\ref{subsec: gasv_model} for details). The velocity map shows the characteristic pattern of a rotating disk upon which other components are superposed. There are two clear deviations from the rotation pattern: (i) a redshifted blob located at the nucleus and (ii) a blueshifted region to the north-west of the nucleus. Multicomponent fits and visual inspection of the corresponding spectra show that the nuclear redshifted blob is associated with a strong redward asymmetry in the line profile, see spectrum N in Fig.\ref{fig: NGC1386}. The blueshifted region north-west of the nucleus persists in maps obtained with multi-Gaussian fits and is discussed in \textsection\ref{subsec: gasv_model}. 

The velocity dispersion map shows very clearly that the strongest line broadening occurs in the nuclear region and is roughly centered on the nucleus. A narrow band of high velocity dispersion is also observed either side of the nucleus, extending roughly 2$^{\prime\prime}$ to the south-east, 3$^{\prime\prime}$ to the north-west and 0\farcs5 above and below it. This band is approximately perpendicular to the axis of the extended line emission, which is brightest in two lobes that are observed north and south of the nucleus (Fig.\ref{fig: [NII]1c}). The greatest broadening, approximately 300 km s$^{-1}$, is observed $\approx$ 0\farcs5 east of the nucleus. The bright lobes of line emission coincide with regions of relatively low velocity dispersion with the smallest values ($\sigma \approx$ 50 km s$^{-1}$) occurring approximately along the direction north-south (roughly coincident with the lobes axis). 

\begin{figure*}[p]
\begin{center}$
\begin{array}{ccc}
\includegraphics[trim=6cm 1.5cm 4cm 0cm, clip=true, scale=0.67]{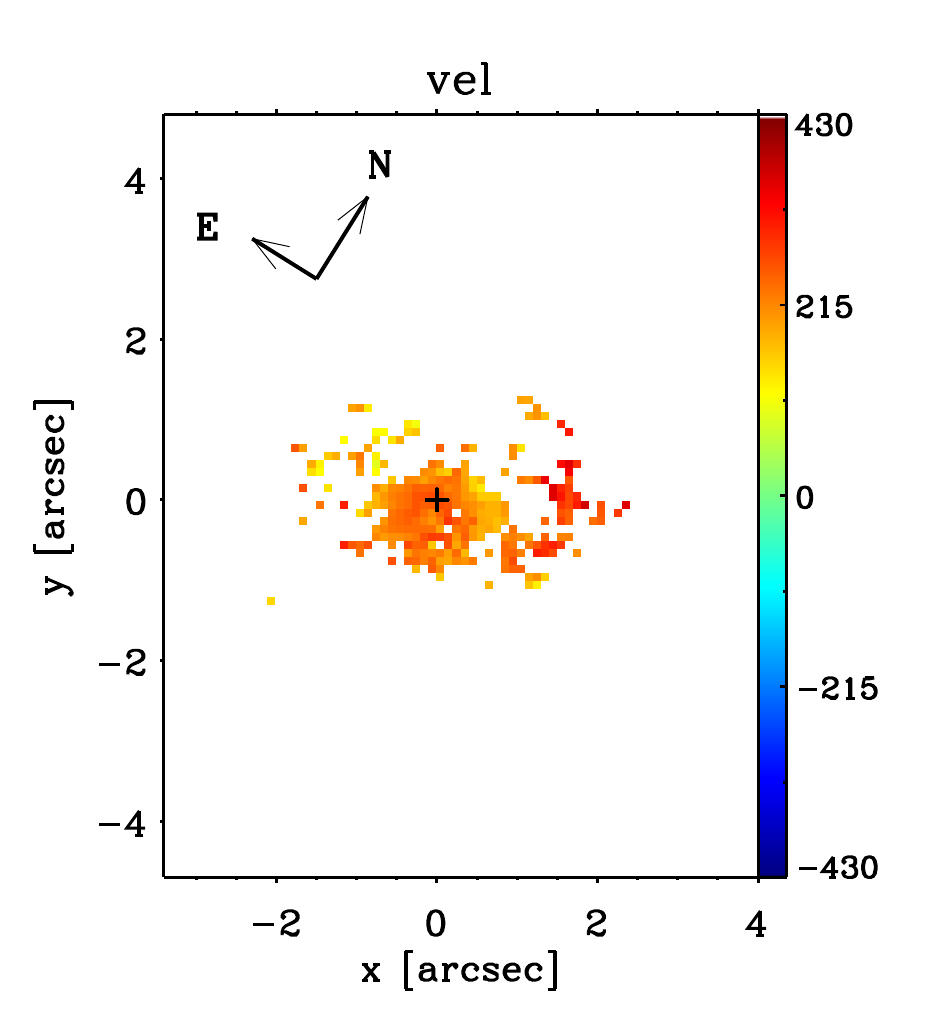} 	&  \includegraphics[trim=7cm 1.5cm 4cm 0cm, clip=true, scale=0.67]{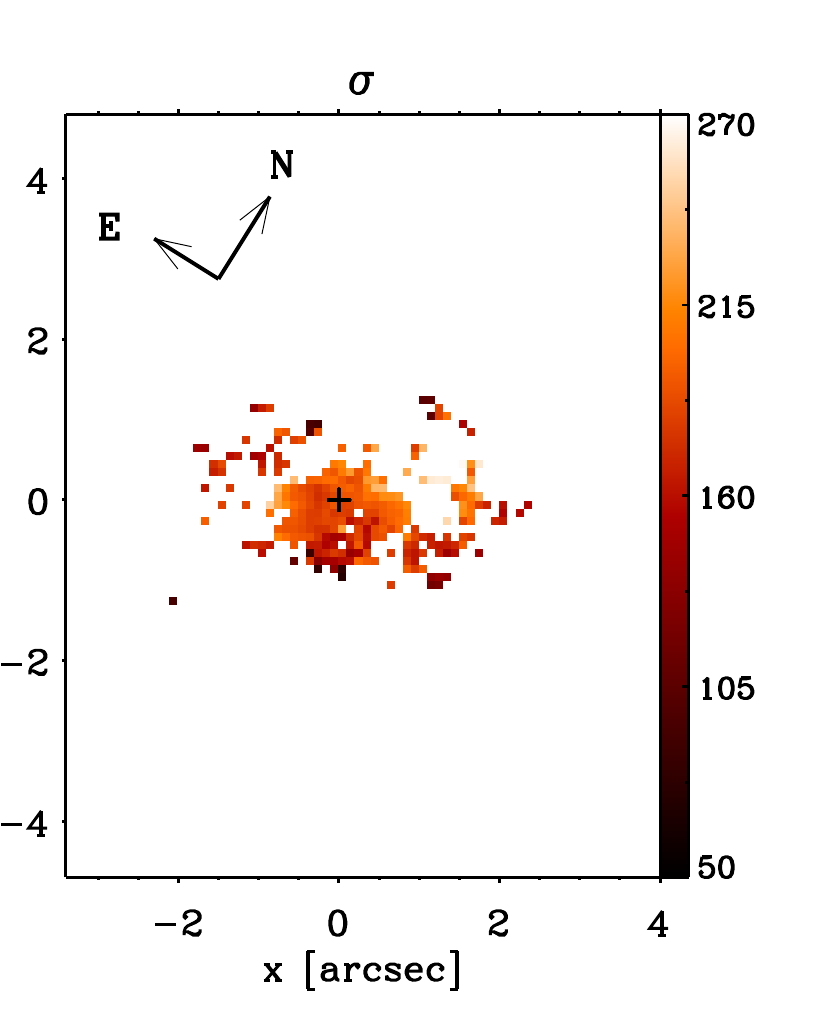}	& \includegraphics[trim=7cm 1.5cm 4cm 0cm, clip=true, scale=0.67]{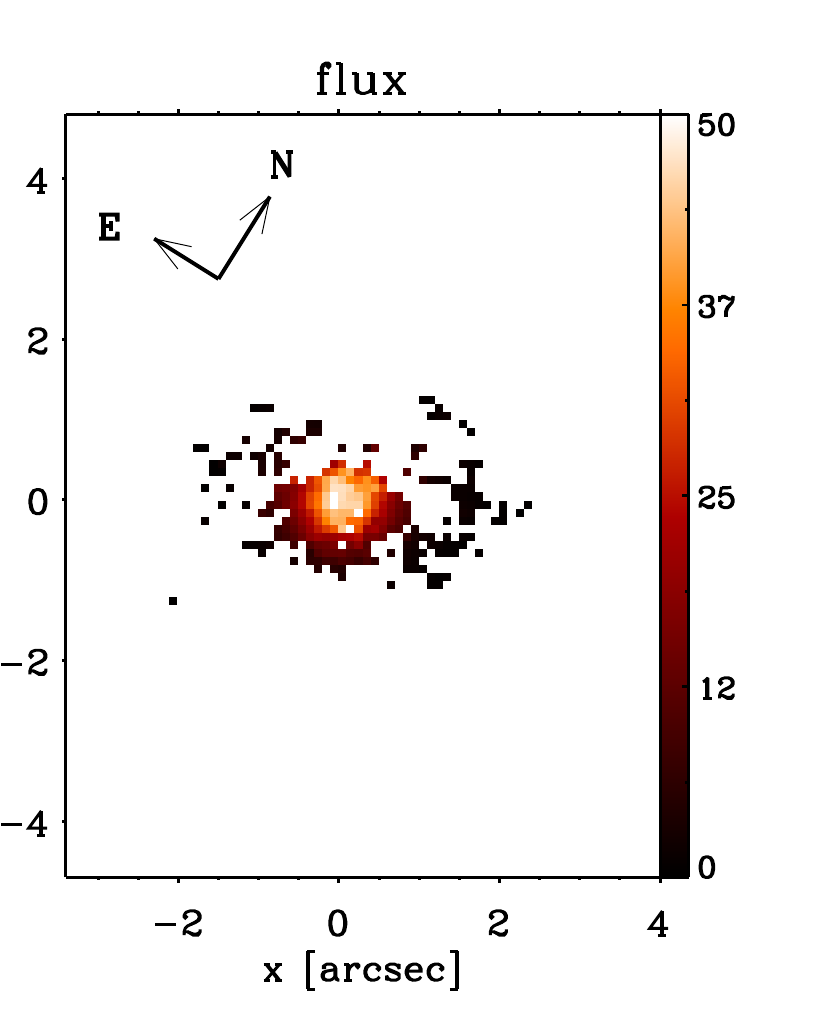} 	\\
\includegraphics[trim=6cm 1.5cm 4cm 1cm, clip=true, scale=0.67]{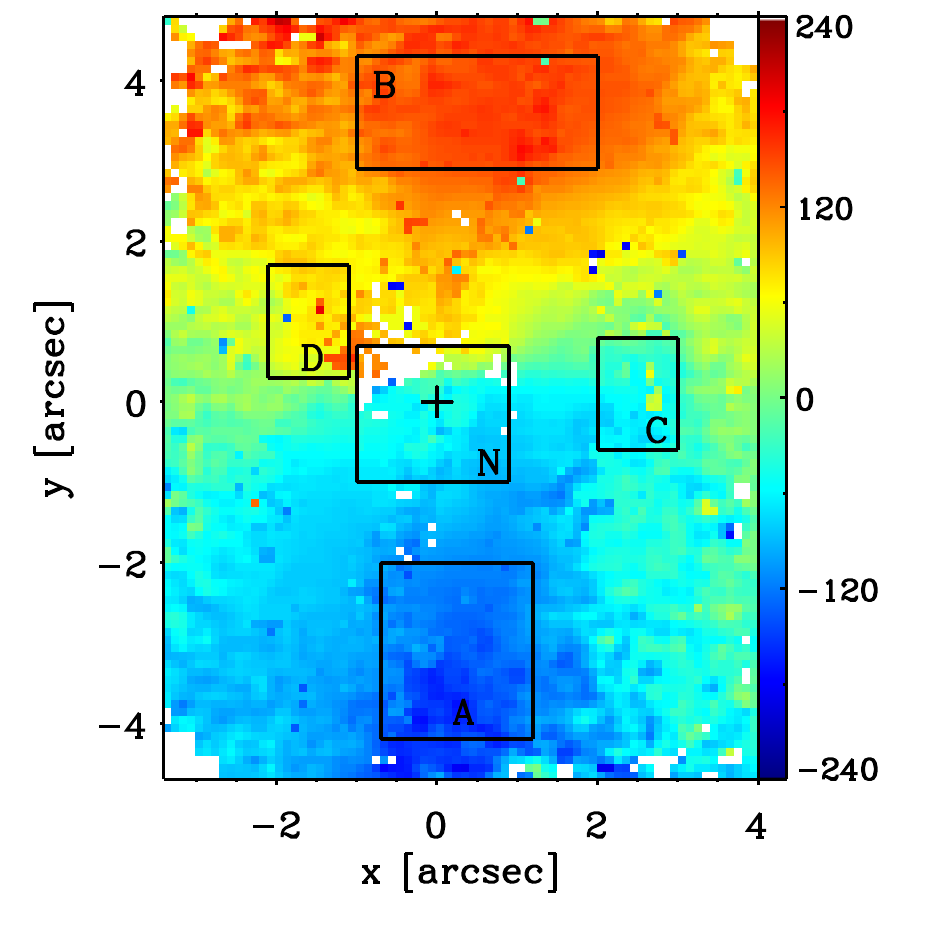} 	& \includegraphics[trim=7cm 1.5cm 4cm 1cm, clip=true, scale=0.67]{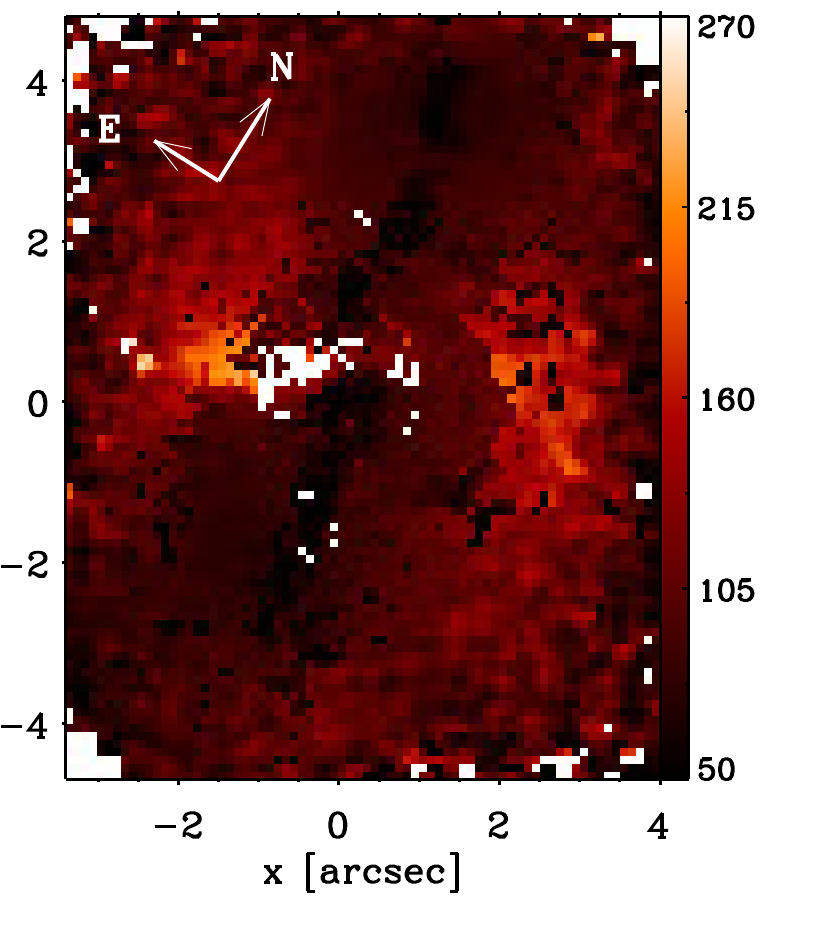}  &  \includegraphics[trim=7cm 1.5cm 4cm 1cm, clip=true, scale=0.67]{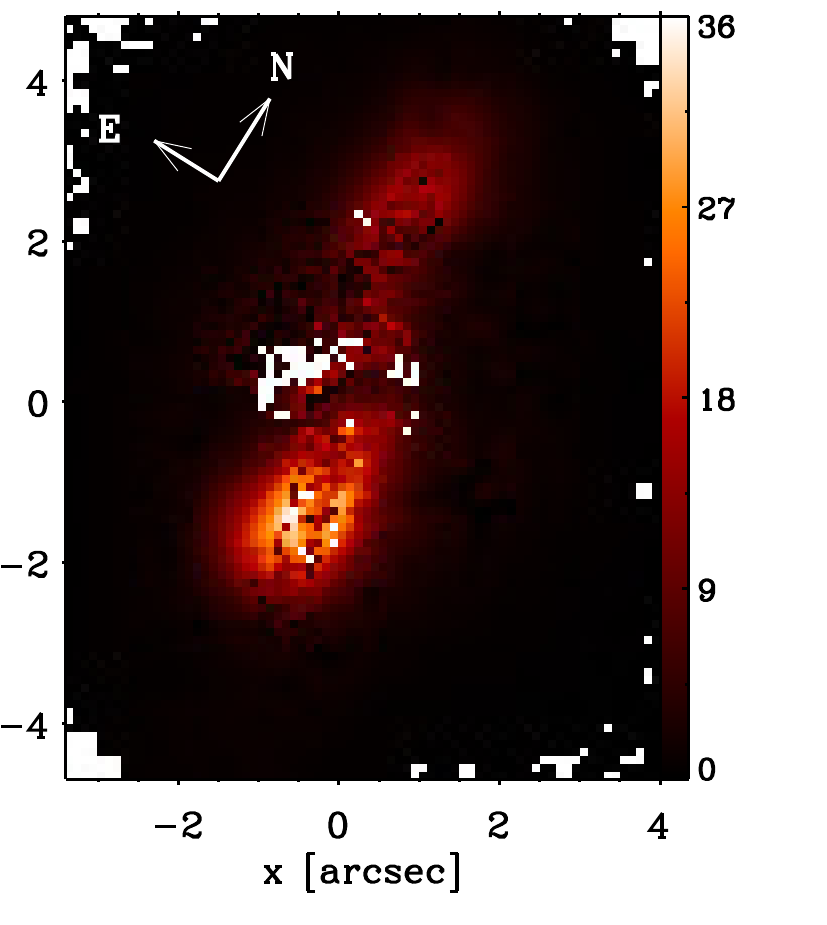}	\\
\includegraphics[trim=6cm 1.5cm 4cm 1cm, clip=true, scale=0.67]{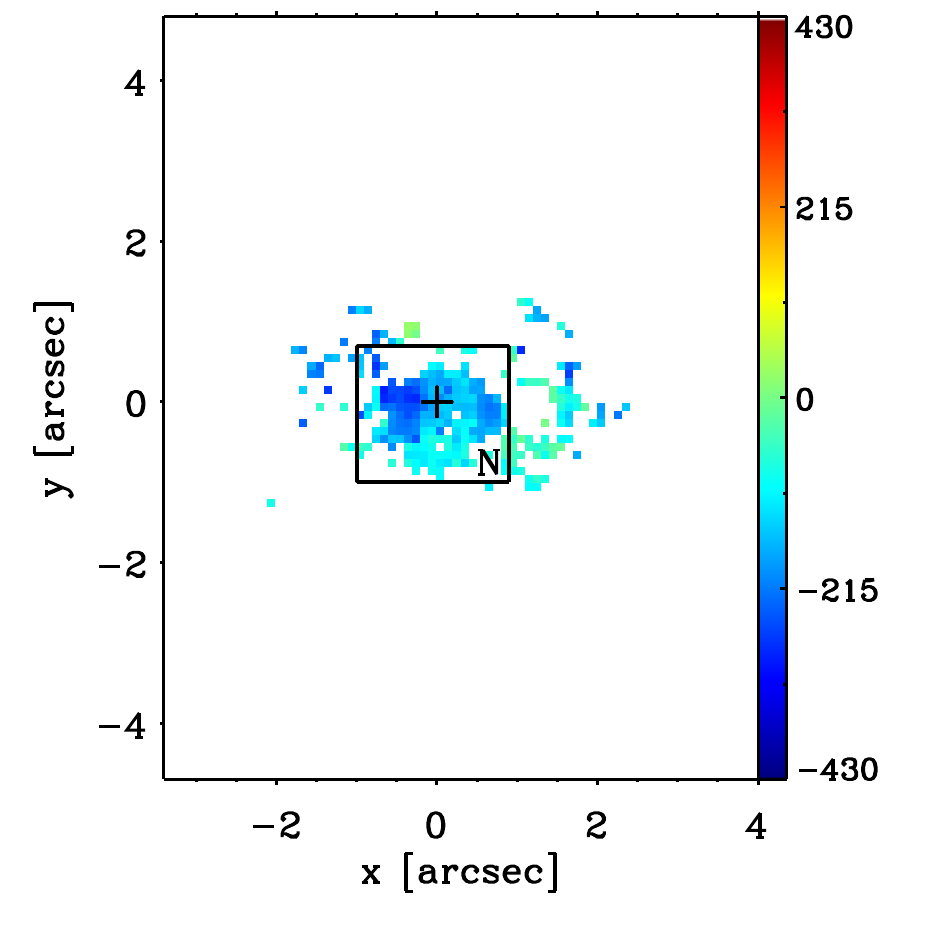} 	& \includegraphics[trim=7cm 1.5cm 4cm 1cm, clip=true, scale=0.67]{sig_C_ref01}	 & \includegraphics[trim=7cm 1.5cm 4cm 1cm, clip=true, scale=0.67]{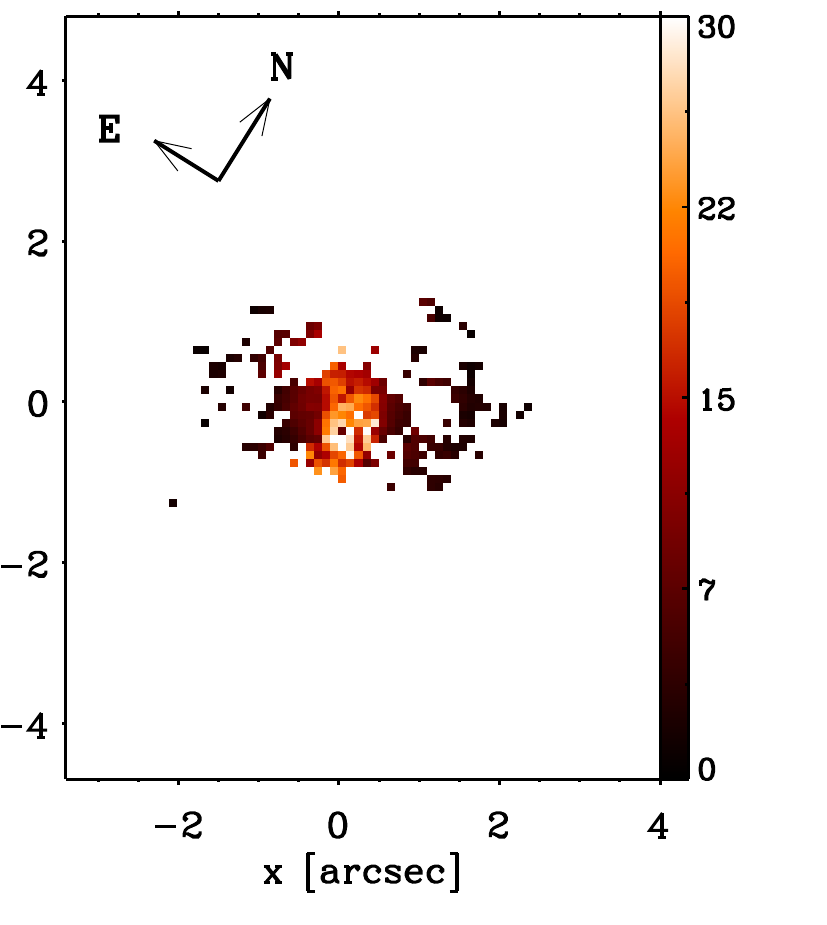}	\\
\end{array}$
\end{center}
\caption{[NII]$\lambda$6583 maps. \textit{Left:} velocity [km s$^{-1}$]. \textit{Middle:} velocity dispersion [km s$^{-1}$]. \textit{Right:} flux [10$^{-16}$ erg/s/cm$^{2}$/spaxel]. \textit{Top:} redshifted broad component. \textit{Center:} narrow component. \textit{Bottom:} blueshifted broad component. Though the narrow component flux map shows the brightest emission to be localized along the north-south direction, low-level emission is present over the entire FOV. Given our fitting constraints, the velocity dispersion maps of the blue- and red-shifted broad components are the same. Labelled boxes represent the subregions that were rebinned to extract the representative spectra shown in Fig.\ref{fig: subr_spec}. As evident from the bottom left panel, subregion N is representative of the spaxels where the broad components dominate. A systemic velocity $v_\mathrm{sys}$ = 786 km s$^{-1}$ was derived from the modeling of the [NII] kinematics and was subtracted from the velocity maps. Maps in the central row are a composite of the narrow component derived from three-, two- and single-Gaussian fit. Maps corresponding to the broad component of the two-Gaussians fit are not shown as the broad component in this case is believed to represent a blend of the blue- and red-shifted broad components included in the three-Gaussians fit. The broad component was included in the two-Gaussians fit only to permit accurate fitting of the narrow component. The black cross at (0,0) marks the continuum peak.}
\label{fig: [NII]maps}
\end{figure*}

Centroid velocities and velocity dispersions for the narrow and the broad components derived from multi-Gaussian fits to the [NII] profiles are shown in Fig.\ref{fig: [NII]maps}. The centroid velocity, velocity dispersion and flux maps for the narrow component ($\sigma\approx 90$ km s$^{-1}$) are derived from a combination of the three-, two- and single-Gaussian fits. The velocity map is very similar to that derived from the single-Gaussian fit alone, the main difference being the absence of the redshifted blob at the nucleus. The blueshifted region north-west of the nucleus is still apparent.

The centroid velocity, velocity dispersion
and flux maps for the broad components ($\sigma\approx 200$ km s$^{-1}$) 
are derived entirely from the three-Gaussian fit\footnote{The broad component used
in the two-Gaussian fit is not used; it may represent blending of red- and blue-shifted broad components.}. 
The red- and blue-shifted broad components are approximately co-spatial and dominate the line
emission within the central $\approx 1^{\prime\prime}$. However, it is notable that the brightest emission
in the red-shifted broad component is shifted slightly ($\approx 0\farcs5$) north of the continuum
brightness peak. The blue-shifted broad component is weaker overall, with the brightest
emission slightly south-west ($\approx 0\farcs5$) of the continuum peak.  
Evidence for line splitting (in the sense that two broad components are required to fit the line profiles)
in scattered spaxels extends well beyond the bright central region. 
In fact, weak double broad components can be found out to a distance of about 2$^{\prime\prime}$ east and west of the nucleus, that is, in a similar orientation to that of the region of enhanced velocity dispersion derived from the single component fit.

The H$\alpha$, [SII] and [OI] lines are characterized by the same features observed in [NII], i.e. two red- and blue-shifted broad components concentrated in the nuclear region and strong narrow-lines along the north-south direction. 

The high velocity dispersion present in the narrow component map along the direction SE-NW, around the points (-1\farcs5, 0\farcs5) and (2\farcs5, 0$^{\prime\prime}$) in the central panel of Fig.\ref{fig: [NII]maps}, can be attributed to the blending of the broad and narrow components all the way to the edge of the FOV. Visual inspection, see for example spectrum C in Fig.\ref{fig: NGC1386}, suggests a substantial decrease in the intensity of the narrow core with respect to the broad base. Therefore, in that region, a single Gaussian fit would be mainly tracing the velocity dispersion of the broad base.  
Despite a number of experiments, it was not possible to perform a fit producing a smooth transition in the velocity dispersion map and a perfect separation of the components.
\vskip10pt

\begin{figure*}[p]
\begin{center}$
\begin{array}{ccc}
\includegraphics[trim=6cm 1.5cm 4cm 0cm, clip=true, scale=0.65]{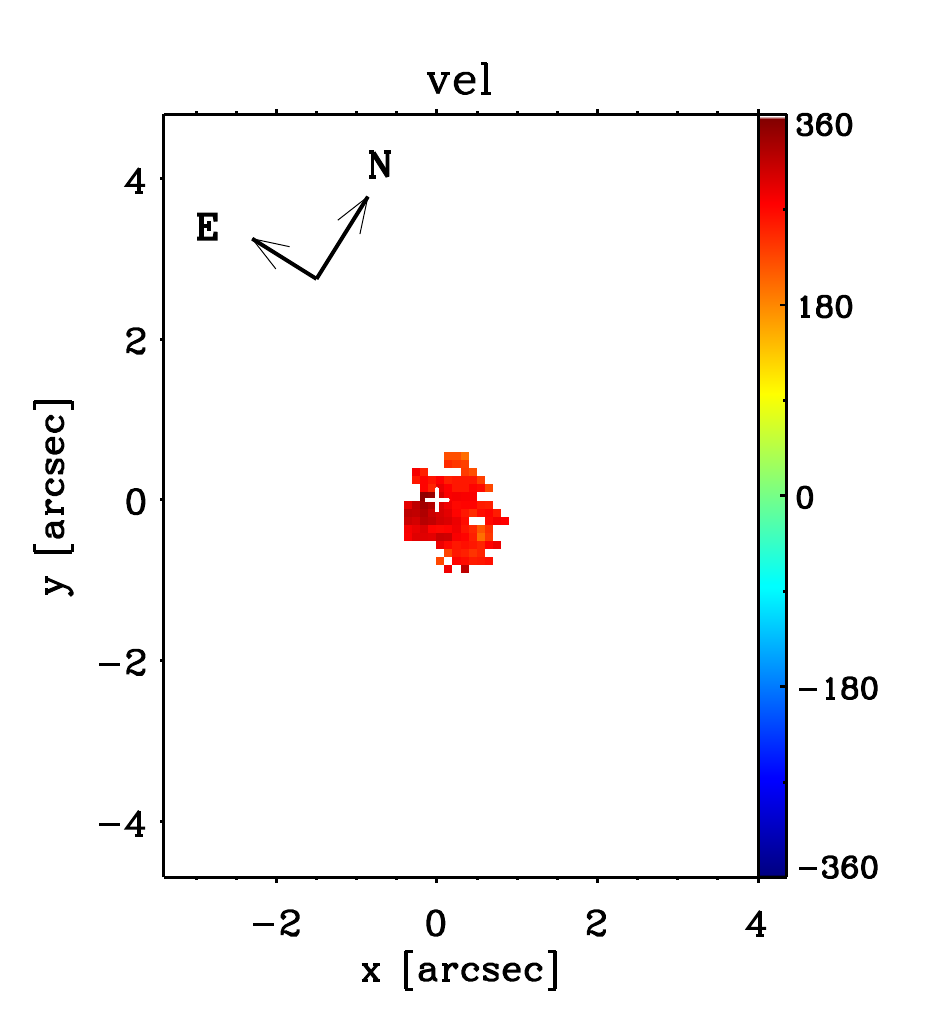}	&	\includegraphics[trim=7cm 1.5cm 4cm 0cm, clip=true, scale=0.65]{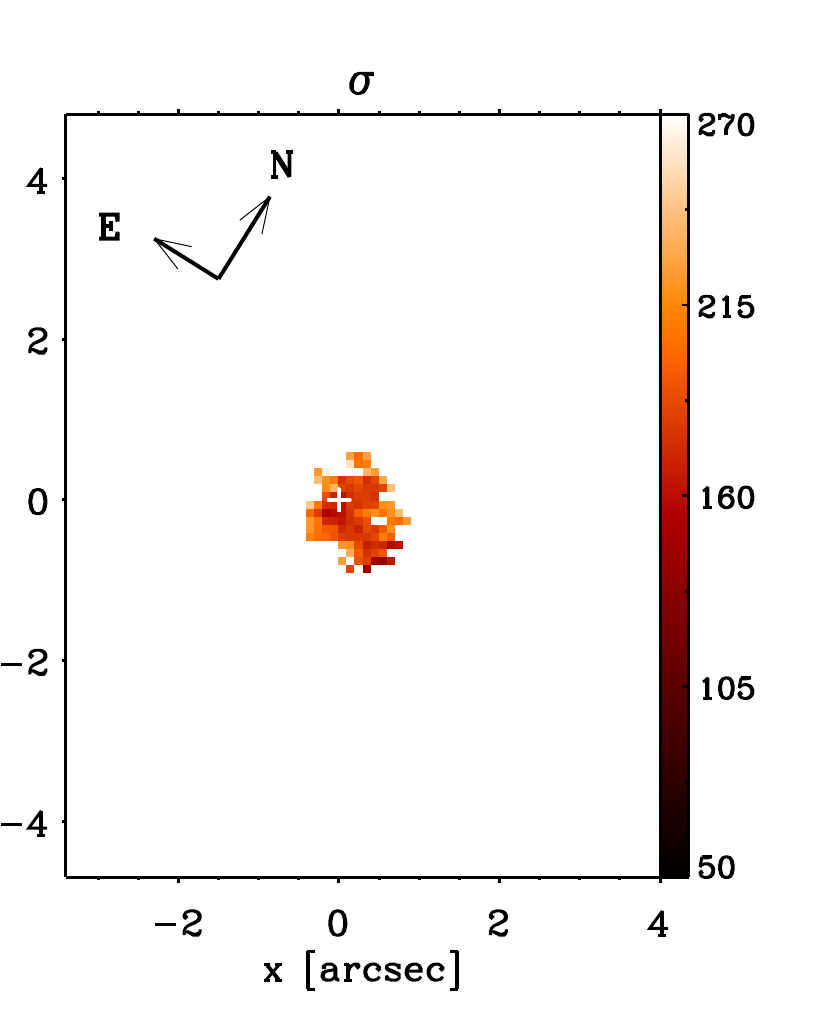} 	& 	\includegraphics[trim=7cm 1.5cm 4cm 0cm, clip=true, scale=0.65]{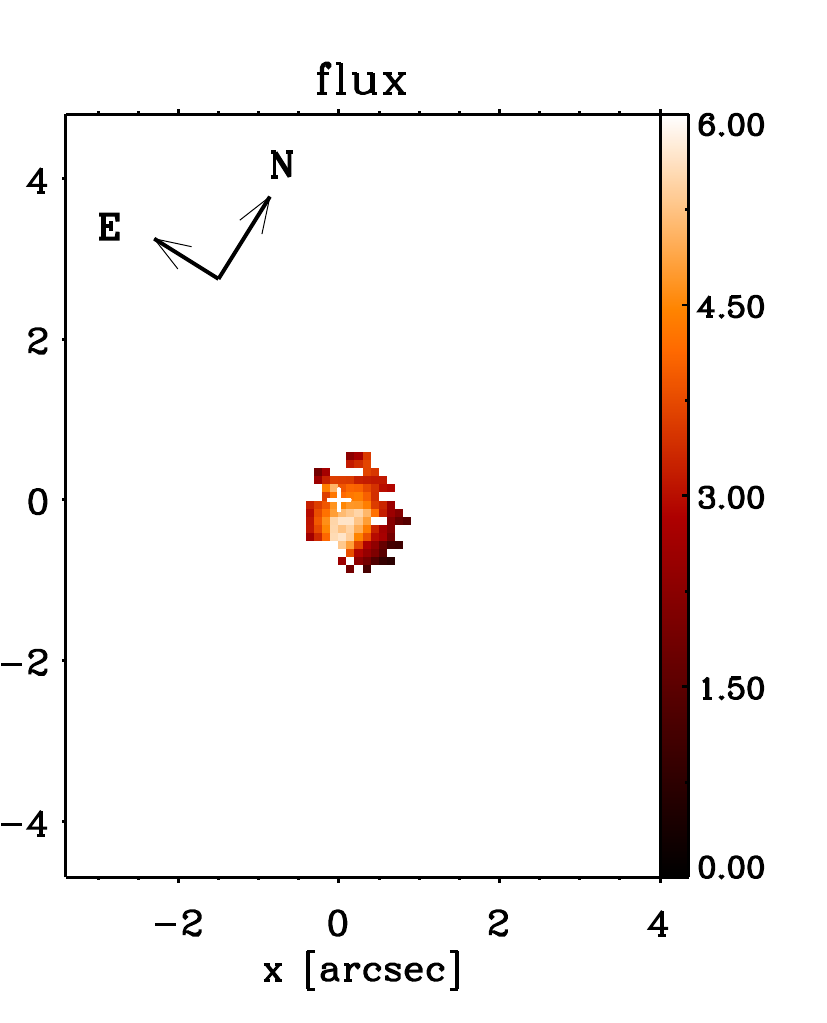}\\
&&\\
\includegraphics[trim=6cm 1.5cm 4cm 1cm, clip=true, scale=0.65]{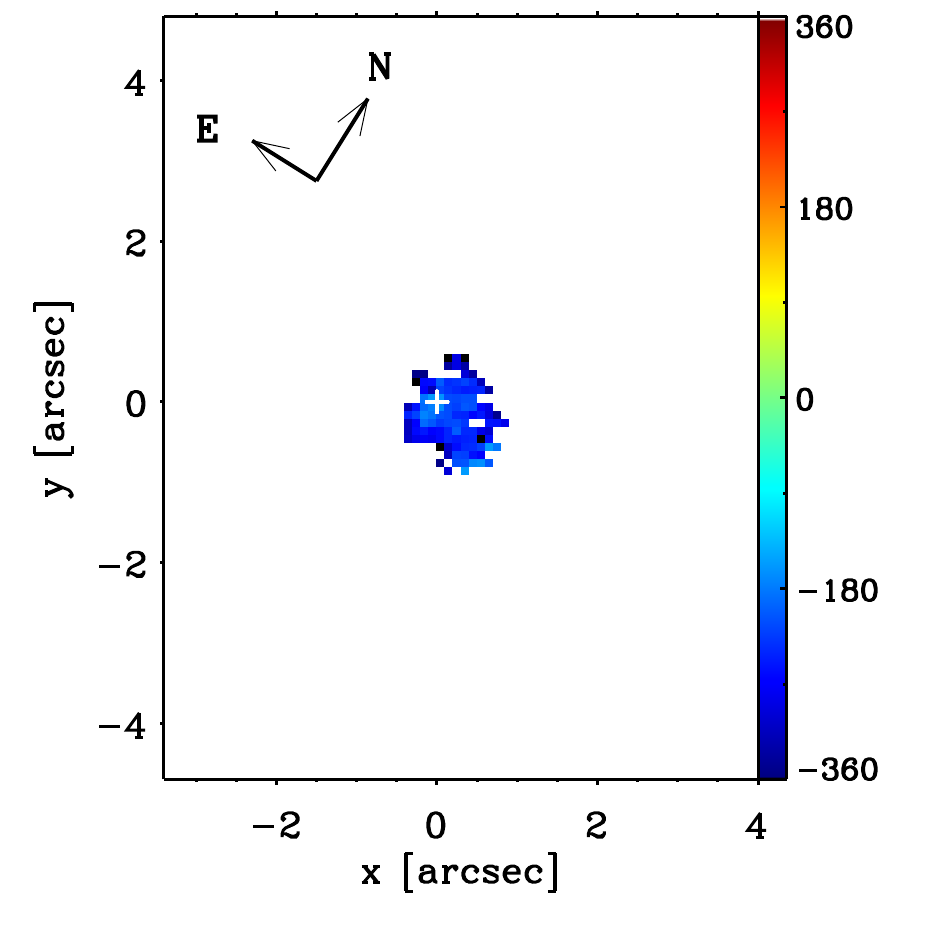} 	&	\includegraphics[trim=7cm 1.5cm 4cm 1cm, clip=true, scale=0.65]{sig1}		&	\includegraphics[trim=7cm 1.5cm 4cm 1cm, clip=true, scale=0.65]{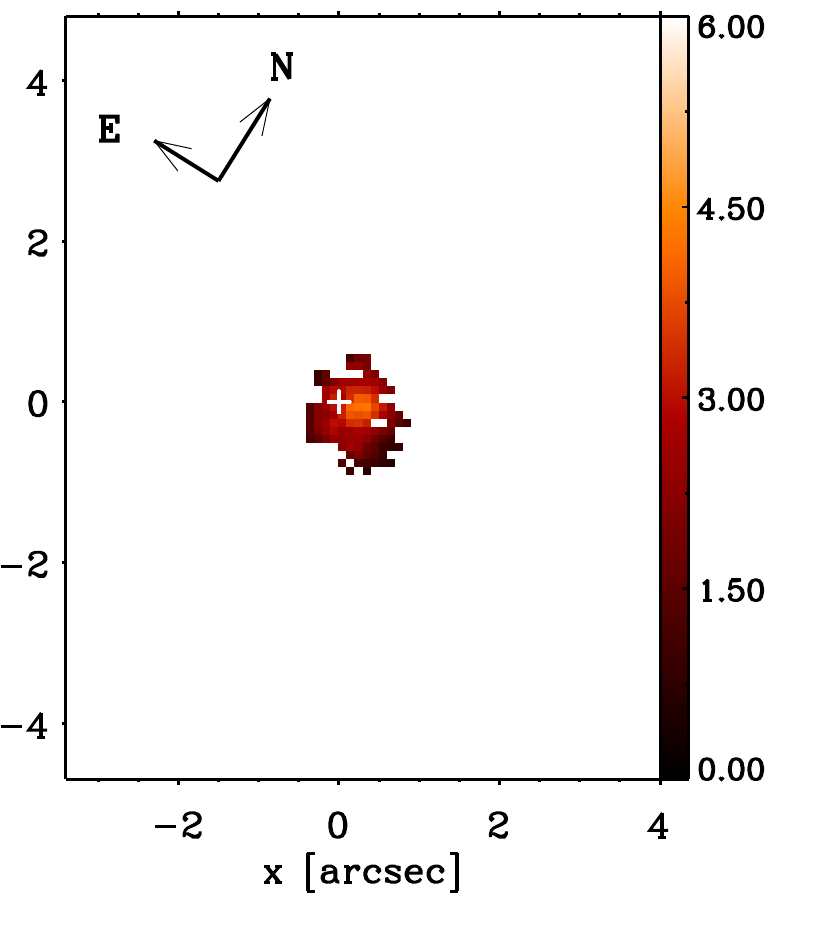}\\
\end{array}$
\end{center}
\caption{[FeVII]$\lambda$6087 maps. \textit{Left:} velocity [km s$^{-1}$]. \textit{Middle:} velocity dispersion [km s$^{-1}$]. \textit{Right:} flux [10$^{-16}$ erg/s/cm$^{2}$/spaxel]. \textit{Top:} redshifted component. \textit{Bottom:} blueshifted component. Note that velocity dispersions are the same by assumption. A systemic velocity $v_\mathrm{sys}$ = 786 km s$^{-1}$ was derived from the modeling of the [NII] kinematics and was subtracted from the velocity maps.}
\label{fig: Fe_maps}
\end{figure*}

Velocity and velocity dispersion maps for the [FeVII] line are presented in Fig.\ref{fig: Fe_maps}. As in the case of the [NII] broad components, the blue- and red-shifted components of the [FeVII] line are approximately co-spatial, and the emission is concentrated within the inner 1$^{\prime\prime}$. From visual inspection of the datacube and the fits performed on individual spaxels, we consider that there is marginal evidence that the [FeVII] emission is elongated along the north-south direction. However the S/N beyond the central $\approx$0\farcs5 is too low to allow a robust fit to the lines.

% ==========================================================================================
% Maps for [NII]: ==================

% ============================================================
% ============================================================
\subsubsection{Uncertainties} \label{sec: uncertainties_gas}

Velocity errors for the narrow component of the [NII] + H$\alpha$ complex are highest in regions where two component fits were used:
values range from 2 km s$^{-1}$ in the region where the narrow component is strong and a broad base is clearly present, to $>$ 50 km s$^{-1}$ in a few regions where the broad base is weak. Errors are significantly smaller in regions where a single Gaussian fit was used. In particular, low values (2-5 km s$^{-1}$) are found along the direction north-south, where the emission in the narrow component is very strong and lines are narrow and single peaked. Larger values (up to 20 km s$^{-1}$) are characteristic of the outer regions, where the S/N decreases. 
Errors in the velocity dispersion of the narrow component have a similar behavior but the range of values obtained from the simulation is smaller: typical errors for the inner 2$^{\prime\prime}$ vary in the range 10-20 km s$^{-1}$. 

Errors in the velocity of the broad component increase from the center outward, with the central arcsecond characterized by values close to 30 km s$^{-1}$ increasing, at the very edge of the fitted region (approximately 2$^{\prime\prime}$ from the nucleus), up to 150 km s$^{-1}$ for the blue component and close to 80 km s$^{-1}$ for the red component. The difference is due to the lower S/N of the blue broad component.

Errors derived for the [SII] velocity and velocity dispersion are very similar to those obtained for H$\alpha$ and [NII]. Only two differences are evident: the first is that errors in the velocity of the blue broad component do not show the usual radial gradient, probably because of the weakness of the line. The second is that errors in the velocity dispersion of the red broad component are smaller than the corresponding errors derived from the [NII] with values ranging between 5 and 35 km s$^{-1}$ and typical values of order 20 km s$^{-1}$ (for comparison, errors derived from the [NII] line range between 5 and 60 km s$^{-1}$ showing a steeper radial gradient increasing outward).

The [OI] velocity and velocity dispersion maps are affected by errors of order 1 km s$^{-1}$ in the center of the south-western lobe, where the S/N is high, increasing to 10 km s$^{-1}$ in the outskirt. The nucleus, where the S/N for [OI] is often not enough to allow a three component fit, is characterized by errors of order 20 km s$^{-1}$. The north-eastern lobe, where the signal is lower than in the south-western lobe, has errors ranging from 4 km s$^{-1}$, in the central region, to 20 km s$^{-1}$ along the rim of the lobe where S/N decreases.

% ===============

% [Fe VII]: ==================

% ===============
% ===============
\subsubsection{Channel maps}

\begin{figure*}[p]
\begin{center}$
\begin{array}{cccc}
\includegraphics[trim=4cm 0cm 6cm 0cm, clip=true, scale=.425]{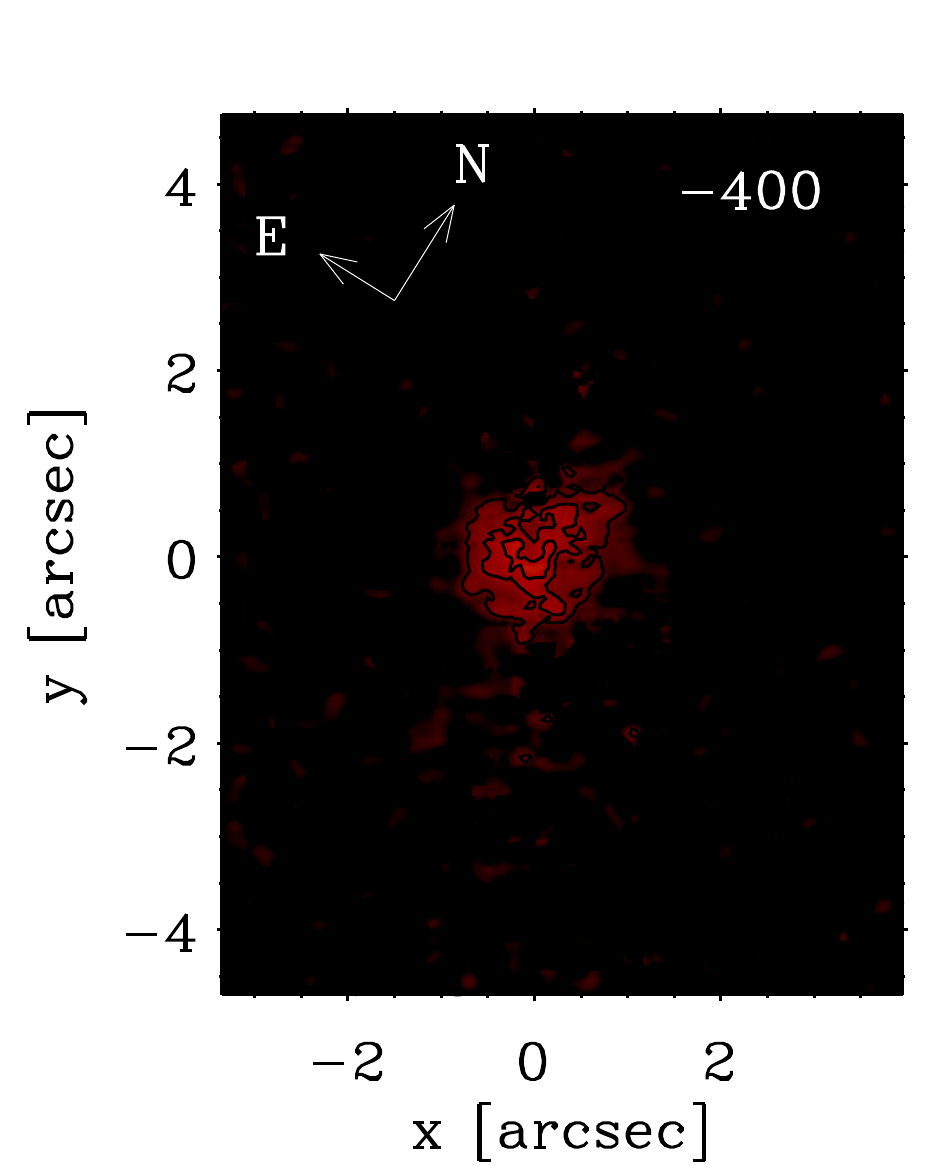} 	& \includegraphics[trim=5cm 0cm 6cm 0cm, clip=true, scale=.425]{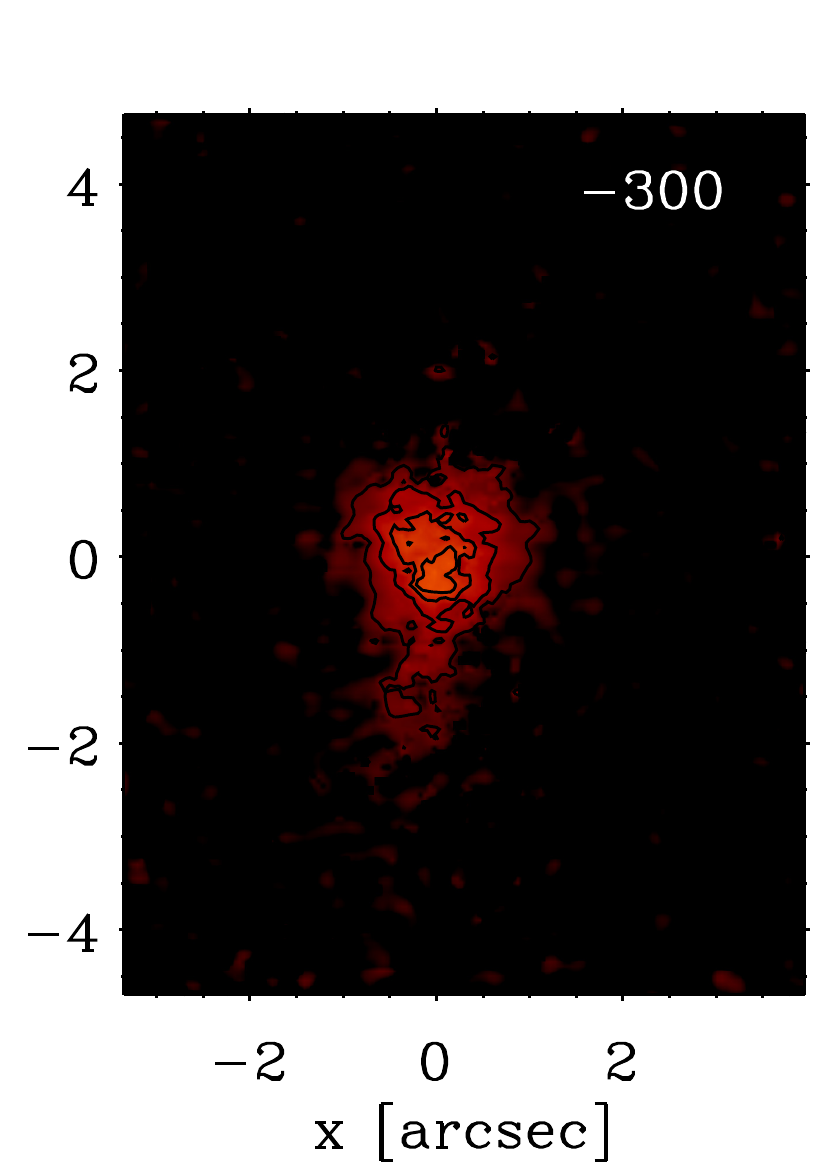}&\includegraphics[trim=5cm 0cm 6cm 0cm, clip=true, scale=.425]{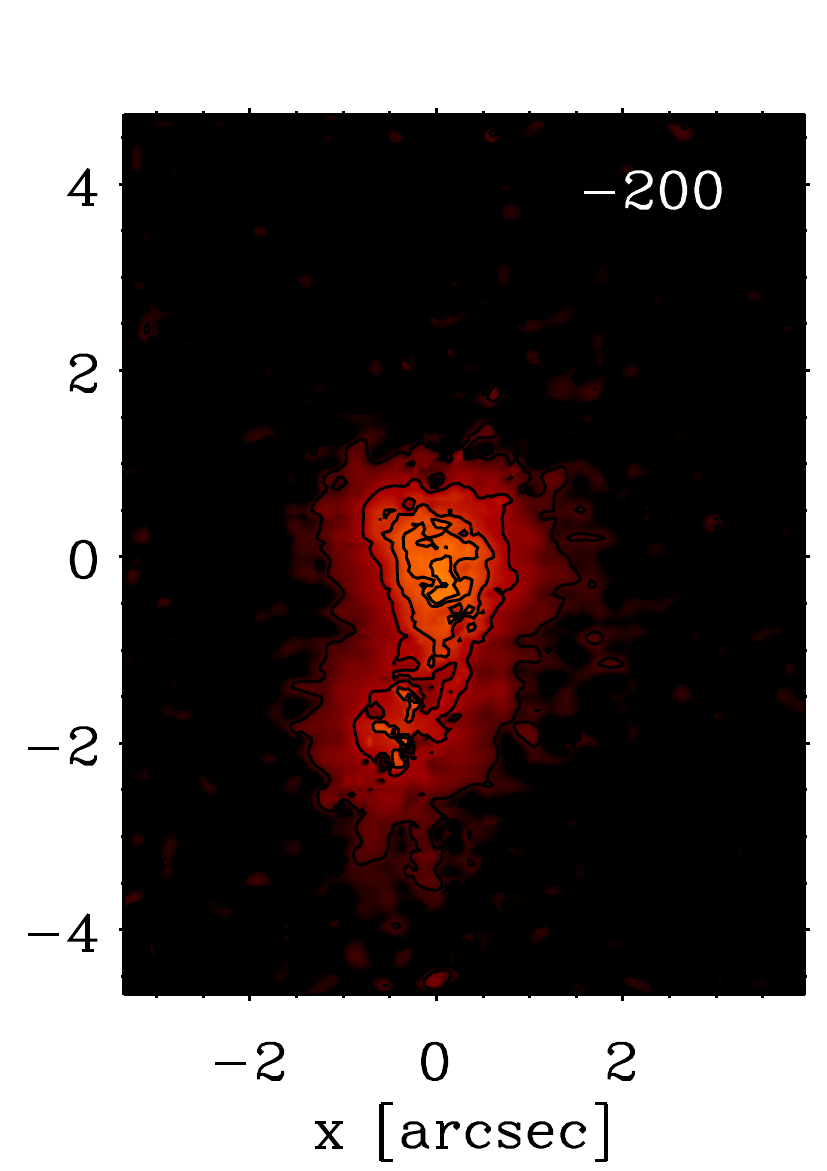} 	& \includegraphics[trim=5cm 0cm 6cm 0cm, clip=true, scale=.425]{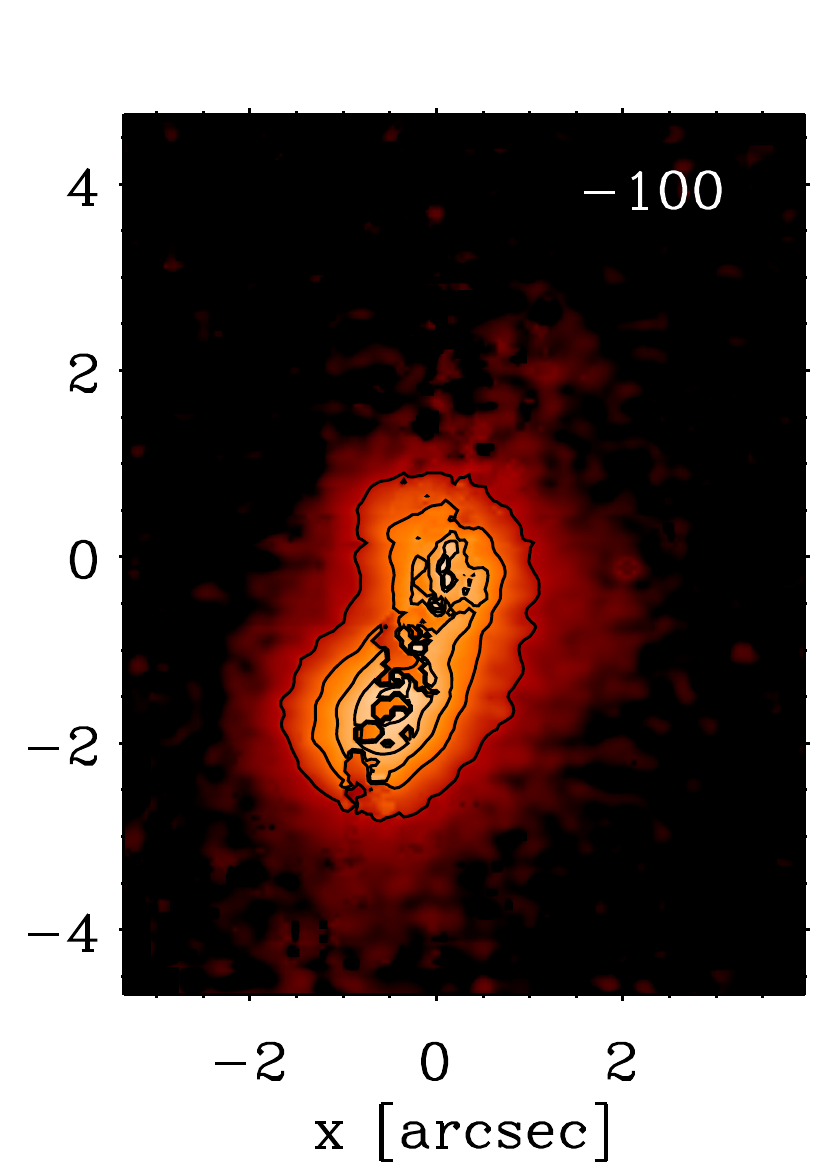}\\
\includegraphics[trim=4cm 0cm 6cm 0cm, clip=true, scale=.425]{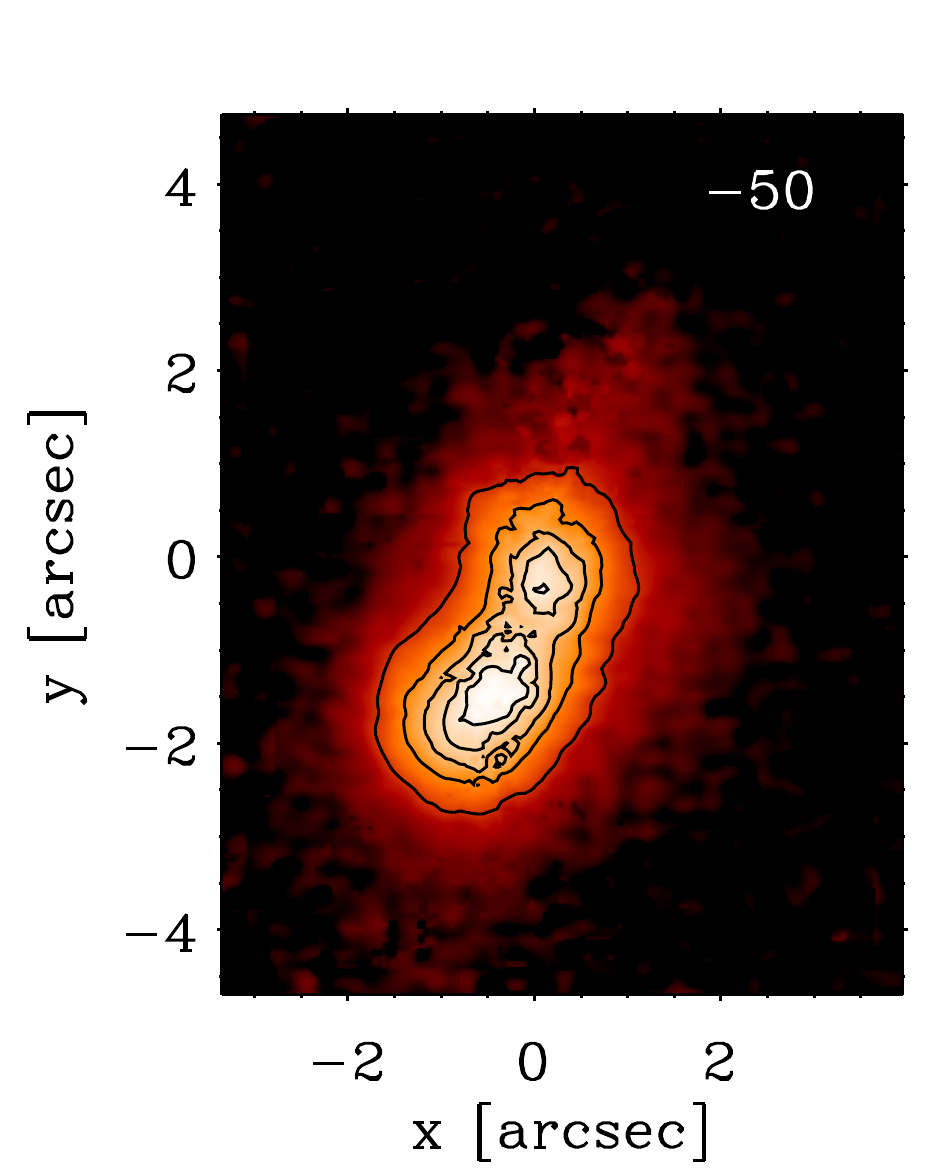}  	& \includegraphics[trim=5cm 0cm 6cm 0cm, clip=true, scale=.425]{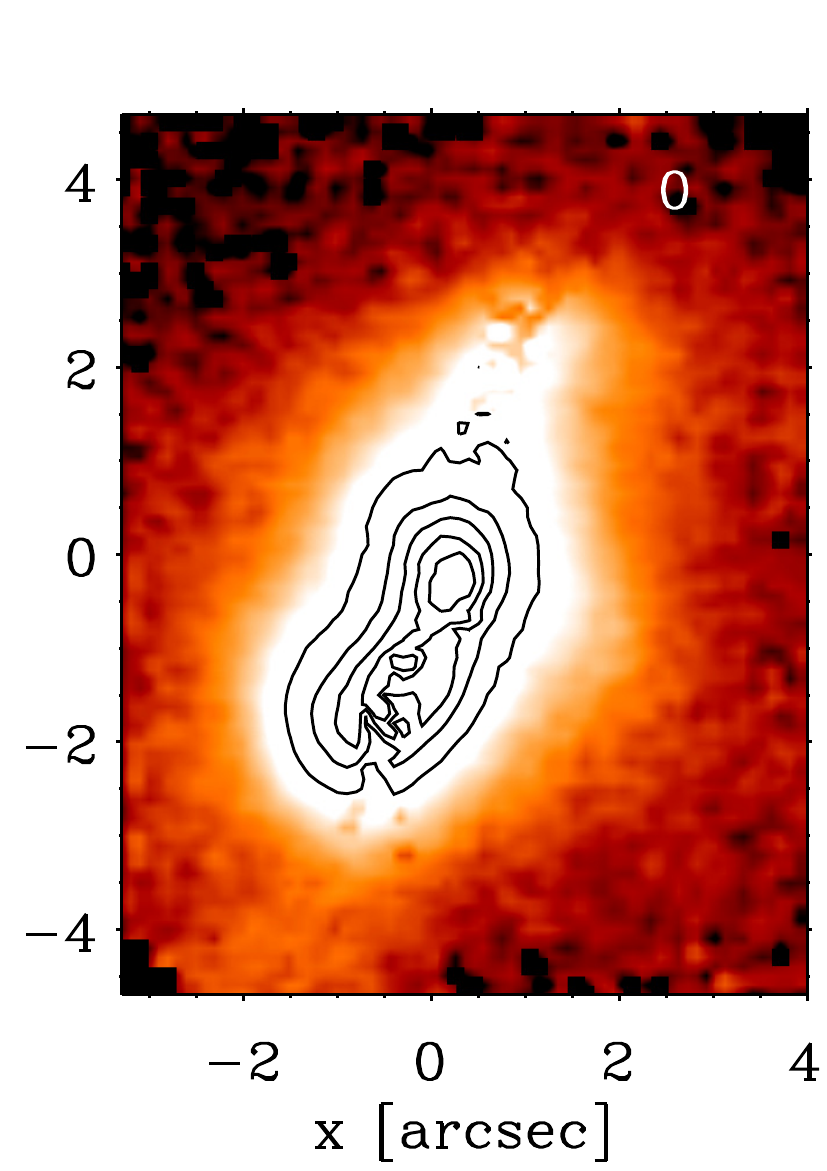} 	&  \includegraphics[trim=5cm 0cm 6cm 0cm, clip=true, scale=.425]{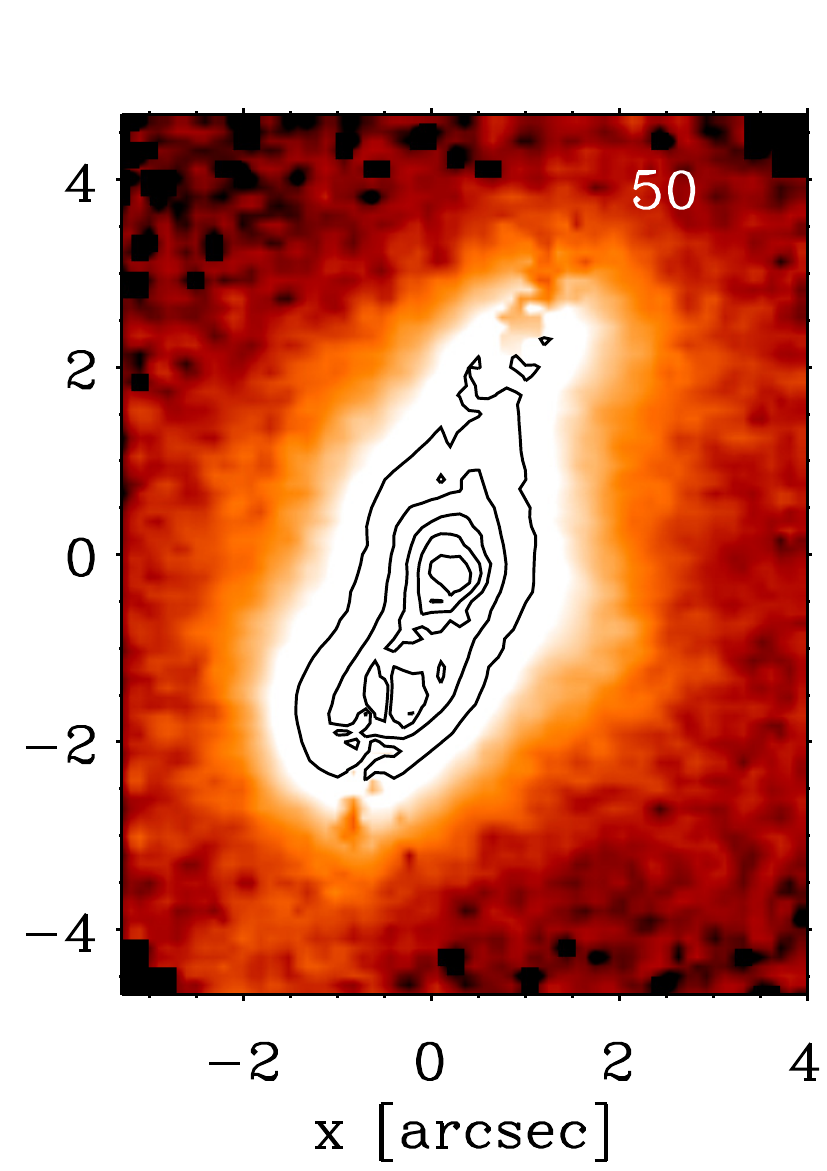} 	& \includegraphics[trim=5cm 0cm 6cm 0cm, clip=true, scale=.425]{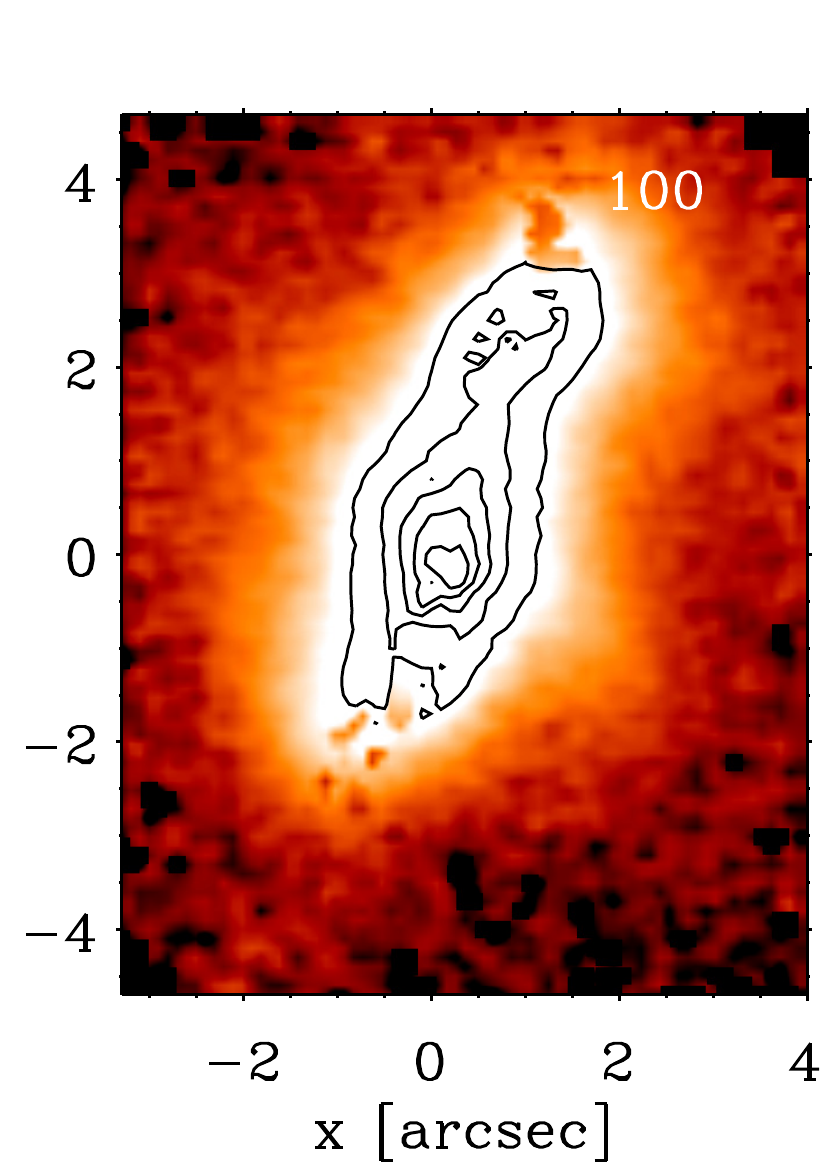}\\
\includegraphics[trim=4cm 0cm 6cm 0cm, clip=true, scale=.425]{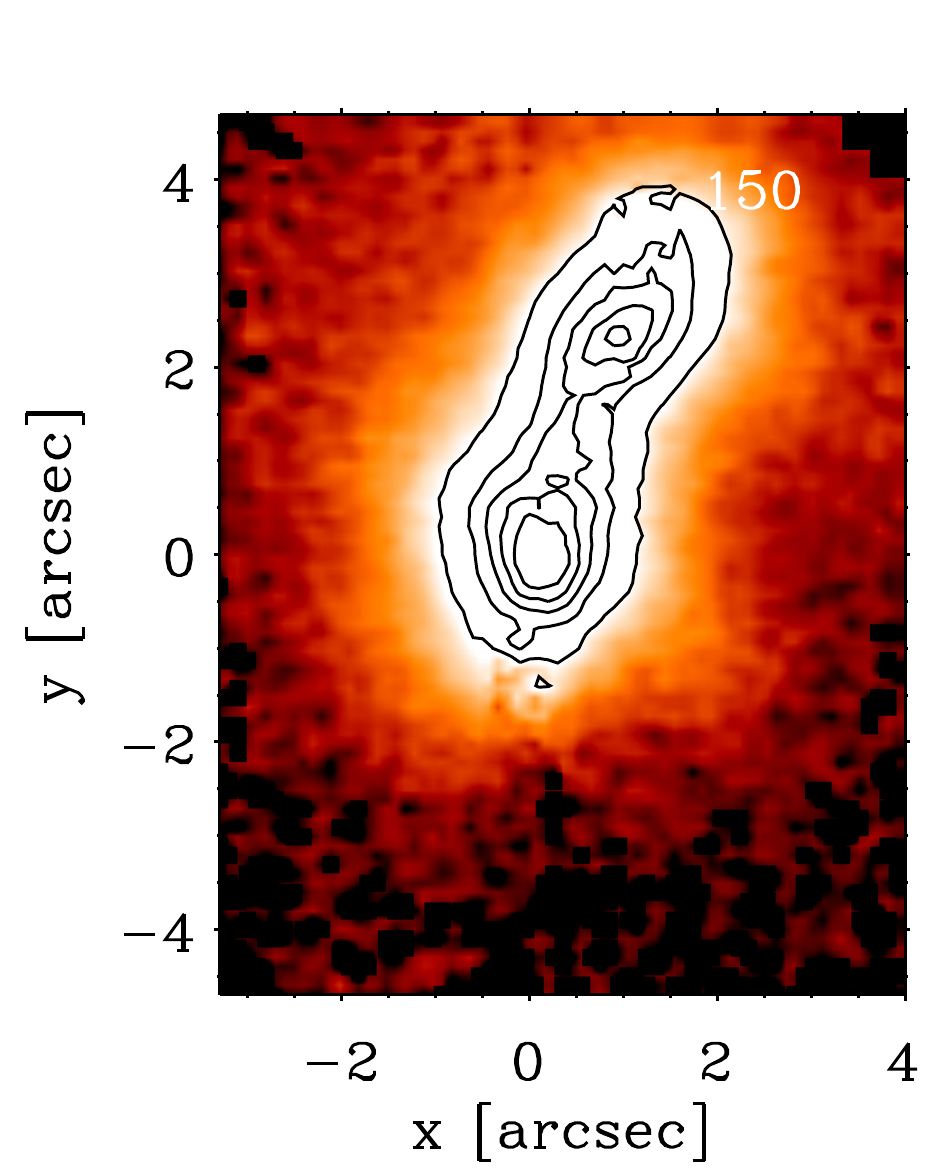}	& \includegraphics[trim=5cm 0cm 6cm 0cm, clip=true, scale=.425]{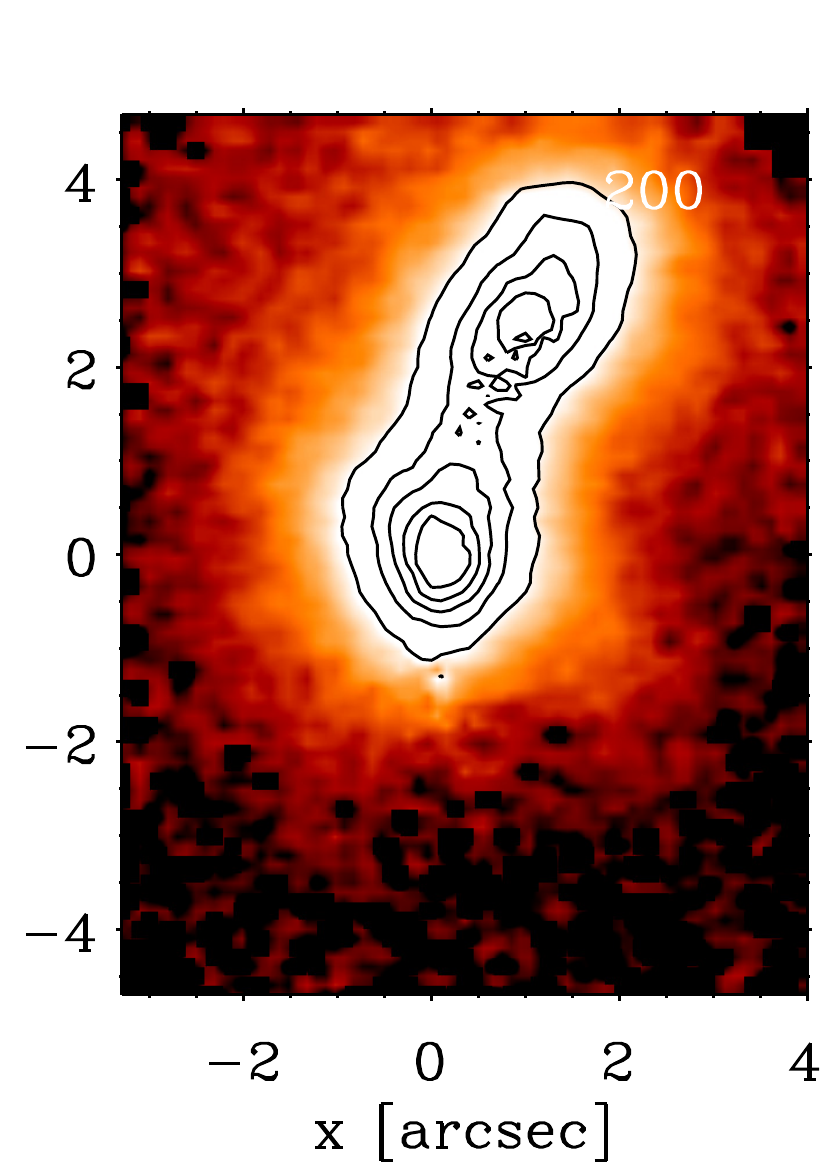} & \includegraphics[trim=5cm 0cm 6cm 0cm, clip=true, scale=.425]{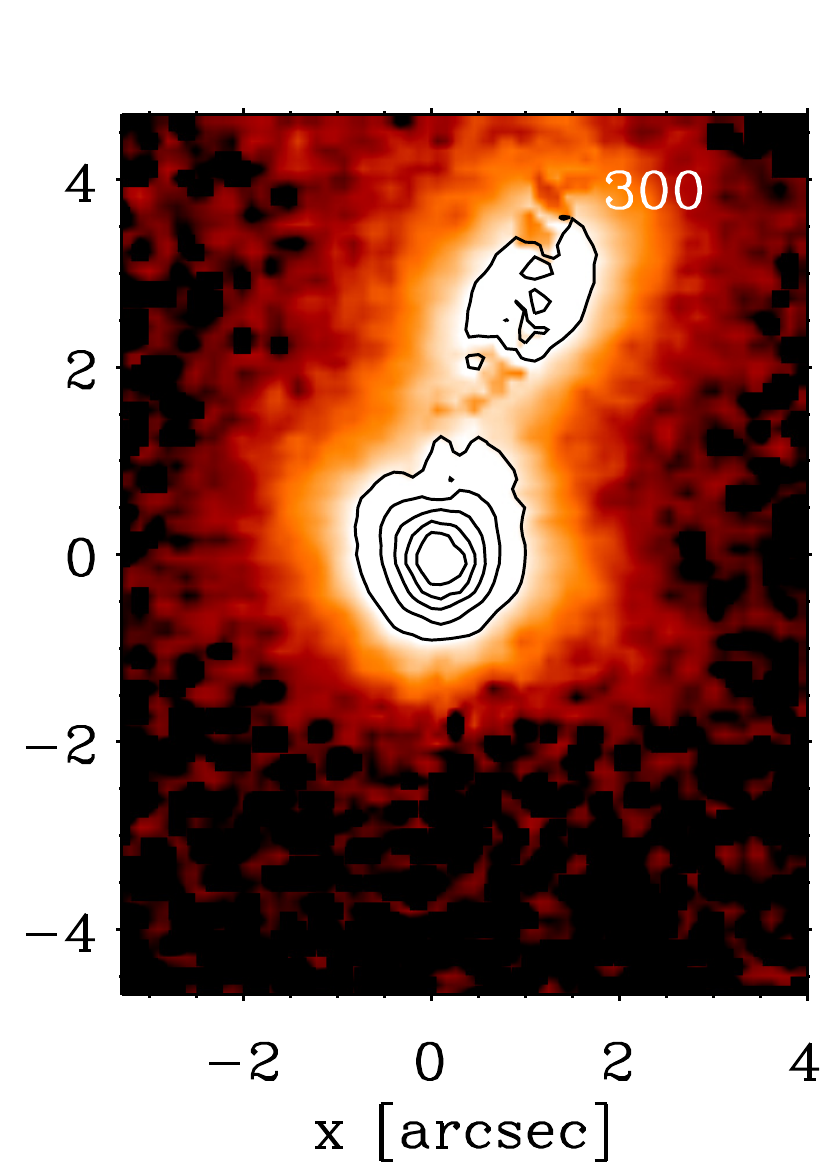} 	& \includegraphics[trim=5cm 0cm 6cm 0cm, clip=true, scale=.425]{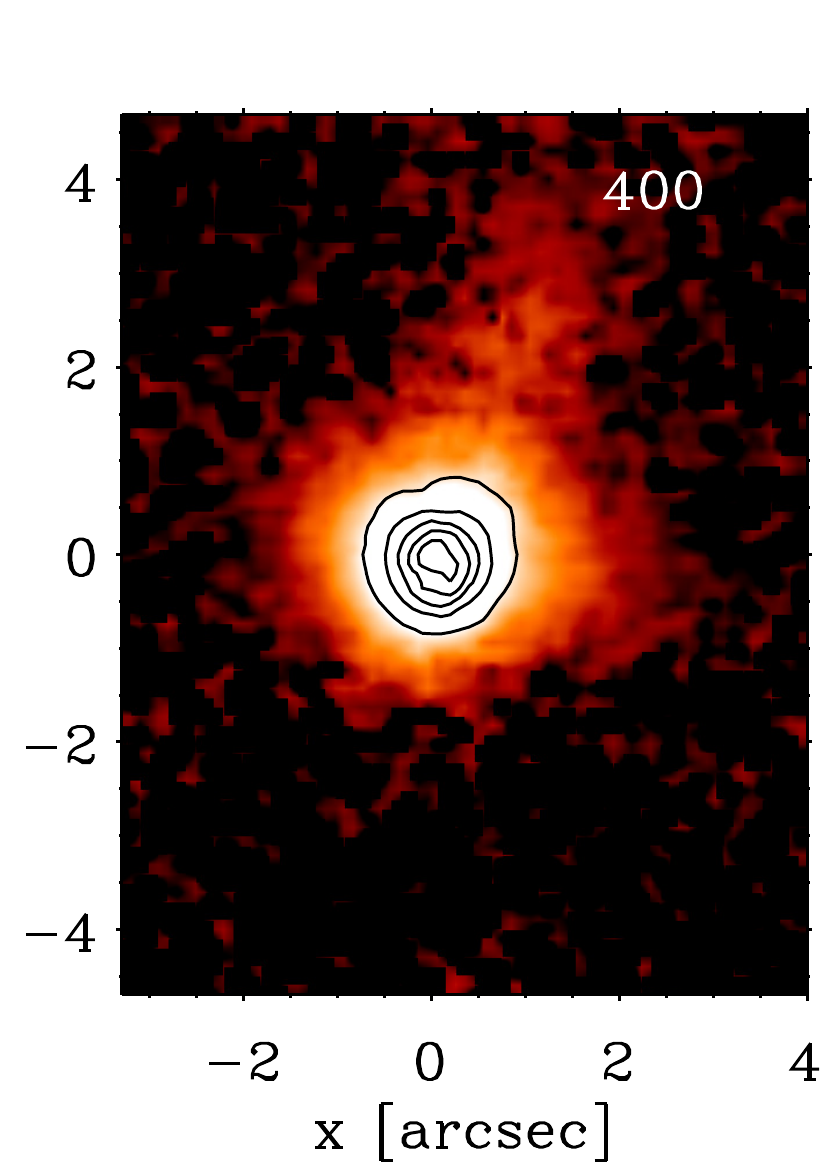}\\
\includegraphics[trim=4cm 0cm 6cm 0cm, clip=true, scale=.425]{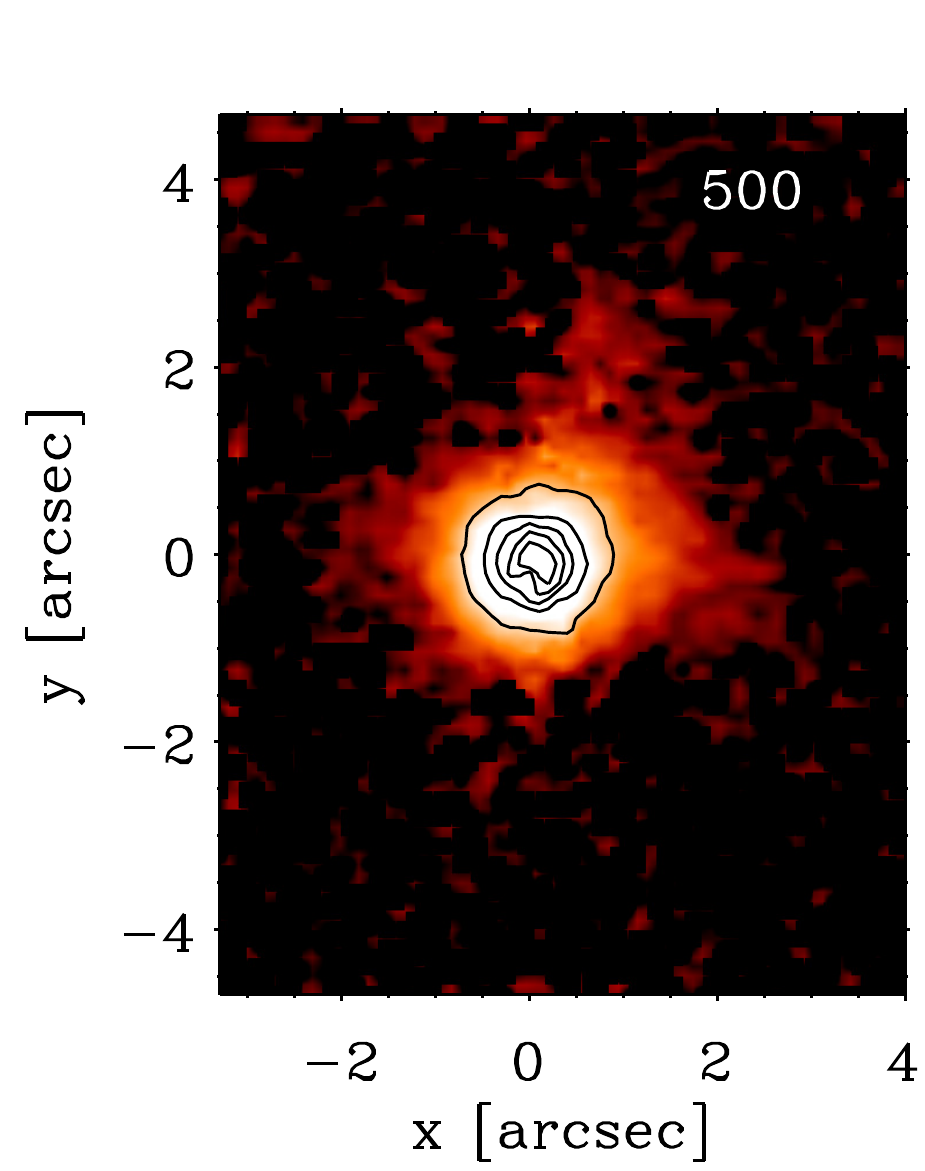} 	& \includegraphics[trim=5cm 0cm 6cm 0cm, clip=true, scale=.425]{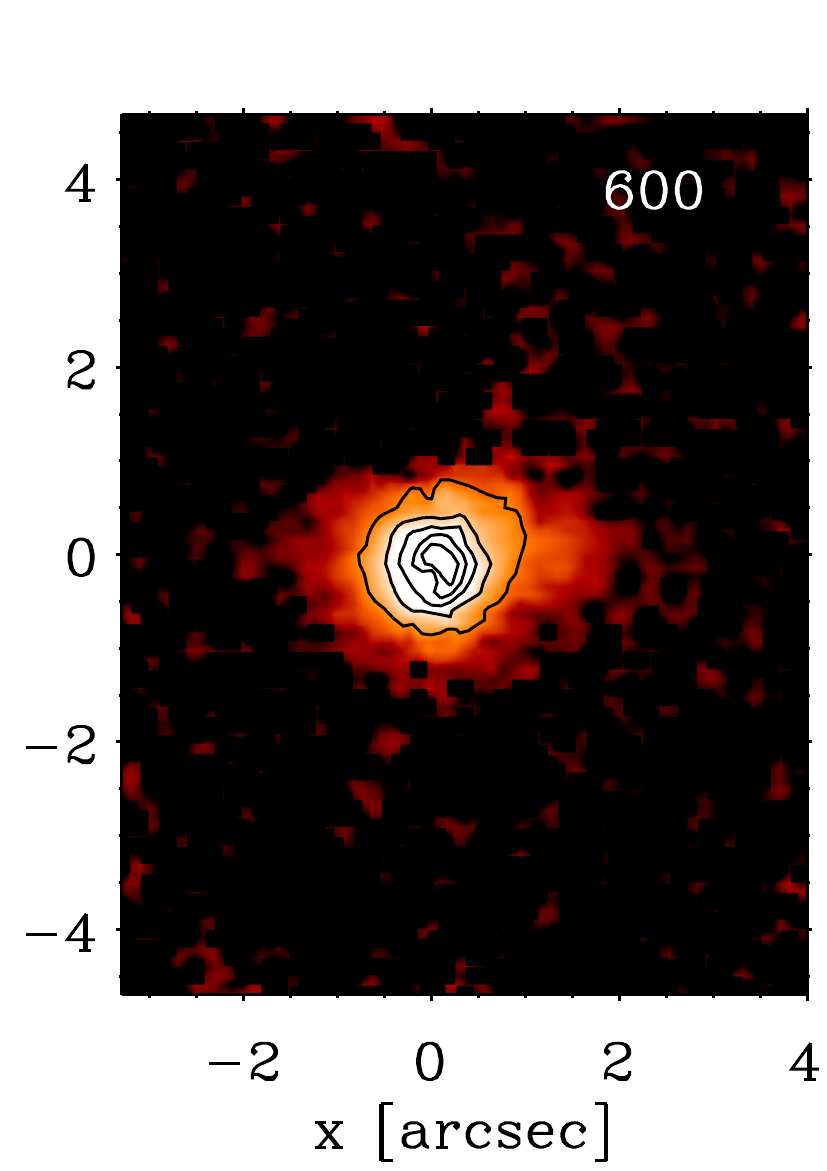} & \includegraphics[trim=5cm 0cm 6cm 0cm, clip=true, scale=.425]{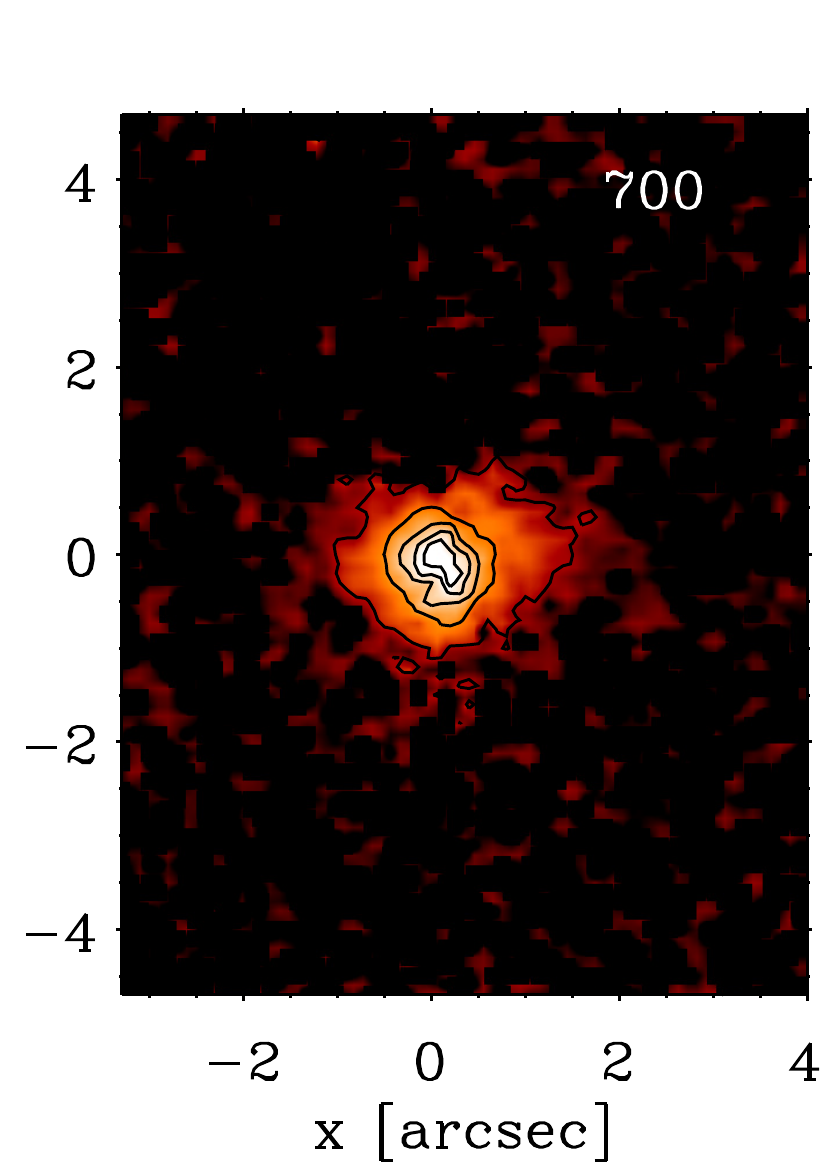} 	& \includegraphics[trim=5cm 0cm 4cm 0cm, clip=true, scale=.425]{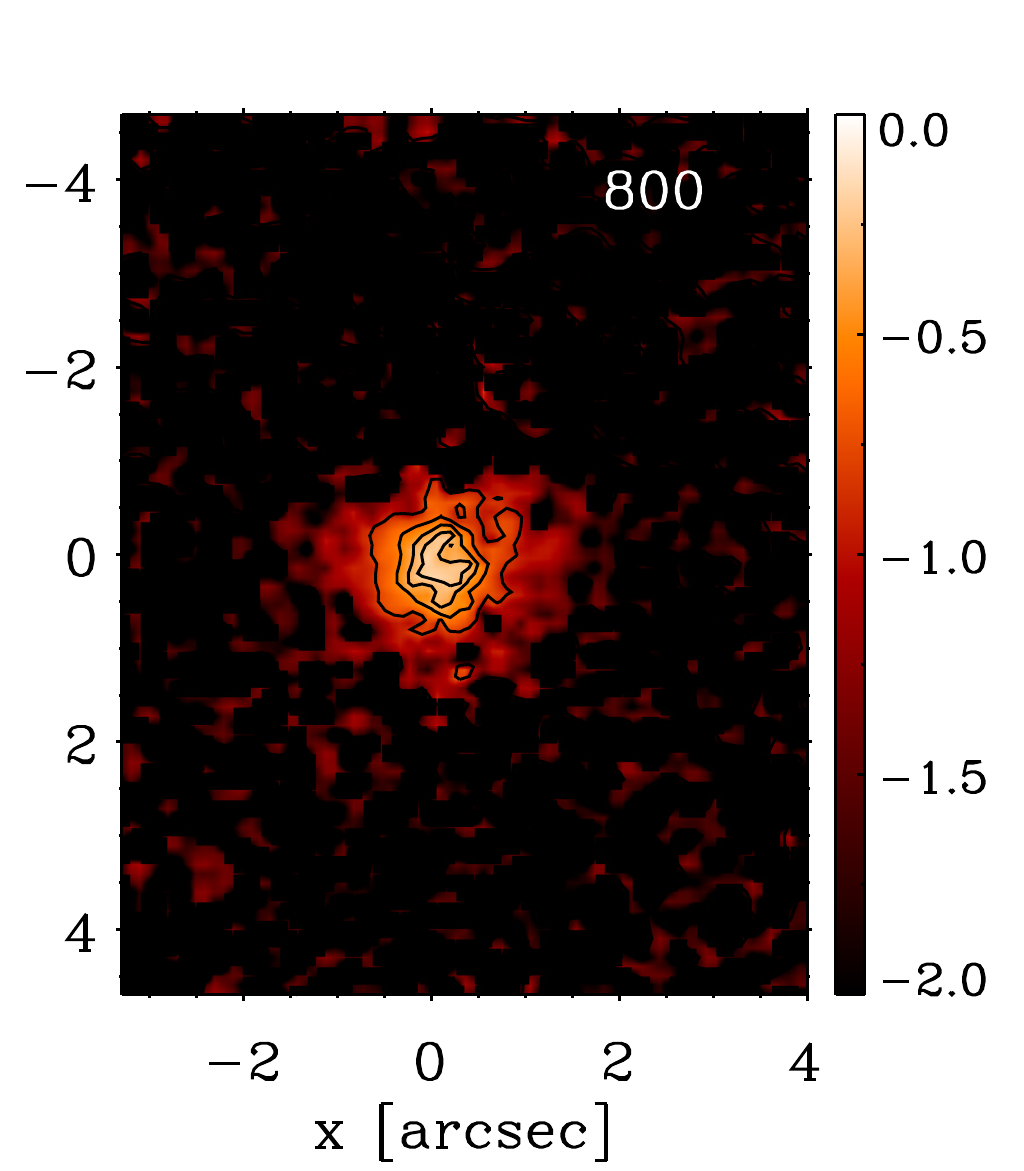}\\
\end{array}$
\end{center}
\caption{Channel maps derived from the [NII] emission lines. Corresponding velocities are shown in the top right corner of each map in km s$^{-1}$. Flux units are Log $10^{-16}$ erg/s/cm$^{2}$/spaxel. To avoid contamination from H$\alpha$, maps corresponding to negative velocities have been derived from the blue side of the [NII]$\lambda$6548 line, whilst maps with positive velocities have been derived from the red side of [NII]$\lambda$6583 line. A systemic velocity $v_\mathrm{sys}$ = 786 km s$^{-1}$ was derived from the modeling of the [NII] kinematics and subtracted from the maps. Contours are equally spaced flux contours.}
\label{fig: channelm}
\end{figure*}

Channel maps were extracted in velocity bins of 50 km s$^{-1}$ along the [NII] profile after subtracting a systemic velocity v$_\mathrm{sys}$ = 786 km s$^{-1}$ derived from kinematical modeling of the [NII] velocity field. In order to minimize H$\alpha$ contamination, negative velocities were derived from the blue side of the [NII]$\lambda$6548 line, whilst positive velocities were derived from the red side of the [NII]$\lambda$6583 line. The maps are shown in Fig.\ref{fig: channelm}. 

As [NII]$\lambda$6548 is a factor of three weaker than [NII]$\lambda$6583, and given that the broad blueshifted component is intrinsically weaker than the the redshifted one, the blueshifted emission cannot be mapped beyond the $-400$ km s$^{-1}$ channel, whereas the redshifted emission can be mapped to $+800$ km s$^{-1}$.

Channel maps at high velocities $\leq -200$ km s$^{-1}$ and $\geq$ 400 km s$^{-1}$ are clearly dominated by circum-nuclear emission. Interestingly, the emission in these channel maps is co-spatial despite the wide range of velocities involved. For velocities within the range $-200:400$ km s$^{-1}$ the line emission seems to result from at least two kinematic features: (i) the strong emission at the nucleus which is isolated at extreme velocities and is due to the broad spectral components, (ii) a contribution from the narrow component associated with a large-scale rotating disk. The large-scale contribution takes the form of off-center lobes of emission which are evident from v $\geq -200$ km s$^{-1}$ in the south to v $\leq$ 300 km s$^{-1}$ in the north.

% =======================================================
% =======================================================
\subsection{Stellar kinematics} 
\subsubsection{Velocity and velocity dispersion}

In order to derive the stellar kinematics we fitted the stellar continuum and absorption lines using the Penalized Pixel Fitting technique \citep[pPFX,][]{CappellariEmsellem04} and a subset of the \cite{Vazdekis99} stellar templates that were computed for an old stellar population with spectral type in the range O-M and metallicity -0.7 $\leq$ [Fe/H] $\leq$ +0.2.
Stellar centroid velocity and velocity dispersions were derived, for each spaxel, after masking the Na I $\lambda
\lambda$5890,5896 absorption lines (which are affected by interstellar gas), the most prominent emission lines, and the O$_{2}$ $\lambda$6875 sky B band.
Given the spectral resolution of the stellar templates, 1.8 \AA, which is approximately the same as that of our observations, 1.9 \AA\ at 6600 \AA, we did not apply any additional correction to the velocity dispersion.

\begin{figure*}[tbh]
\begin{center}$
\begin{array}{cc}
\includegraphics[trim=6cm 1.5cm 4cm 1cm, clip=true, scale=0.63]{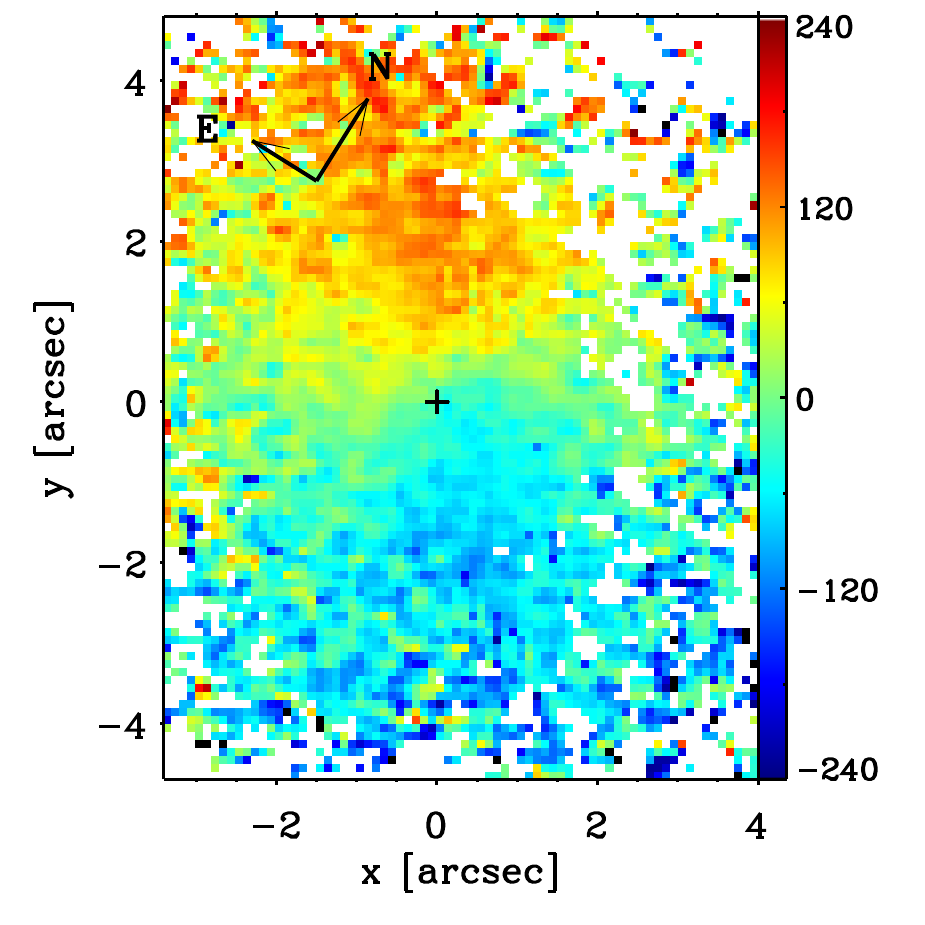} & \includegraphics[trim=6cm 1.5cm 4cm 1cm, clip=true, scale=0.63]{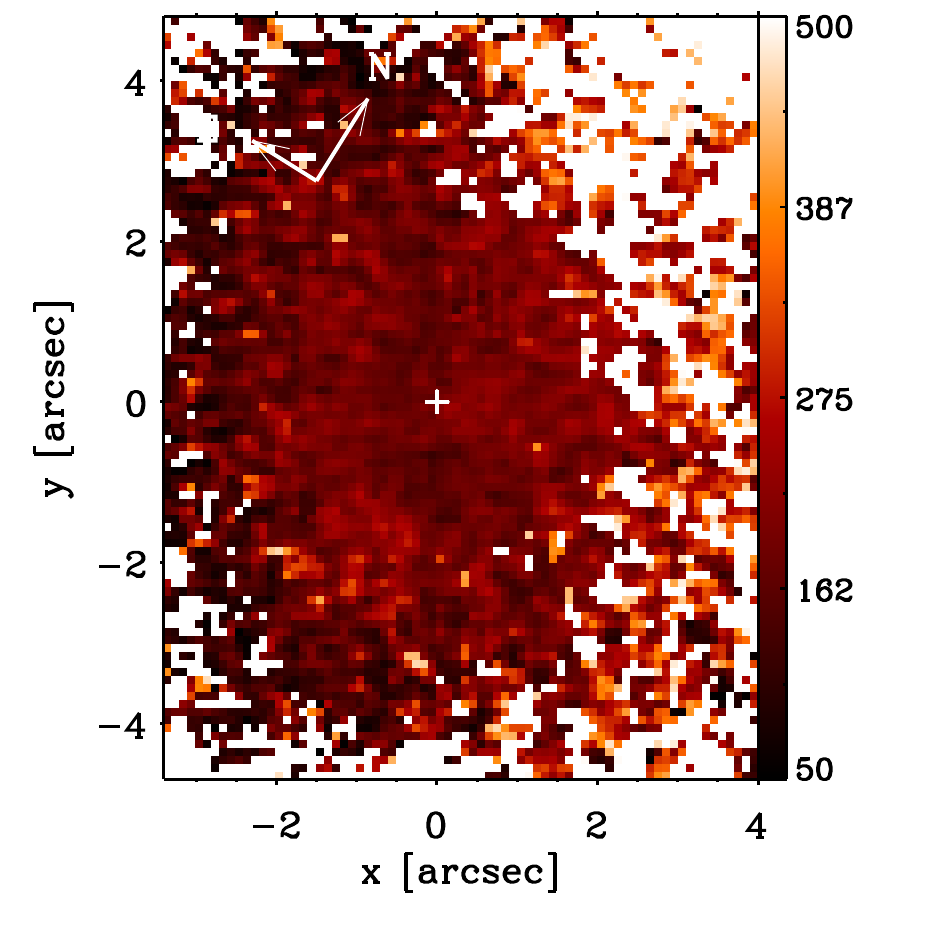}\\
\end{array}$
\end{center}
\caption{\textit{Left:} velocity map as derived from the stellar absorption features excluding the [NaD] doublet. \textit{Right:} stellar velocity dispersion. Units: km s$^{-1}$. A systemic velocity $v_\mathrm{sys,*}$ = 743 km s$^{-1}$, as derived from modeling of the stellar kinematics, was subtracted from the velocity field.}
\label{fig: stellark}
\end{figure*}

Although many pixels north of the nucleus were flagged during the fit we can see that, similarly to the gas kinematics, the stellar velocity field (left panel in Fig.\ref{fig: stellark}) displays a rotation pattern in which the north-eastern side is receding. 
The stellar velocity dispersion (right panel in Fig.\ref{fig: stellark}) reaches values close to 250 km s$^{-1}$ near the center decreasing to 100 km s$^{-1}$ toward the edges of the FOV. The small systematic increase, evident from south-east to north-west, is an artifact due to the decreasing S/N. The mean velocity dispersion within circles of increasing radii is fairly constant with $\bar \sigma \approx 195$ km s$^{-1}$ within an aperture of radius 2\farcs5 (190 pc). 
\vskip5pt

Previous estimates of the central stellar velocity dispersion have been determined by \citet{DresslerS83}, \citet{NelsonW95} and \citet{garciariss2005}.
\citeauthor{DresslerS83} obtained $\sigma$ = $195 \pm 16$ km s$^{-1}$ from spectra taken with the du Pont Las Campanas 2.5 m spectrographs in the spectral region 4800-5400 \AA\ using a 2$^{\prime\prime} \times$4$^{\prime\prime}$ slit (FWHM\footnote{Full width at half maximum = 2.36 $\sigma$.} $\approx$ 180 km s$^{-1}$). This value is consistent with our findings. 

With a similar set up (aperture of 2\farcs2$\times$3\farcs6 and optical spectra including the Mg $b\ $$\lambda$5175 absorption line), \citeauthor{NelsonW95} used the Royal Greenwich Observatory Spectrograph on the 3.9 m Anglo Australian Telescope (FWHM $\approx$ 80 km s$^{-1}$) obtaining a significantly smaller value of $\sigma = 120 \pm 30$ km s$^{-1}$. \citeauthor{garciariss2005} used the Richey-Cretien spectrograph on the 4m Mayall telescope at Kitt Peak National Observatory with a slit width of 1\farcs5. From the calcium triplet they derived the stellar velocity dispersion using two methods obtaining $\sigma_{1} = 123 \pm 3$ km s$^{-1}$ and $\sigma_{2} =133 \pm 3$ km s$^{-1}$.

The origin of the discrepancy amongst the set of values summarized here is unclear.

% =======================================================
% =======================================================
\subsubsection{Uncertainties}

Typical uncertainties in the stellar velocity field range from 10 km s$^{-1}$, in the nuclear region, up to 15 km s$^{-1}$, in the outer part of the field. Errors on the stellar velocity dispersion are slightly larger, ranging from 15 km s$^{-1}$ near the nucleus, up to 25 km s$^{-1}$ toward the outskirts of the FOV.

% =======================================================
% Table with line ratios and electron densities. The text section comes later.
% It was placed here to make it show up into a given page.

\begin{table*}
\center
\caption{LINE RATIOS AND ELECTRON DENSITIES}  \label{tab: ratios_n}
\scalebox{1}{
\begin{tabular}{c l l l l c c}
\hline
\hline
Subr		 		& \multicolumn{1}{c}{[NII]6583/H$\alpha$}	& \multicolumn{1}{c}{[OI]6366/H$\alpha$} 		&  \multicolumn{1}{c}{[SII](6716 + 6731)/H$\alpha$}	&	[SII]6716/6731		& \multicolumn{1}{c}{n$_{e}$ cm$^{-3}$} 	 \\
	 			&  							& 								&									&					& 				\\
\hline
& & &   \\
\multicolumn{6}{c}{narrow} \\
\hline
& & &   \\
	A 			& 2.45 $\pm$ 0.58				& 0.45 $\pm$ 0.09					& 1.07 $\pm$ 0.18						&	1.23 $\pm$ 0.18	&  200$_{-35}^{+40}$\\
	B 			& 1.78 $\pm$ 0.01				& 0.22 $\pm$ 0.02					& 0.80								&	1.24 $\pm$ 0.01	&  190$_{-14}^{+16}$\\
	C 			& 1.55 $\pm$ 0.06				& 0.46 $\pm$ 0.03					& 0.63 $\pm$ 0.2			 			&	1.2 $\pm$ 0.04		& 234$_{-45}^{+50}$	\\
	D 			& 1.43 $\pm$ 0.09				& 0.36 $\pm$ 0.05					& 0.65 $\pm$ 0.11			 			&	1.6   $\pm$ 0.03	& 195$_{-34}^{+38}$ \\
	N 			& 1.65 $\pm$ 0.01		 		& 0.13 $\pm$ 0.01					& 0.85 $\pm$ 0.01						&	0.96	$\pm$ 0.02	& 704$_{-54}^{+59}$\\	
& & & \\
\multicolumn{6}{c}{blue broad} \\
\hline
& & &   \\
	N 			& 1.12 $\pm$ 0.01				& 0.15 $\pm$ 0.01				 	& 0.24 $\pm$ 0.02			 			&	2.81 $\pm$ 0.81$^{\dagger}$	& 		&  \\
& & & \\
\multicolumn{6}{c}{red broad} \\	
\hline
& & &   \\
	N 			& 1.29		 			 	& 0.17						 	& 0.57 $\pm$ 0.01			 			&	0.95 $\pm$ 0.02	& 752$_{-46}^{+51}$		&  \\
& & & \\
\hline
\end{tabular}}
\tablecomments{See the velocity map derived from the narrow component in Fig.\ref{fig: [NII]maps} for subregion IDs. Omitted errors are smaller than 0.01. $^{\dagger}$Exceeds high density limit. \vskip5pt}
\end{table*}

% =======================================================
% =======================================================
\subsection{Line fluxes}
\label{subsec: line_flux}

The spatial distribution of the integrated [NII]$\lambda$6583 line flux obtained from the single Gaussian fit to the [NII] line over the whole FOV is shown in the right panel of Fig.\ref{fig: [NII]1c}. The strongest emission is observed within a circular region of radius $r \approx 1^{\prime\prime}$ around the nucleus. The other main features are two elongated regions (the ``lobes") extending about 3$^{\prime\prime}$ north and south of the nucleus in a direction similar to that of the compact radio emission reported by \citet{nagarWMG99} and \citet{MundellEtAl09}. Weaker [NII] emission is present over the entire FOV at a level of 3 $\times$ 10$^{-17}$ erg s$^{-1}$ cm$^{-2}$ spaxel$^{-1}$.

The spatial distribution of the integrated line flux obtained from the multi-Gaussian fit is shown in the right column of Fig.\ref{fig: [NII]maps} for the broad and narrow components of the [NII]$\lambda$6583 emission line. 

As evident from the figures, the multi-Gaussian decomposition shows that the narrow component originates from the lobes while the broad component is essentially confined to the nucleus.

\begin{figure*}[t]
\begin{center}$
\begin{array}{ccccccccc}
\includegraphics[trim=6cm 1.7cm 5.65cm 1cm, clip=true, scale=0.65]{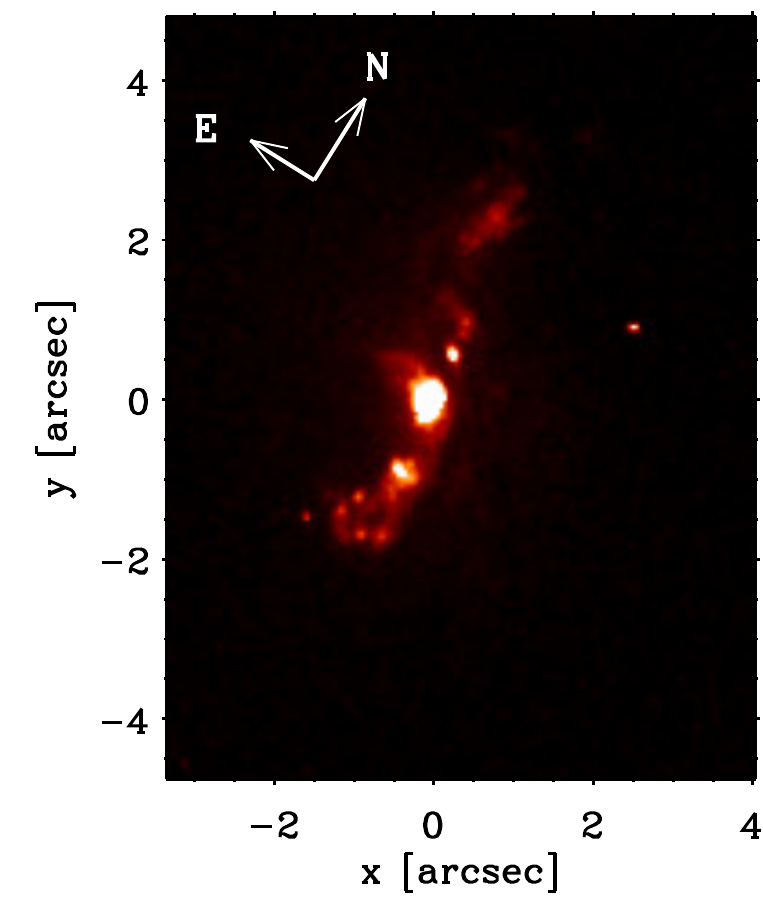} 	&&  & & \includegraphics[trim=7cm 1.7cm 5.65cm 1cm, clip=true, scale=0.65]{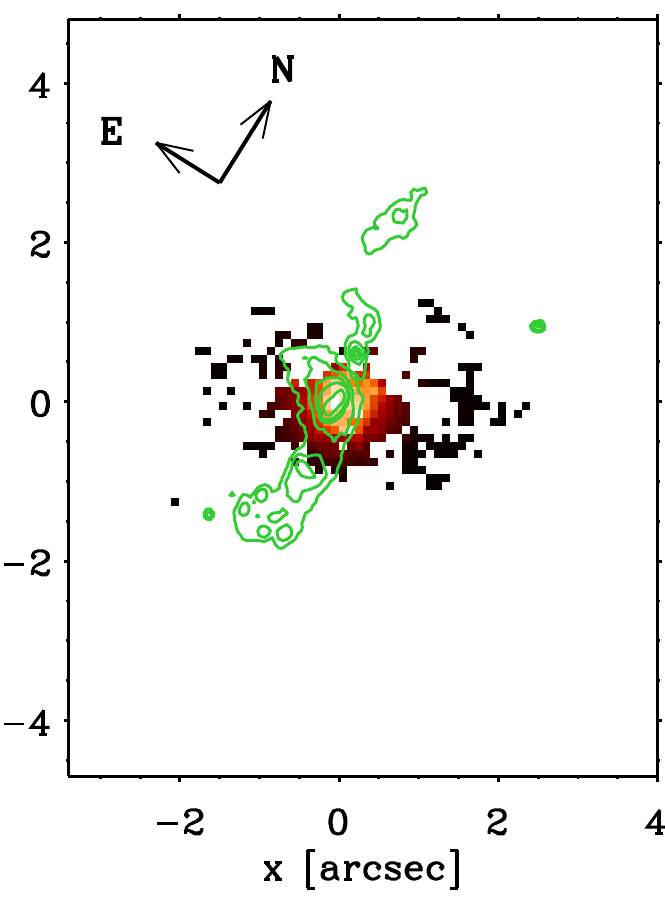} & & & & \includegraphics[trim=7cm 1.7cm 5.65cm 1cm, clip=true, scale=0.65]{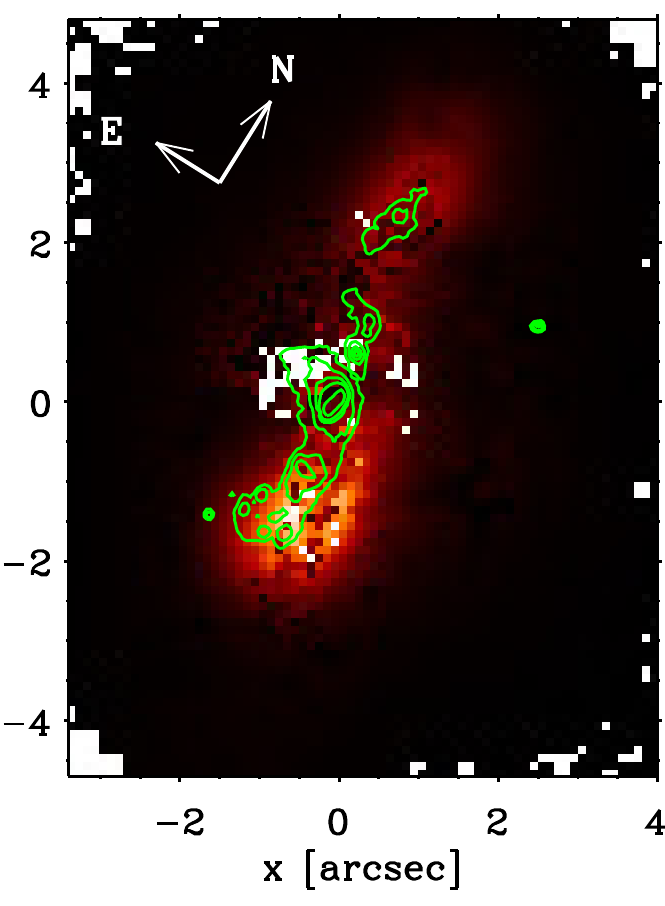}\\
\end{array}$
\end{center}
\caption{\textit{Left:} [OIII]$\lambda$5007 image from HST observations (WFPC2-PC/F502N, proposal ID: 6419, PI: A. Wilson, \citealt{ferruit2000}). \textit{Center:} [OIII] contours overlaid on the [NII] redshifted broad component flux. \textit{Right:} [OIII] contours overlaid on the [NII] narrow component flux. The images were aligned assuming a coincidence between the peak of the [OIII] emission and the continuum peak derived from our observation.}
\label{fig: oxigen_overplots}
\end{figure*}

Allowing for the lower spatial resolution of our data,
the spatial distribution of the [NII] flux is broadly consistent with the emission line
morphologies observed in HST WFPC2 narrow band images in
H$\alpha$ + [NII] and [OIII]$\lambda$5007 \citep[][see also Fig.\ref{fig: oxigen_overplots}]{ferruit2000}. The HST images, however, resolve substructure in the lobes, revealing several bright knots embedded in more diffuse structures, with the two brightest knots being located approximately 0\farcs5 north and 1$^{\prime\prime}$ south of the nucleus, respectively. In the HST images, the [OIII] emission has approximately same distribution as the H$\alpha$ + [NII] emission, although the knots are more prominent in [OIII] whereas the more diffuse structures are more prominent in H$\alpha$ + [NII].

The HST images also reveal a ``plume'' of emission (present in both H$\alpha$ + [NII] and [OIII]) that 
extends approximately 1$^{\prime\prime}$ to the east and north of nucleus, and which appears to be distinct from
the north-south lobes. This feature has no clear counterpart in our flux maps, but it corresponds to a handful of spaxels characterized by extreme line-blending, and it may be
related to the region of enhanced velocity dispersion seen in Fig.\ref{fig: [NII]1c}. Indeed, in 
the version of the H$\alpha$ + [NII] image presented by \citeauthor{ferruit2000} in their Fig. 5, a low surface brightness bar-like structure is present, crossing the nucleus from east to west, perpendicular to the lobes,
and extending approximately 2$^{\prime\prime}$ on each side. The plume may represent the inner part of this structure, which is approximately co-spatial with the region of enhanced velocity dispersion, and the velocity
residuals discussed below in Section \ref{subsec: gasv_model}.

The bright central region of the H$\alpha$ + [NII] flux map is also resolved by the HST images,
into an elongated structure approximately 0\farcs5 long and oriented slightly west of north
(see Fig.4 in \citeauthor{ferruit2000}, the feature is also visible in the structure map in Fig.\ref{fig: NGC1386}). Close inspection shows that it is composed of two distinct 
components, to the north and south, respectively, which have higher excitation 
(as measured by the [OIII]/H$\alpha$+[NII] line ratio) than the surrounding more diffuse emission.
These HST components appear to be partially resolved in the flux maps derived from
the broad kinematic components, corresponding to the asymmetric brightness distribution, relative to the
continuum peak, of the broad red- and blue-shifted components of the line profiles.
\vskip10pt

The flux distribution of the blue- and red-shifted components of the [FeVII]$\lambda$6087 emission line is shown in the right column of Fig.\ref{fig: Fe_maps}. As for the broad components of other lines, the redshifted component is brighter than the blueshifted one.

% Maps for [NII] with oxigen: ...........................

% =======================================================
% =======================================================
\subsection{Gas excitation}
\label{sec: gas_ex_result}

When we compute the [NII]/H$\alpha$ ratio we find that the narrow component has values close to 1.6 approximately within the lobes and at the nucleus while the broad components have lower values, i.e. $\approx$ 1.1 for the broad blue component and $\approx$ 1.2 for the broad red component.

\begin{figure}[p]
\begin{center}$
\begin{array}{c}
\includegraphics[trim=0cm 0cm 0cm 0.5cm, clip=true, scale=.64]{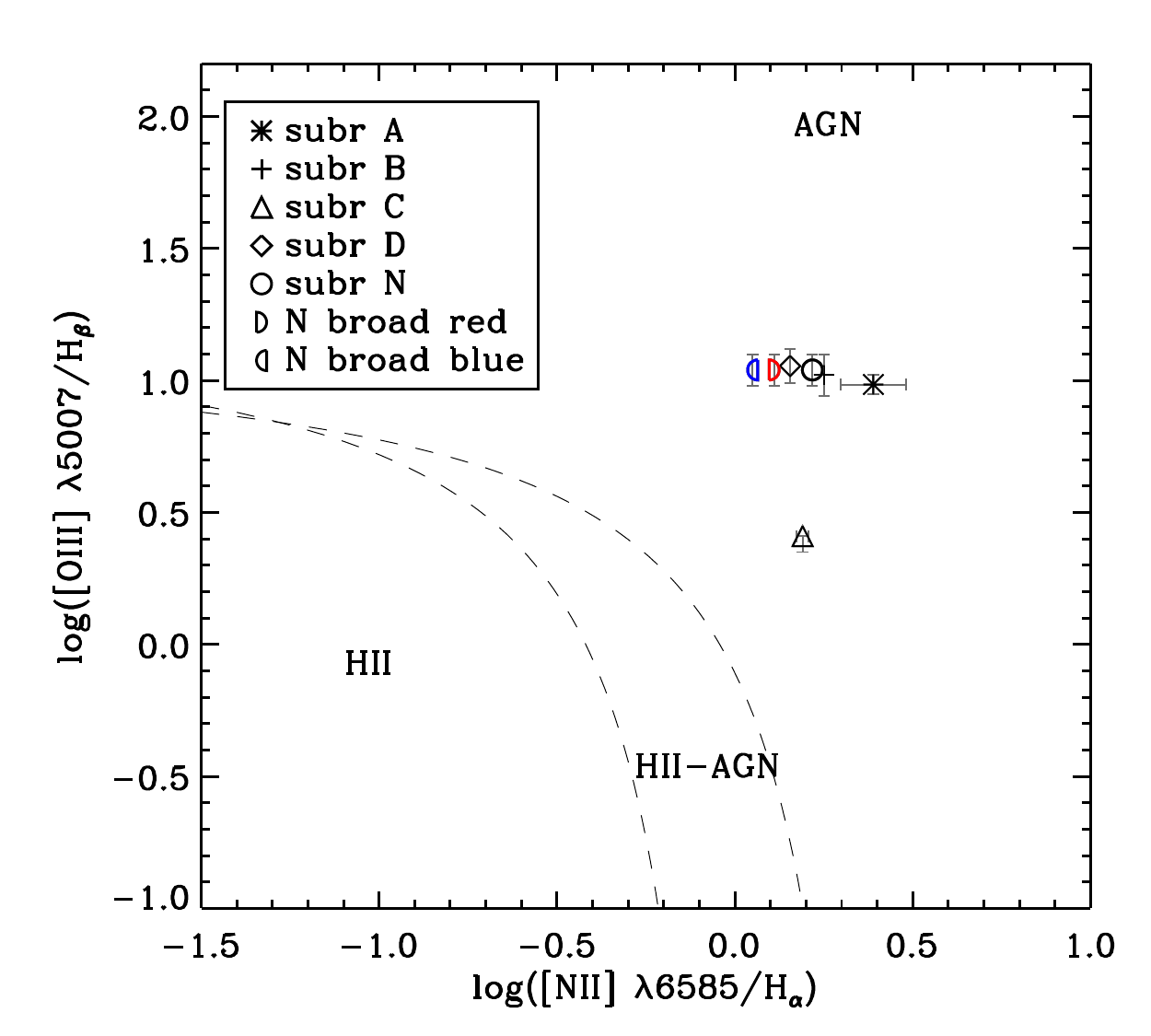}\\ 
 \includegraphics[trim=0cm 0cm 0cm 0.5cm, clip=true, scale=.64]{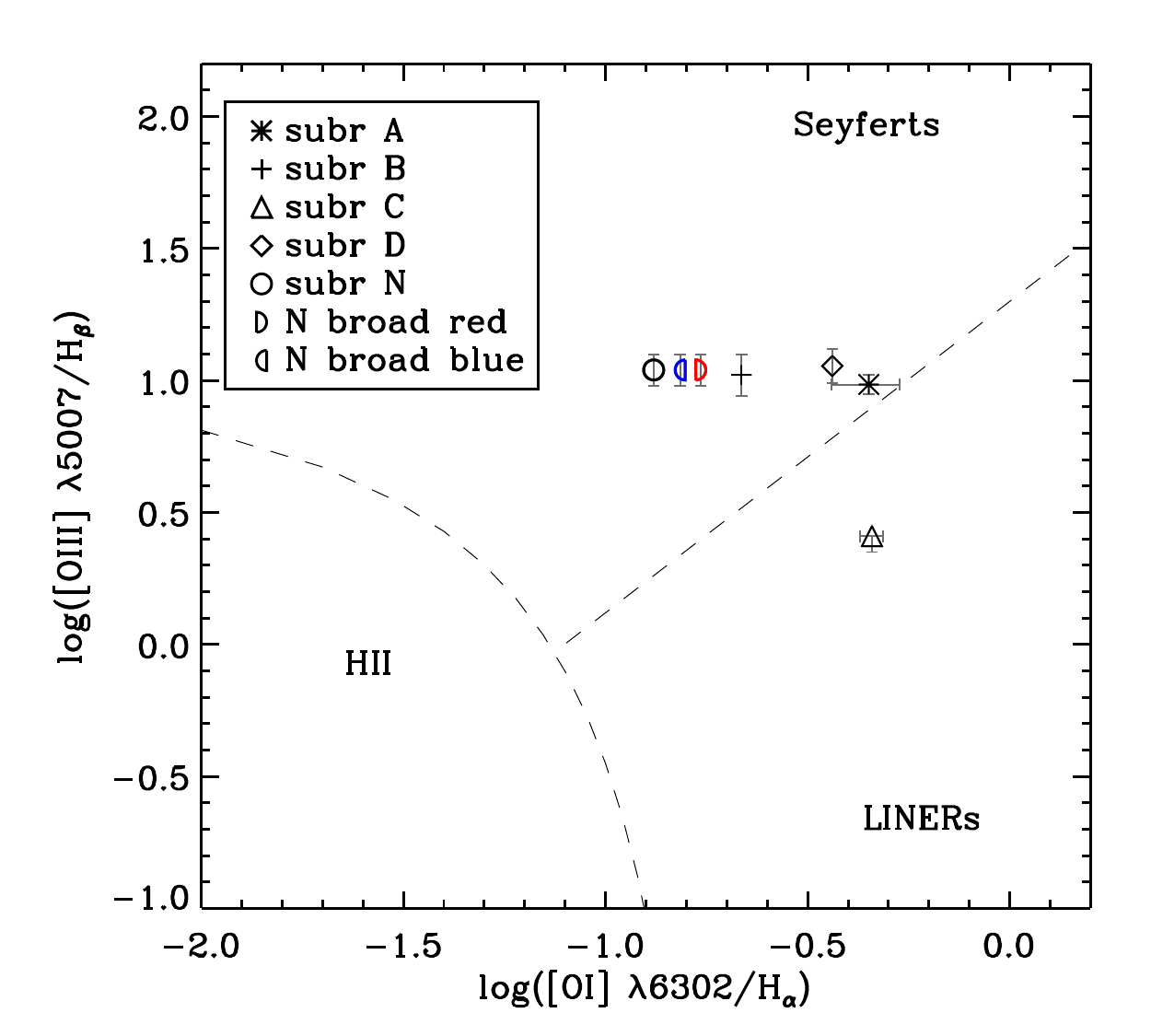}  \\
\includegraphics[trim=0cm 0.25cm 0cm 0.5cm, clip=true, scale=0.64]{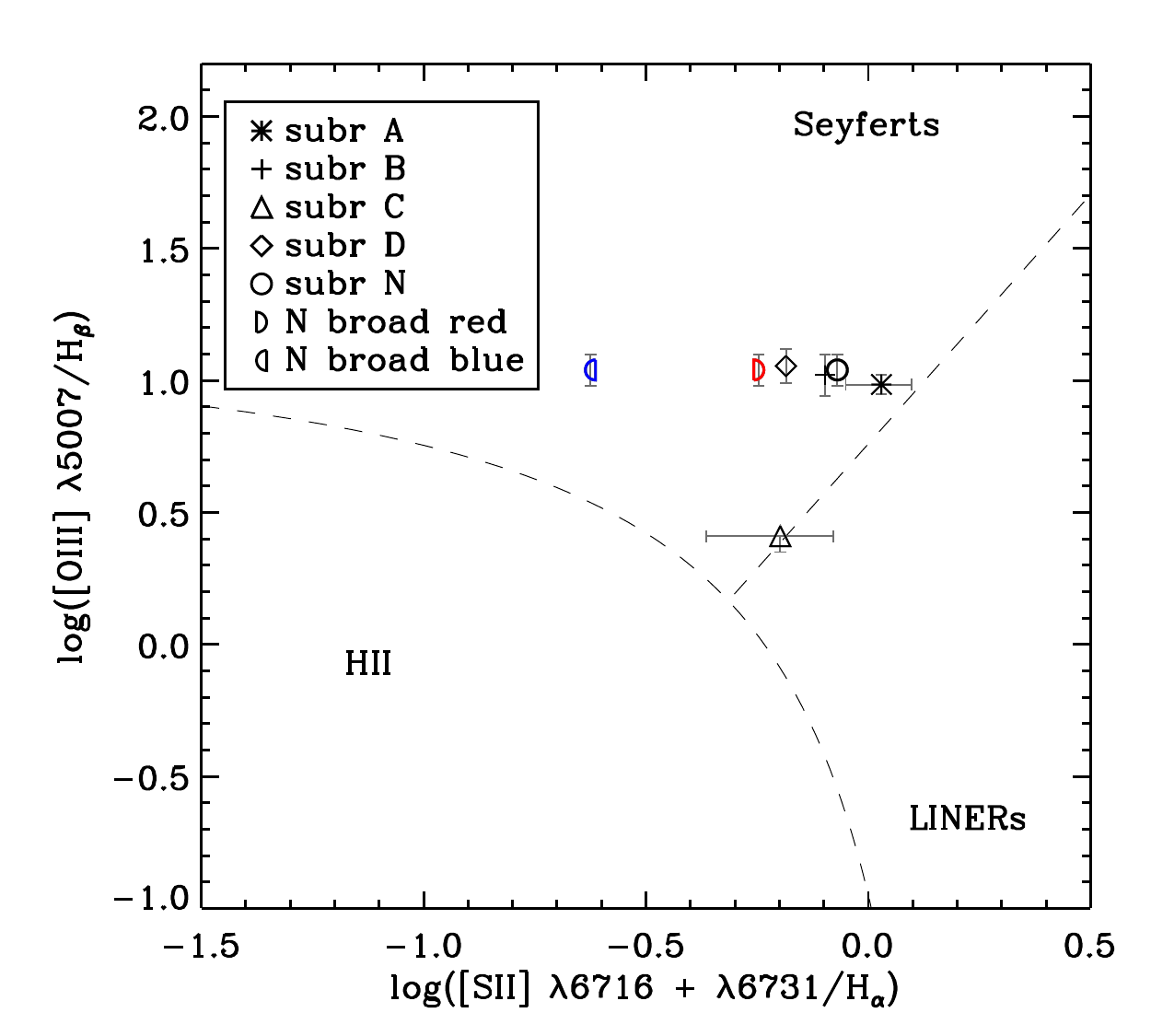} \\
\end{array}$
\end{center}
\caption{BPT diagrams for the subregions specified in Fig.\ref{fig: [NII]maps}. Values for the ratio [OIII] $\lambda$5007/H$\beta$ have been estimated from Fig.10 in \citet{WeaverWilson91}. Vertical error bars represent the range of values measured by \citeauthor{WeaverWilson91} within the regions of interest. Typical errors along the abscissa are smaller than the symbol size with the exception of points from subregion A and C. The dashed boundary lines are taken from \citet{KGKH06}.}
\label{fig: bpt}
\end{figure}

Line ratios derived from the simultaneous fit to the lines in the subregion spectra are given in Table \ref{tab: ratios_n} and plotted in the Baldwin-Phillips-Terlevich (BPT) diagrams \citep{BPT81} shown in Fig.\ref{fig: bpt}. As our spectral coverage does not include the [OIII] and H$\beta$ emission lines, we estimated the [OIII]/H$\beta$ ratio from Fig.10 in \citet{WeaverWilson91}. Over the whole FOV, the line ratios are well within the AGN photoionization region with the spectrum of subregion C showing ratios typical of LINERs and all of the others falling into the Seyfert region. 

This is consistent with results obtained by \citet{bennert06} who performed long slit observations of NGC 1386, along the north-south axis of the [NII] lobes and the extended [OIII] emission, finding line ratios typical of HII regions only beyond 6$^{\prime\prime}$ from the nucleus.

% =======================================================
% =======================================================
\subsection{Electron density}

\begin{figure*}[t]
\begin{center}$
\begin{array}{ccc}
\includegraphics[trim=6cm 1.5cm 4cm 0cm, clip=true, scale=0.675]{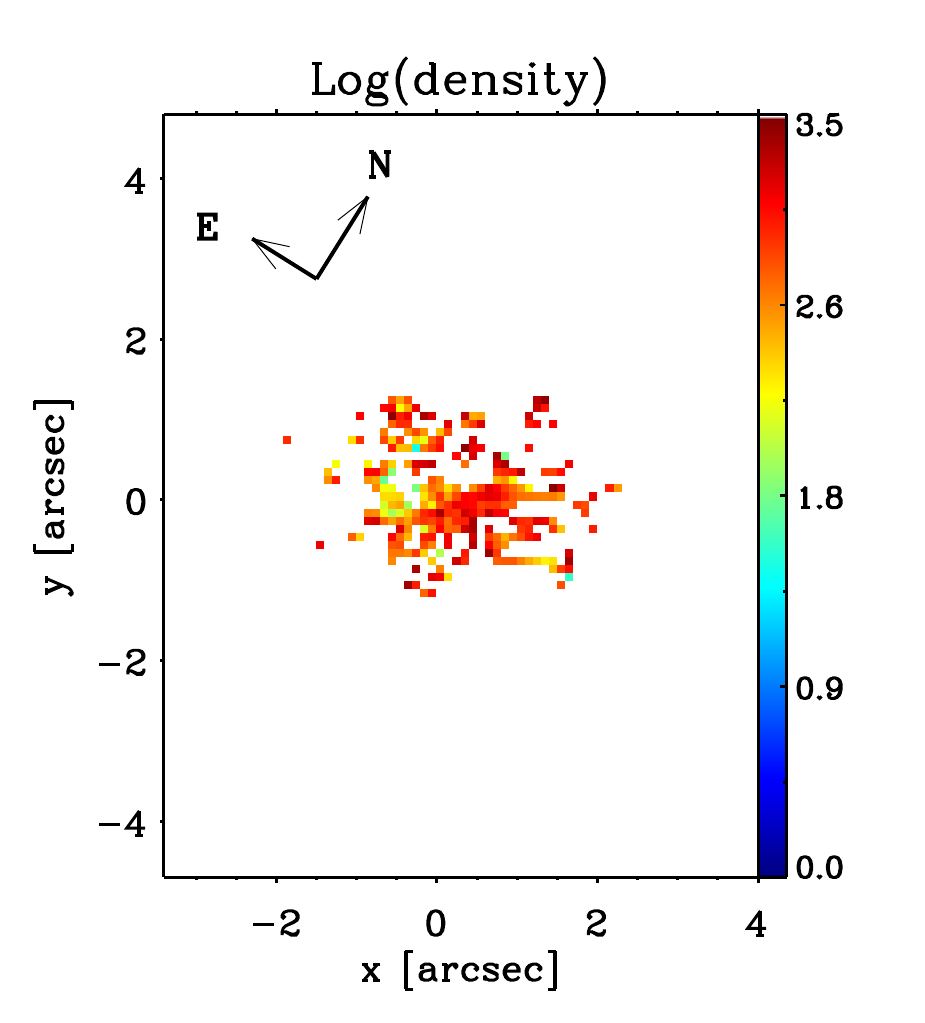} & \includegraphics[trim=7cm 1.5cm 4cm 0cm, clip=true, scale=0.67]{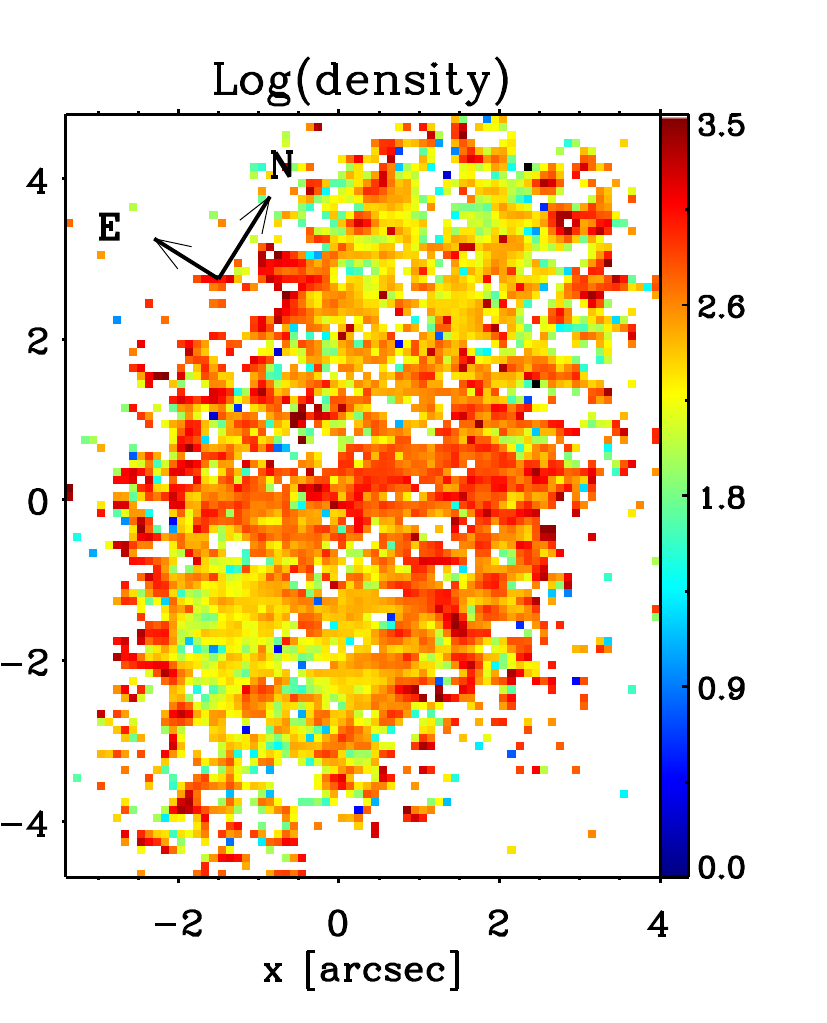} 	& \includegraphics[trim=7cm 1.5cm 4cm 0cm, clip=true, scale=0.67]{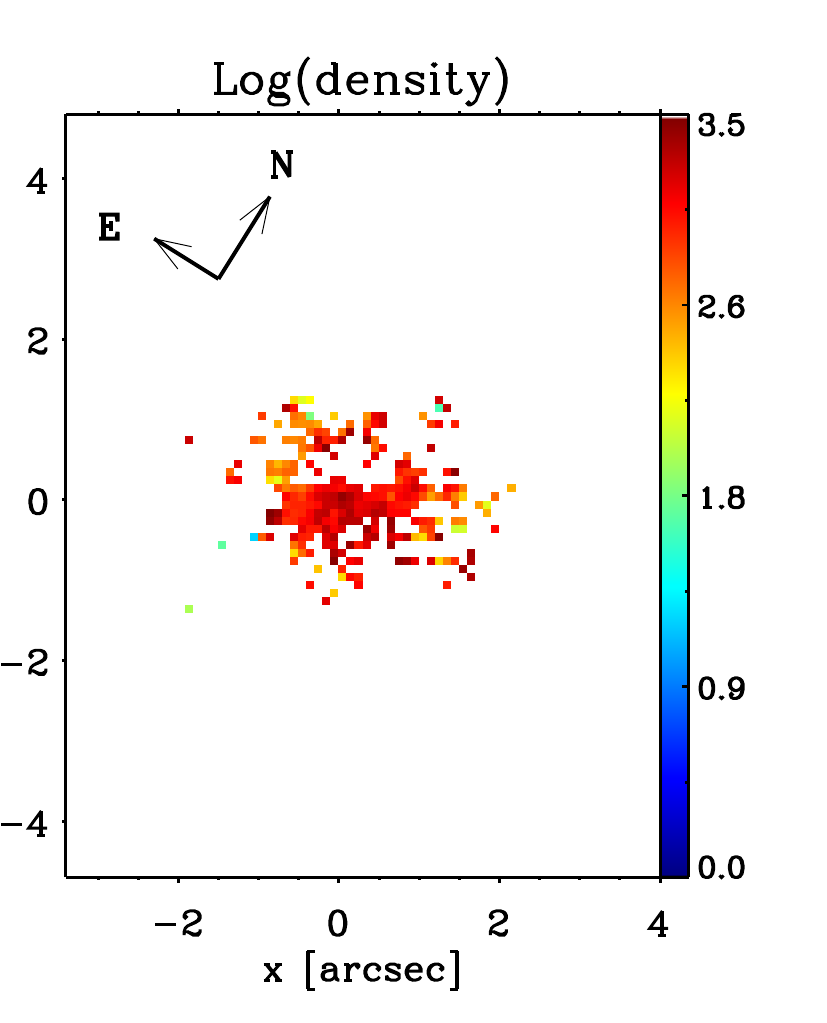}\\
\end{array}$
\end{center}
\caption{Electron density maps derived from the ratio [SII]$\lambda$6716/$\lambda$6731. \textit{Left:} blueshifted broad component. \textit{Center:} narrow component. \textit{Right:} redshifted broad component. Units: cm$^{-3}$. The color scale is arbitrarily adjusted to highlight the features of the maps.}
\label{fig: density}
\end{figure*}

Electron densities were derived from the intensity ratio [SII]$\lambda$6716/$\lambda$6731 assuming a temperature of 10$^{4}$K. The resulting maps are presented in Fig.\ref{fig: density}. For the narrow component, the highest densities are found along the direction SE-NW, in particular in the inner 1$^{\prime\prime}$ around the nucleus, reaching values $\approx$ 800 cm$^{-3}$. Beyond this region, to the north and south of the nucleus, the density of the emitting gas ranges between 150 and 300 cm$^{-3}$. 

High densities are also evident in the broad component maps. Median values of $\approx$ 800 cm$^{-3}$ are found for the blue broad component and $\approx$ 1100 cm$^{-3}$ for the red broad component. 

Values for the electron density derived from the spectra extracted from the subregions identified in Fig.\ref{fig: [NII]maps} are given in the last column of Table \ref{tab: ratios_n}. 
It was not possible to derive the electron density for the blueshifted broad component of subregion N since the measured value of the [SII]$\lambda$6716/$\lambda$6731 ratio (2.8) is well above the high density limit (approximately $1.4$, \citealt{Osterbrock_book89}); the high value seems to result from a poor fit due to heavy blending of the lines. 
\vskip10pt

Bearing in mind the different nature and modeling of the data, a consistent decrease in the density as a function of the distance from the nucleus was also derived from VLT-FORS1 long slit spectroscopy by \citet{bennert06}.

% BPT diagrams: ==================

% Density maps: ==================

% ============================================================
% ~~~~~~~~~~~~~~~~~~~~~~~~~~~~~~~~~~~~~~~~~~~~~~~~~~~~~~~~~~~~
% ============================================================
\section{Discussion} \label{sec: disc}

% ============================================================
% ============================================================
\subsection{The stellar velocity field}
\label{subsection: stellarv}

We fitted the stellar velocity field with a kinematic model describing circular orbits in a plane \citep{VDKruitAllen78, BertolaBDSS91}: 

\begin{align} \nonumber
\label{eq: vmod}
&v_\mathrm{mod}(r,\psi) 	 = v_\mathrm{sys} +\\ 
		& \frac{Ar\ \mathrm{cos}(\psi - \psi_{0})\ \mathrm{sin}\theta \ \mathrm{cos}^{p}\theta}{\{r^{2}[\mathrm{sin}^{2}(\psi - \psi_{0}) + \mathrm{cos}^{2}\theta\ \mathrm{cos}^{2}(\psi - \psi_{0})] + c^{2}_{0}\mathrm{cos}^{2}\theta\}^{p/2}}.
\end{align}

This yields a velocity curve that increases linearly at small radii and becomes proportional to $r^{(1-p)}$ at large radii.
The parameter $v_\mathrm{sys}$ is the systemic velocity, $A$ is the amplitude of the rotation curve, $r$ and $\psi$ are the radial and angular coordinates of a given pixel in the plane of the sky, $\psi_{0}$ is the position angle of the line of nodes (measured with respect to the $x$-axis of the image frame, increasing counterclockwise; $ \psi_{0} - 65^{\circ}$ gives the value in degrees east of north), $\theta$ is the disk inclination ($\theta$ = 0 for face-on disks). The parameter $p$ measures the slope of the rotation curve where it flattens, in the outer region of the galaxy, varying in the range 1-1.5. Finally, $c_{0}$ is a concentration parameter, which gives the radius at which the velocity reaches 70\% of the amplitude A. 

Because of the limited size of the region analyzed, the parameter $p$ is poorly constrained by the data, therefore we assume $p=1$, which corresponds to an asymptotically flat rotation curve at large radii. The stellar velocity curve already seems to flatten at $r \approx 1^{\prime\prime}$ supporting this choice for $p$. 
The inclination $\theta$ is also poorly constrained. Assuming that the gas lies on a thin circular disk, we adopted the value $\theta = 65^\circ$ given by NED and derived from the 2MASS K band axis ratio of the large-scale structure of the galaxy. The same value is also recovered from PAH emission within 15$^{\prime\prime}$ of the nucleus \citep{RdPRS14}.

\begin{figure*}[htb]
\begin{center}$
\begin{array}{clclc}
\includegraphics[trim=6cm 1.7cm 4cm 0cm, clip=true, scale=0.63]{vel_sys} & & \includegraphics[trim=7cm 1.7cm 4cm 0cm, clip=true, scale=0.63]{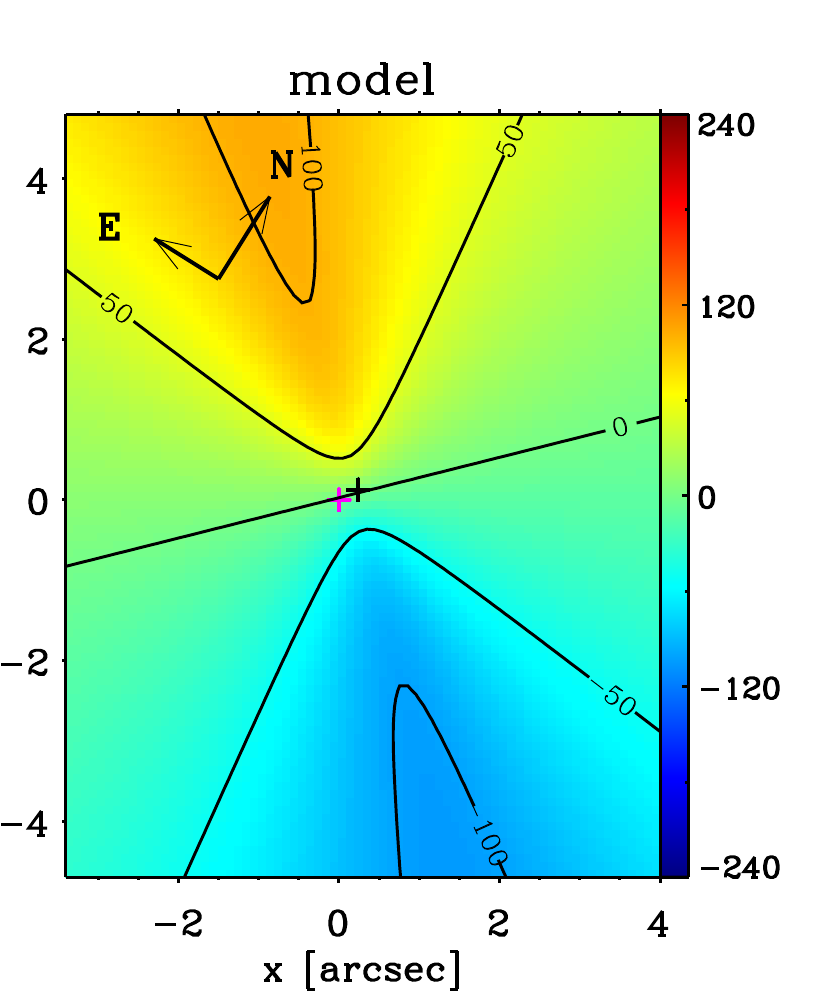} & & \includegraphics[trim=7cm 1.7cm 4cm 0cm, clip=true, scale=0.63]{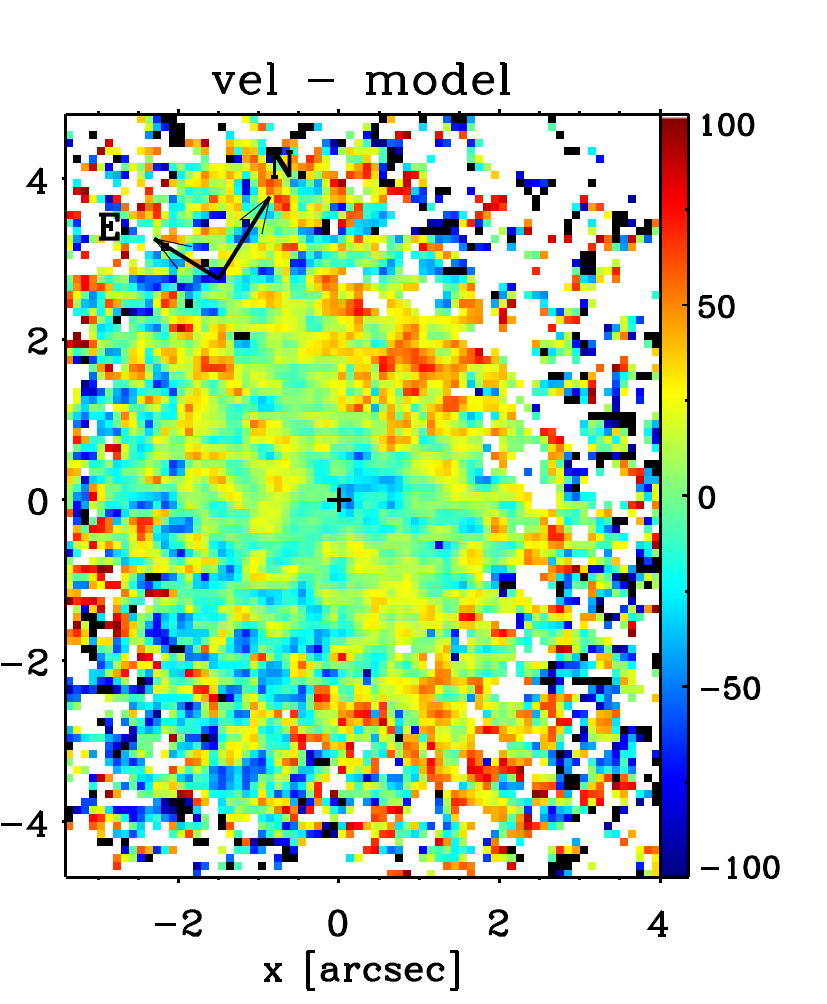}  \\
\end{array}$
\end{center}
\caption{\textit{Left:} velocity map [km s$^{-1}$] as derived from the stellar absorption features excluding the [NaD] doublet. \textit{Center:} rotation model fit. \textit{Right:} residual.}
\label{fig: rotationms}
\end{figure*}

We used a Levenberg-Marquardt least-squares algorithm to fit the rotation model to the velocity map with the fit determining values for the parameters ($A$, $v_\mathrm{sys}$, $\psi_{0}$, c$_{0}$) and the center of the rotation field (x0, y0). These values are listed in Table \ref{tab: vmod_star} where results are reported for a fit in which the kinematical center is fixed at the continuum peak and for a fit in which the center is left a free parameter. 
Maps of the observed and best fit model velocity field are shown in Fig.\ref{fig: rotationms} together with the residual map for the fit performed with a fixed center.

The fit performed with the kinematical center (x$_{0}$,y$_{0}$) left as a free parameter returns a shift from the photometric nucleus which is 0\farcs2 north and 0\farcs1 west of the continuum peak. Considering our angular resolution (0\farcs9), this is negligible and we conclude that the stellar kinematical center coincides (within the uncertainties) with the photometric nucleus (N).

The sparsity of data to the north of the nucleus in the velocity map (Fig.\ref{fig: rotationms}) might introduce a bias in the position angle of the line of nodes, $\psi_{0}$, toward large values (eastward of north). To test this possibility, we repeated the fit to a symmetrized version of stellar velocity map, which was created by reflecting the south-eastern side of the map about the vertical axis passing through the center of the image. All of the fitted parameters, and in particular $\psi_{0}$, showed negligible variations suggesting that the recovered value of $\psi_{0}$ is not affected by the spatial distribution of the flagged pixels.

% ============================================================
% ============================================================
\subsubsection{Stability of the solution} 

\textbf{Dependence on the noise:} to estimate the uncertainties in the recovered parameters due to the observed noise level, we fitted the velocity maps recovered from the Monte Carlo simulation described in section \textsection\ref{sec: uncertainties}. Standard deviations for the distribution of the recovered parameters were computed and are given in Table \ref{tab: vmod_star}. The estimated uncertainties for the amplitude of the velocity curve, A, and the systemic velocity, $v_\mathrm{sys}$, were smaller than 1 km s$^{-1}$.

\textbf{Dependence on the initial guess:} to test the dependence of the recovered parameters on the initial guess, we performed 200 fits of the stellar velocity map where the initial guess for each parameter was randomly drawn from a uniform distribution. To avoid unreasonable initial guesses that would generate obviously bad fits, the limits of the initial guess distribution for each parameter were determined from visual inspection of the map to be fitted. The range of values allowed for the initial guess is given in the rightmost column of Table \ref{tab: vmod_star}.

The procedure was performed with both the center as a free parameter and the center fixed at the nucleus. In both cases all the fits converged exactly to the same solution and none of the coefficients reached the boundaries allowed for the fit.

\textbf{Dependence on the value adopted for p:} when $p$ is fixed at values larger than 1 (implying a velocity curve which decreases at large radii), the amplitude increases steadily, reaching a value of 300 km s$^{-1}$ when $p = 1.5$. A similar increase is found for the parameter $c_{0}$ (as expected from its definition) while the other fitted parameters, v$_{sys}$ and $\phi_{0}$, are almost insensitive to the value adopted for $p$.

\begin{table}[t]
\caption{FITS TO THE OBSERVED STELLAR VELOCITY FIELD}  
\begin{tabular}{l lll}
\hline
\hline
	 Parameter				&  				& Notes					&	Initial guess\\
\hline
& &    \\

 \multicolumn{4}{c}{Fixed center}\\
 \hline
 & & \\
 	A [km s$^{-1}$] ..........		& 118 $\pm$ 1		& 						& 100:300\\
	$v_\mathrm{sys,\star}$ [km s$^{-1}$] ...	& 743 $\pm$ 1		& geocentric$^{\ddagger}$	& 500:900\\
	$\psi_{0}$ [deg] ............... 		& 103 $\pm$ 0.2	& 						& 60:135\\
	c$_{0}$ [arcsec] ............		& 1.02 $\pm$ 0.03	& 						& 0:5\\
	p ..........................			& 1				& fixed \\
	$\theta$ [deg] .................		& 65				& fixed\\
	x$_{0}$ [arcsec] ...........		& 0				& fixed at N \\
	y$_{0}$ [arcsec] ...........		& 0				& fixed at N\\
	
& &    \\

 \multicolumn{4}{c}{Free center}\\
 \hline
& &    \\
	A [km s$^{-1}$] ..........		& 117 $\pm$ 1		& 						& 100:300\\
	$v_\mathrm{sys,\star}$ [km s$^{-1}$] ...	& 744 $\pm$ 1		& geocentric				& 500:900\\
	$\psi_{0}$ [deg] ............... 		& 104 $\pm$ 0.2	& 						& 60:135\\
	c$_{0}$ [arcsec] ............		& 0.85 $\pm$ 0.04	& 						& 0:5\\
	p ..........................			& 1				& fixed \\
	$\theta$ [deg] ..................		& 65				& fixed\\
	x$_{0}$ [arcsec] ............		& + 0.2 $^{\dagger}$	& from N					& -1:1\\
	y$_{0}$ [arcsec] ............		& + 0.1 $^{\dagger}$	& from N 					& -1:1\\

 & & \\
\hline
\end{tabular}
\label{tab: vmod_star}
\tablecomments{$^{\dagger}$ The formal uncertainty estimated for this coefficient is much smaller than 0\farcs1. $^{\ddagger}$ The heliocentric systemic velocity is $v_\mathrm{sys,\star}$ = 754 km s$^{-1}$.}
\end{table}

% ============================================================
% ============================================================
\subsection{The gas velocity field}
\label{subsec: gasv_model}

The velocity map derived from the narrow component covers the whole FOV and shows a pattern typical of rotation plus some distortion (see Fig.\ref{fig: [NII]maps}, first column, second row). In order to isolate non-rotational motions and also to determine the properties of the rotation, we fitted the kinematic model described by eq.\ref{eq: vmod} to the gas velocity field. The model, in this case, represents gas rotating in a plane with circular orbits.

Pixels with errors in excess of 50 km s$^{-1}$ were masked prior to performing the fit. The derived parameters are given in Table \ref{tab: vmod_gas_totl} for three realizations of the model: in the first case, model \textit{1GK}, the position angle of the line of nodes was fixed at the value recovered from the stellar kinematics and the center was fixed at the continuum peak (which coincides with the kinematical center derived from the stellar velocity field); in the second case, model \textit{2GK}, the position angle of the line of nodes was a free parameter and the center was fixed; in the third realization, model \textit{3GK}, both the position angle of the line of nodes and the center were free to vary. As for the fit to the stellar kinematics, for all models the parameters $p$ and $\theta$ were held fixed at values of 1 and $65^{\circ}$ respectively.

Assuming $p = 1$ implies a flat rotation curve at large radii. \citet{SchHnk03} obtained single slit observations of NGC 1386, and used the H$\alpha$ and [NII]$\lambda$6583 emission lines to derive the rotation curve out to 15$^{\prime\prime}$ from the nucleus, along the galaxy major axis. From their Fig.4 it is evident that the rotation curve is still rising within the region probed in our analysis. It is therefore clear that neither $p$ nor the amplitude, $A$, can be well constrained by our data. Nevertheless, the main goal of fitting a rotation curve to the observed velocity field is to isolate radial motions close to the nucleus, rather than to characterize the galaxy-scale rotation curve (which would require data from a wider range of radii).

In model \textit{1GK} the amplitude reaches a high value (308 km s$^{-1}$) which is strongly dependent on the position angle assumed for the line of nodes, e.g. a variation of 1$^{\circ}$ in the adopted value for $\psi_{0}$ corresponds to a variation of 10 - 15 km s$^{-1}$ in $A$.

The systemic velocity derived from this model is about 40 km s$^{-1}$ higher than the value obtained from the stellar kinematics.
This difference could be due to the presence of gaseous radial motions superposed on the rotation pattern. 
The channel maps in Fig.\ref{fig: channelm} seem to suggest a systemic velocity, $v_{cm}$, which is about 70 km s$^{-1}$ higher than derived from the modeling of gas kinematics. This would make the channel maps kinematically symmetric with respect to the zero velocity map. Nevertheless, $v_{cm}$ is clearly not consistent with the double peaked velocity distribution of the narrow component (i.e. the distribution of the velocity values is not symmetric with respect to $v_{cm}$).    

\begin{figure*}[!h]
\begin{center}$
\begin{array}{clclc}
\includegraphics[trim=6.2cm 1.7cm 4cm 0cm, clip=true, scale=0.63]{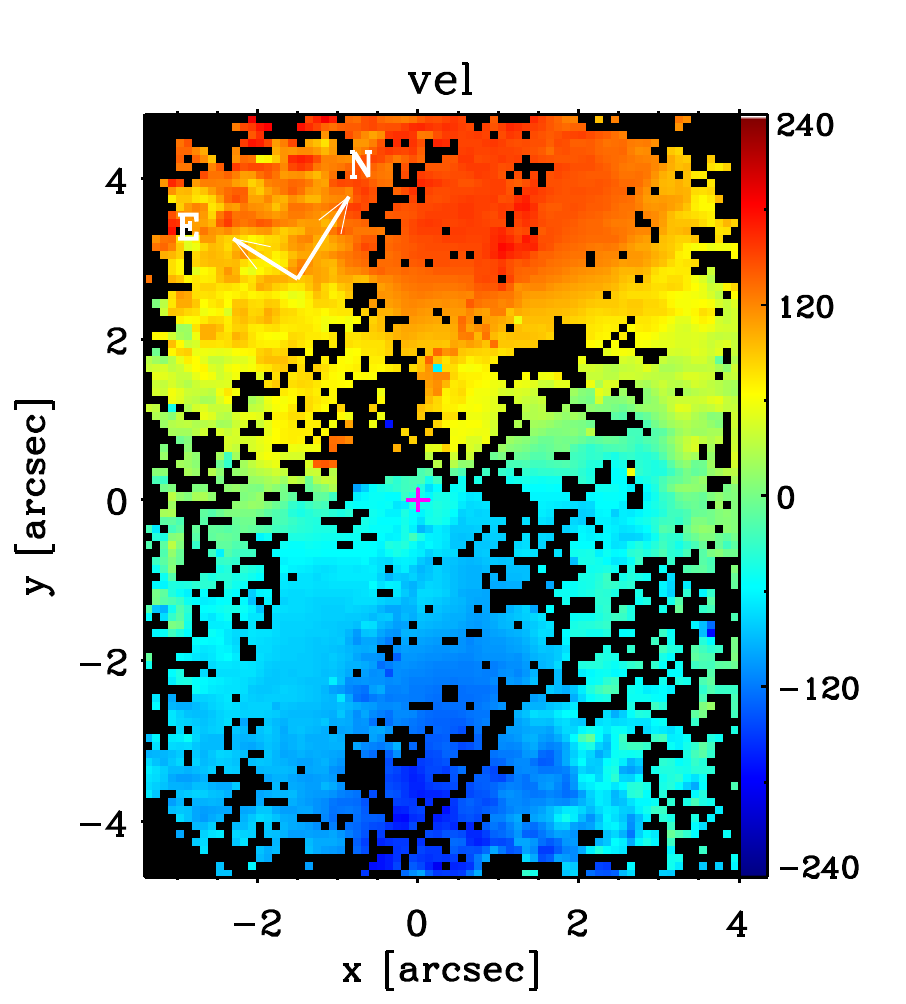} & &\includegraphics[trim=7cm 1.7cm 4cm 0cm, clip=true, scale=0.63]{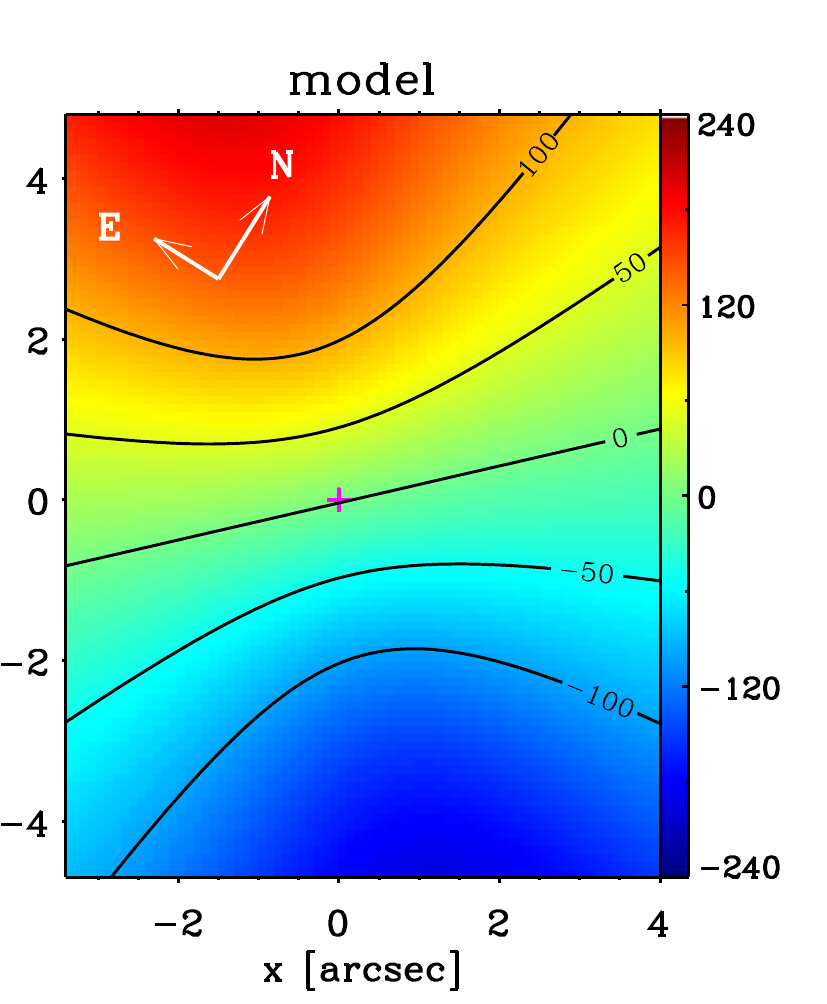} & &\includegraphics[trim=7cm 1.7cm 4cm 0cm, clip=true, scale=0.63]{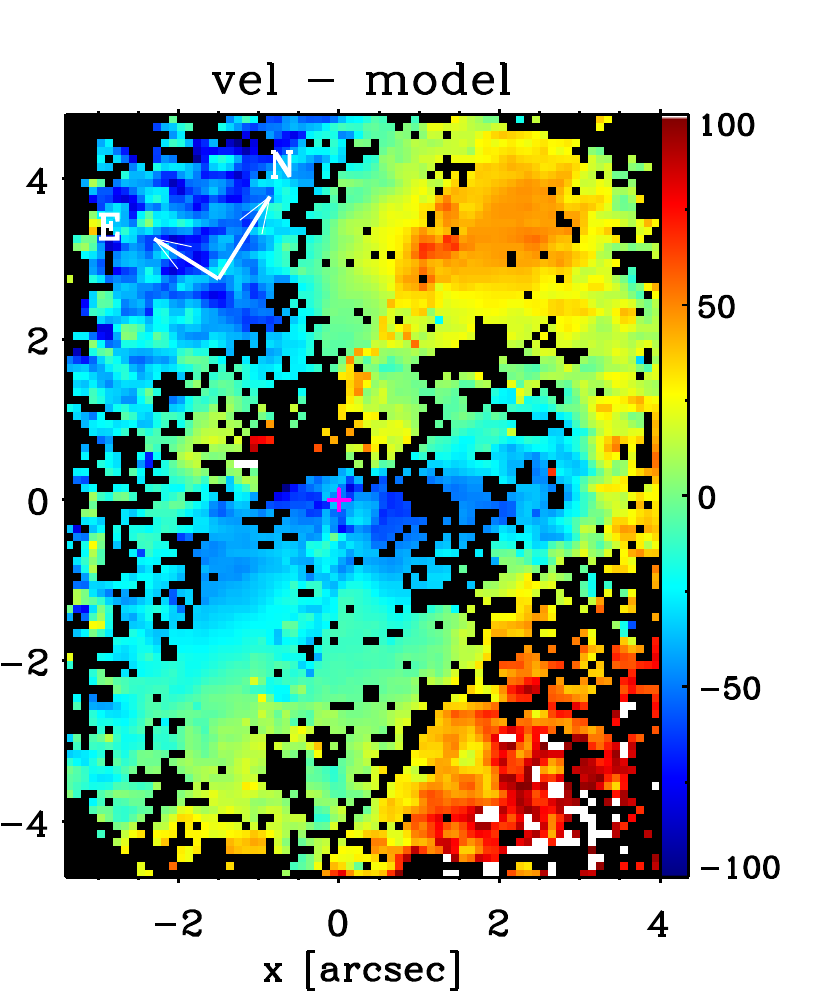}  \\
&&\\
\includegraphics[trim=6.2cm 1.5cm 4cm 1cm, clip=true, scale=0.63]{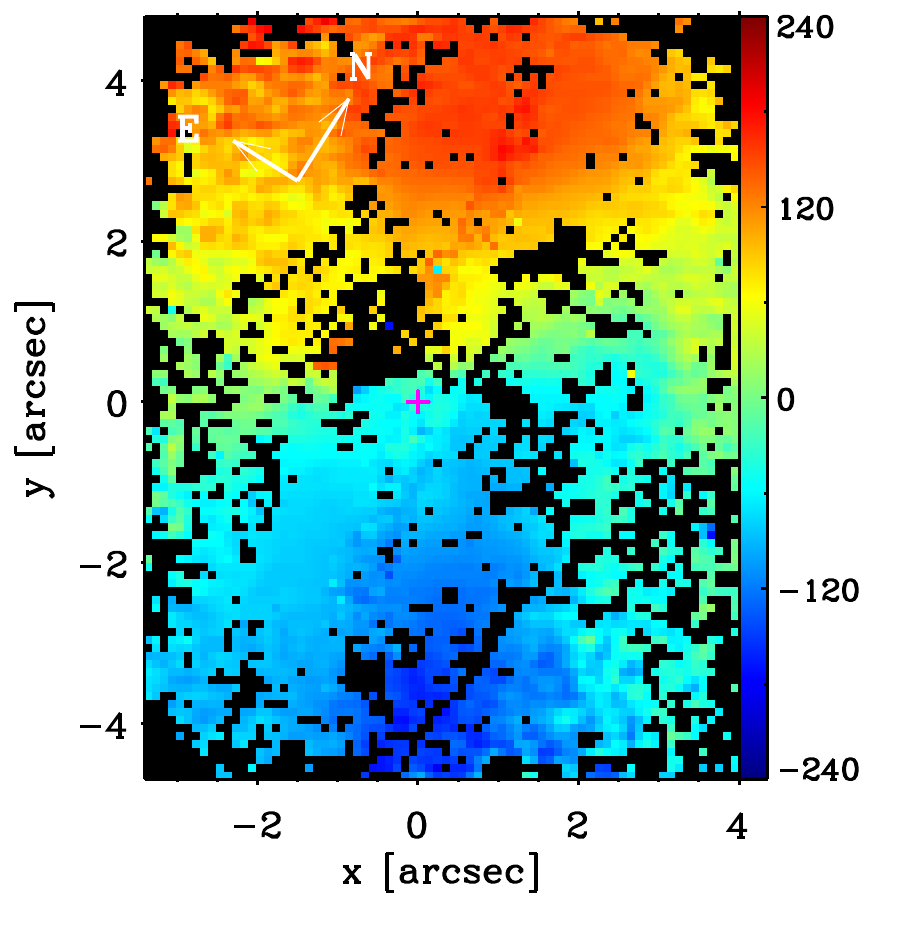} & &\includegraphics[trim=7cm 1.5cm 4cm 1cm, clip=true, scale=0.63]{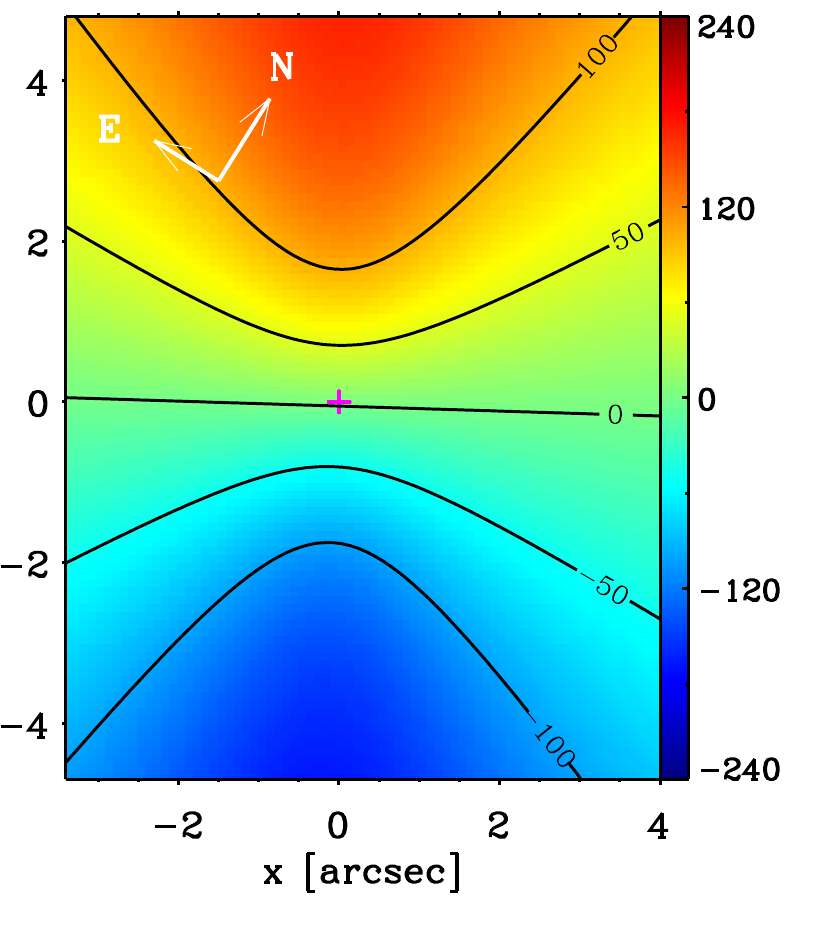} & &\includegraphics[trim=7cm 1.5cm 4cm 1cm, clip=true, scale=0.63]{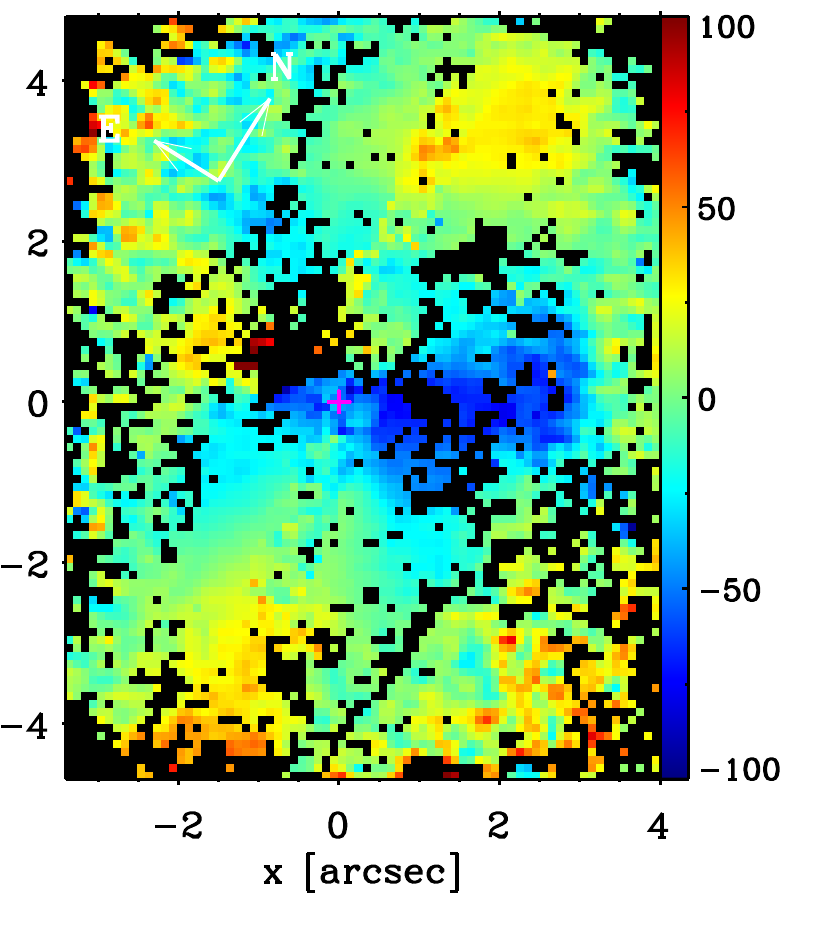}  \\
&&\\
\includegraphics[trim=6.2cm 1.7cm 4cm 1cm, clip=true, scale=0.63]{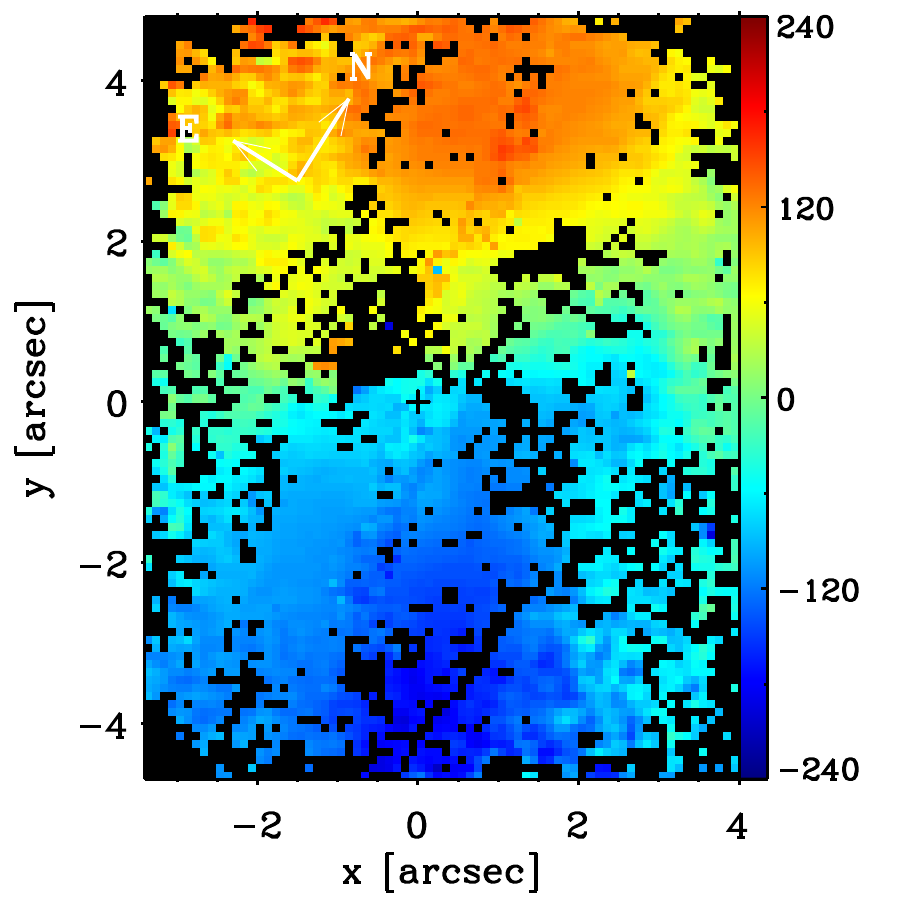} & &\includegraphics[trim=7cm 1.7cm 4cm 1cm, clip=true, scale=0.63]{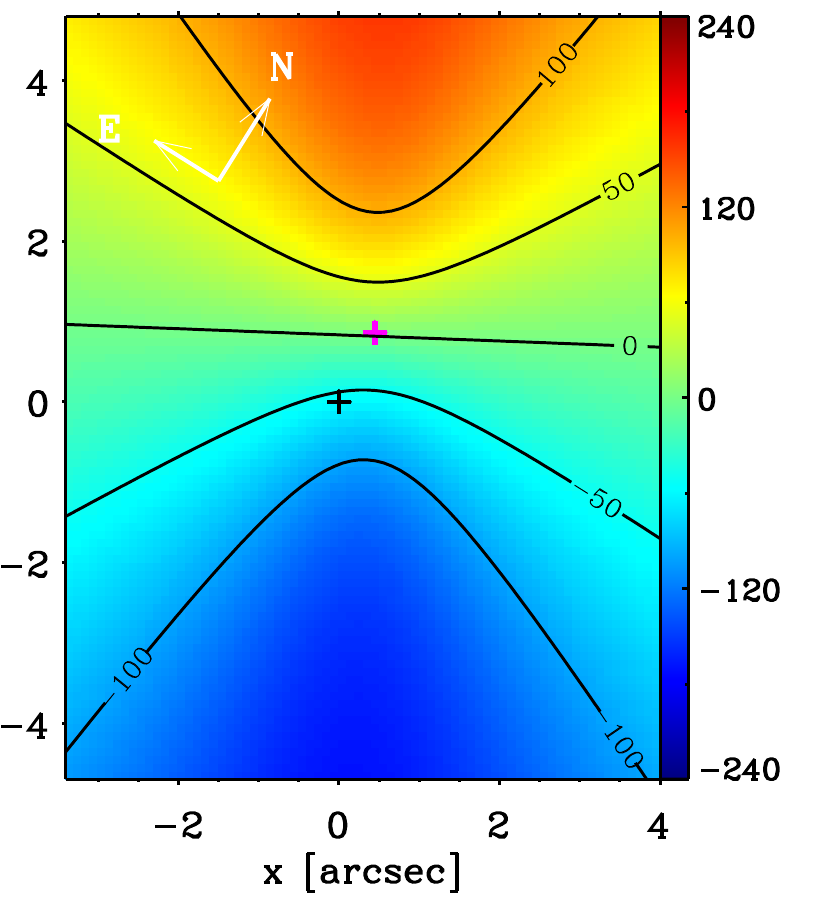} & &\includegraphics[trim=7cm 1.7cm 4cm 1cm, clip=true, scale=0.63]{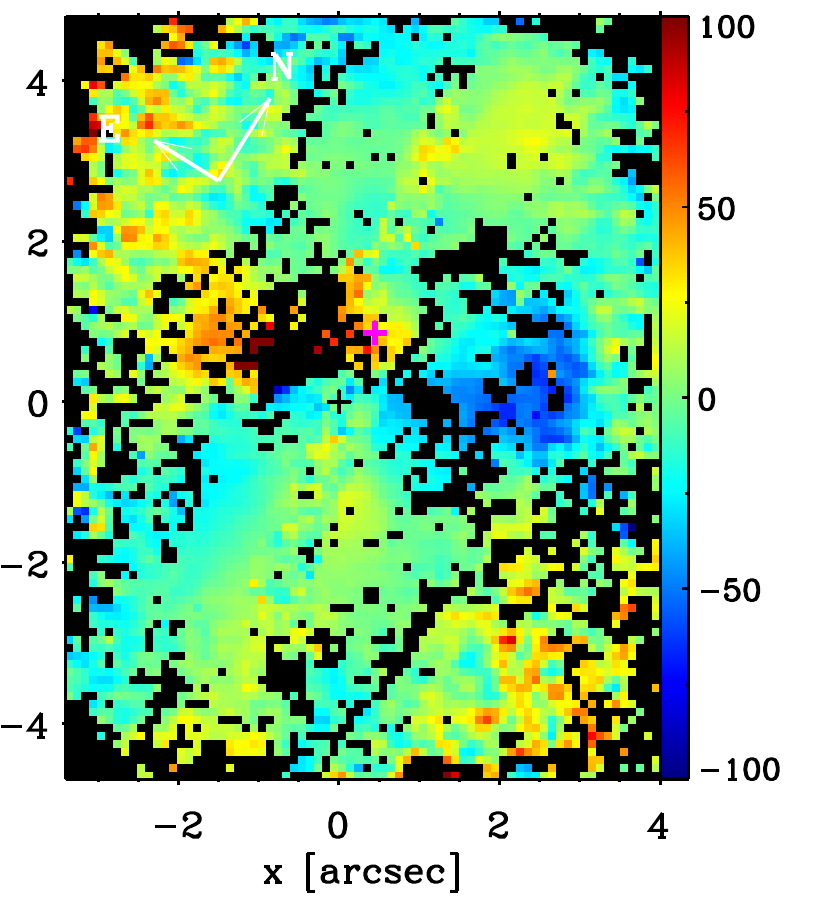}  \\
\end{array}$
\end{center}
\caption{\textit{Top:} model \textit{1GK}, the center is fixed at the continuum peak and the line of nodes is fixed at the value derived from stellar kinematics. \textit{Middle:} model \textit{2GK}, the line of nodes is a free parameter and the center is fixed at the continuum peak. \textit{Bottom:} model \textit{3GK}, line of nodes and center are free parameters. \textit{Left:} velocity map [km s$^{-1}$] as derived from the narrow component of the [NII]$\lambda$6583 emission line; pixels with errors larger than 50 km s$^{-1}$ are masked (in black). \textit{Center:} rotation model fit [km s$^{-1}$]. \textit{Right:} residual [km s$^{-1}$]. Magenta cross: kinematical center. Black cross: continuum peak. Maps are shown after subtraction of the systemic velocity derived from each fit. The fit parameters are given in Table \ref{tab: vmod_gas_totl}.}
\label{fig: rotationm}
\end{figure*}

Models 2GK and 3GK produce relatively small residuals ($\lesssim 30$\,km\,s$^{-1}$) 
over most of the FOV (with the exceptions noted below) indicating that the gas velocity field
is consistent with a rotating disk at a position angle $\psi_{0} = 88^{\circ}$ 
(middle and bottom rows of Fig.\ref{fig: rotationm}).
This differs by 15$^{\circ}$ with respect to the position angle obtained from the fit to the stellar kinematics. 
Offsets of comparable size
have been found in several other Seyfert galaxies and appear to be 
more common in Seyfert than in quiescent galaxies, 
perhaps indicating that weak non-axisymmetric perturbations are more 
prevalent in the former \citep{dumasMEN07}.
\vskip10pt

In model 3GK, where the coordinates of the kinematical center are free parameters in the fit,
the recovered center is offset 1$^{\prime\prime}$ north of the nucleus. Offsets of similar size, 
although in different directions, were also found by \cite{WeaverWilson91}, \cite{SchHnk03} and \cite{bennert06}. It is possible that this offset is related to the
distortion in the velocity field that is clearly visible as the blueshifted residuals (discussed below)
north-west of the nucleus. The distortion is not due to misfits related to the presence of
the two broader components, it seems to be a genuine feature of the narrow component.

In model 2GK, we assume that the kinematical center is that
recovered from the fit to the stellar velocity field, which is consistent with the continuum peak,
and hence the fit was performed with the center fixed at the continuum peak, (0,0). 
It is likely that the gas kinematical center is affected by the presence of inflows or outflows, while the stellar kinematical center should be a more reliable tracer of the true gravitational kinematical center. Therefore, we consider the parameters recovered from model 2GK to be more reliable than those  derived from model 3GK.

The overall pattern of the residuals obtained from models 2GK and 3GK is similar to that produced by model 1GK. However, there are differences in the relative
amplitudes in the sense that for 2GK and 3GK, residuals in the outer regions (beyond 2$^{\prime\prime}$ from the nucleus) 
are weaker relative to those in the inner (within 2$^{\prime\prime}$) regions. 
An interesting feature in the residual maps obtained from these models is a blueshifted region 2$^{\prime\prime}$ - 3$^{\prime\prime}$ to the north-west of the nucleus, where the residual velocities reach amplitudes of 50 - 70 km s$^{-1}$. This region is most prominent in model 2GK, but is clearly present in model 1GK. There is evidence of a redshifted counterpart to the east of the nucleus. Unfortunately, the H$\alpha$ + [NII] lines are heavily blended over much of this region (black pixels) preventing a satisfactory line profile decomposition (Section \ref{sec: line_fitting}). Nevertheless, redshifted residuals are present around the edges of the blended area in all models, strongly suggesting that a redshifted counterpart is indeed present. The combination of blue and red-shifted residuals together indicate the present of a distinct kinematical component spanning the nucleus, approximately along the galaxy minor axis.

Comparison with the velocity dispersion map presented in Fig.\ref{fig: [NII]1c} indicates that this residual is associated
with the region of high velocity dispersion extending from $-2^{\prime\prime}$ to $+3^{\prime\prime}$ along the SE-NW
direction (i.e., roughly along the galaxy minor axis) 
and smoothly transitioning to lower velocity dispersions. As noted in Section \ref{subsec: line_flux},
these kinematic features are approximately co-spatial with
a low surface brightness bar-like structure seen in the HST H$\alpha$ $+$ [NII] image of \citealt{ferruit2000}.

The velocity dispersion map presented in Fig.\ref{fig: [NII]1c} shows that these residuals are associated with a region of high velocity dispersion extending from $-2^{\prime\prime}$ to +3$^{\prime\prime}$ along the SE-NW direction and smoothly transitioning to lower velocity dispersion values.

\vskip10pt
As evident from the velocity maps in Fig.\ref{fig: [NII]maps} the two broad components are systematically blue- and redshifted with respect to the larger-scale narrow component, suggesting the presence of either an inflow or an outflow. Their mean velocities relative to the systemic velocity derived from the kinematics of the narrow component are $-140$ km s$^{-1}$ and 250 km s$^{-1}$ respectively. These broad components appear co-spatial in our flux maps, but it seems reasonable to identify at least the central part of the region emitting them with the two components (northern and southern) of the bright central feature in the HST emission line images of \citet{ferruit2000}. However, evidence for line splitting, in the sense that two broad components are required to fit the line profiles,
extends well beyond the bright, unresolved central region. 
In fact, weak double broad components can be found out to a distance of approximately 2$^{\prime\prime}$ east and west of the nucleus, that is, in a similar orientation to that of the region of enhanced velocity dispersion derived
from the single-component fit.

The [FeVII] line profile is also dominated by blue- and red-shifted components, although with more extreme mean velocities ($-250$ and 270 km s$^{-1}$, Fig.\ref{fig: Fe_maps}). This line originates closer to the AGN, therefore velocities higher than those derived for the low-ionization species are expected. The emission in this line is also centrally concentrated (within 1$^{\prime\prime}$ of the nucleus) and again, it seems reasonable to identify the blue- and red-shifted kinematic components with the central structures visible in the HST image.

% ============================================================
% ============================================================
\subsubsection{Stability of the solution}

\textbf{Dependence on the noise:} as previously described for the stellar kinematics, we fitted the velocity maps recovered from the Monte Carlo simulation described in section \textsection\ref{sec: uncertainties}. Standard deviations for the distribution of the recovered parameters were computed and are given in Table \ref{tab: vmod_gas_totl} as an estimate of the uncertainty. 

\textbf{Dependence on the initial guess:} as for the stellar kinematics, we tested the dependence of the recovered parameters on the initial guesses and again we find that using initial guesses drawn randomly from the range specified in Table \ref{tab: vmod_gas_totl} the solution is always stable converging to the same values.

\textbf{Dependence on the value adopted for p:} the amplitude, $A$, and the concentration parameter, $c_{0}$, are the most sensitive to the value adopted for $p$. We assumed $p = 1$, which is justified by the flattening of the rotation curve at large radii \citep{SchHnk03}. When $p$ is larger than 1, the amplitude immediately reaches high values (i.e. larger than 400 km s$^{-1}$ for model \textit{1GK} and larger than 280 km s$^{-1}$ for models \textit{1GK} and \textit{2GK}). The systemic velocity and $\psi_{0}$ are almost insensitive to $p$.

% ============================================================
% ============================================================
\subsubsection{Comparison with previous work}

Although parameters like the amplitude, $A$, and $c_{0}$ may not be reliable due to the limited spatial extent of our data, we believe that the other parameters are reliably and accurately determined. Perhaps the most important of these parameters is the systemic velocity. Recall that the systemic velocity derived from stellar kinematics should be the most reliable measure of the galaxy recessional velocity as the value derived from gas kinematics may be affected by non-rotational motions, such as inflows or outflows. 

Various estimates of the systemic velocity have been made at lower spatial resolution using the gas kinematics derived from long-slit observations of the H$\alpha$ and [NII]$\lambda$6583 emission lines (e.g. \citealt{WeaverWilson91, StorchiBRASWB96, SchHnk03}). All of these measurements produce values in the range 868-890 km s$^{-1}$, while the heliocentric systemic velocity derived here is $v_\mathrm{sys,gas}$ = 797 $\pm$ 1 km s$^{-1}$ (from model \textit{2GK}). 

A contribution to the discrepancy comes from our decision to fix the gaseous kinematical center at the stellar kinematical center. Indeed, the discrepancy decreases when the center is fitted as a free parameter, but there is still a difference $\Delta$v $>$ 50 km s$^{-1}$ with previous measurements. In addition, the lower resolution of previous studies would have made it more difficult to disentangle the rotating velocity field from the additional components. However it should be noted that \citeauthor{SchHnk03} derive a systemic velocity of 877 km s$^{-1}$ relying solely on the outer portion of their velocity curve, i.e. using data between 8$^{\prime\prime}$ and 15$^{\prime\prime}$, where the curve flattens.
When measuring the velocity at the brightness peak, they obtain a value ($v_\mathrm{sys}$ = 789 $\pm$ 15 km s$^{-1}$) consistent with our findings. 

\vskip10pt
The kinematics of the coronal line [FeVII]$\lambda$6087 in NGC 1386 have previously been investigated by \citet{rodriguezA06}. They performed single slit observations of the galaxy, with the slit aligned along the north-south direction. As in this work, they detected a double peaked emission line with the redshifted component stronger than the blueshifted one. They observed emission up to about 106 pc (1\farcs4) north and south of the nucleus (the spatial scale has been adjusted to our adopted distance). The maps presented in Fig.\ref{fig: Fe_maps} show a smaller spatial extension. Visual inspection of the datacube suggests that emission is present over a scale comparable with that proposed by \citeauthor{rodriguezA06}, however the S/N is not high enough to allow a robust fit to the spectra. As proposed by \citeauthor{rodriguezA06}, the presence of this emission, roughly aligned with the bright lobes visible in the flux maps, and probably associated with the bright nuclear knots revealed by \citet{ferruit2000}, is suggestive of a bipolar outflow along the radiation cone.

% Rotation models for stellar kinematics: =======

% Rotation Models: =======

% Comprehensive map for the coefficients of the rotation curve
\begin{table}[t] %\begin{center}
\caption{FITS TO THE OBSERVED GASEOUS VELOCITY FIELD}  
\begin{tabular}{l lll}
\hline
\hline
	 Parameter				&  						& Notes					& Initial guess	\\
\hline
 & & \\	
 \multicolumn{4}{c}{Geometry from stellar kinematics (model \textit{1GK})}\\
 \hline
 & & \\	
	A [km s$^{-1}$] ...........		& 308 $\pm$ 1				& 						& 0:900	\\
	$v_\mathrm{sys,gas}$ [km s$^{-1}$] ..	& 787 $\pm$ 1		& geocentric$^{\ddagger}$				& 500:900	\\
	$\psi_{0}$ [deg] ................ 	& 103 					& fixed	 				& \\
	c$_{0}$ [arcsec] ............		& 4.93 $\pm$ 0.01			&						& 0:5 	\\
	p ..........................			& 1						& fixed 		\\
	$\theta$ [deg] ..................		& 65						& fixed		\\
	x$_{0}$ [arcsec] ............		& 0  						& fixed at N 				&\\
	y$_{0}$ [arcsec] ............		& 0 						& fixed at N 				& \\

 & & \\
 \multicolumn{4}{c}{Fixed center (model \textit{2GK})}\\
 \hline
 & & \\
 	A [km s$^{-1}$] ...........		& 226 $\pm$ 1				& 						& 180:300	 \\
	$v_\mathrm{sys, gas}$ [km s$^{-1}$] ..	& 786 $\pm$ 1				& geocentric$^{\ddagger}$	& 500:900	 \\
	$\psi_{0}$ [deg] ................ 		& 88 $\pm$ 0.3				& 	 					& 60:100$^{\circ}$ \\
	c$_{0}$ [arcsec] ............		& 3.08 $\pm$ 0.02			& 						& 0:5 \\
	p ..........................			& 1						& fixed \\
	$\theta$ [deg] ..................		& 65						& fixed\\
	x$_{0}$ [arcsec] ............		& 0						& fixed at N\\
	y$_{0}$ [arcsec] ............		& 0						& fixed at N\\
 & & \\
 \multicolumn{4}{c}{Free center (model \textit{3GK})}\\
 \hline
& &    \\

	A [km s$^{-1}$] ...........		& 213 $\pm$ 1				& 						& 180:300	\\
	$v_\mathrm{sys,gas}$ [km s$^{-1}$] ..	& 810 $\pm$ 1				& geocentric$^{\ddagger}$				& 500:900	\\
	$\psi_{0}$ [deg] ................ 		& 88 $\pm$ 0.1				& 	 					& 60:100$^{\circ}$\\
	c$_{0}$ [arcsec] ............		& 2.57 $\pm$ 0.02			&						& 0:5 	\\
	p ..........................			& 1						& fixed 		\\
	$\theta$ [deg] ..................		& 65						& fixed		\\
	x$_{0}$ [arcsec] ............		& + 0.4 $^{\dagger}$ 		& from N 					& -1:1\\
	y$_{0}$ [arcsec] ............		& + 0.9 $^{\dagger}$			& from N 					& -1:1\\

 & & \\

\hline
\end{tabular}
\label{tab: vmod_gas_totl}
%\end{center}
\tablecomments{$^{\dagger}$ The estimated uncertainty for this coefficient is much smaller than 0\farcs1. $^{\ddagger}$The heliocentric systemic velocity is derived by adding 11 km s$^{-1}$ to the geocentric value.}
\end{table}

% ===============================
% ===============================
\section{Interpretation}\label{sec: nuclear_conf}

Our analysis of the IFS data cube has revealed multiple components of line-emitting gas characterized by distinct kinematic properties. In particular, we identify three main kinematic components: (1) a low velocity dispersion (``narrow") component that extends over the entire FOV, (2) a higher velocity dispersion (``broad") component observed within approximately 2$^{\prime\prime}$ of the nucleus, but mostly concentrated within $1^{\prime\prime}$ and apparently coincident with the two nuclear knots resolved in the HST observation of \citet{ferruit2000}, and (3) a region of enhanced velocity dispersion, associated with red- and blue-shifted velocity residuals and with a faint emission structure seen in the \citeauthor{ferruit2000} HST imaging, points to the presence of a distinct kinematic component roughly perpendicular to the galaxy disk.

The velocity field of the narrow component is consistent with gas predominantly rotating in circular orbits with inclination consistent with the galaxy disk. However, fits to the velocity field reveal residuals indicating that significant deviations from rotation about the axis of the large-scale disk are also present.

% ===========================
\subsection{The narrow component - disk rotation}

The presence of multiple components of emission lines is evident from the total flux map presented in Fig.\ref{fig: [NII]1c}, as well as from the flux distributions in the third column of Fig.\ref{fig: [NII]maps}, and the channel maps of Fig.\ref{fig: channelm}. The bright lobes extending along the direction north-south, and the low-level emission present over the entire FOV, are associated with the narrow kinematic component. They show rotational motion and low velocity dispersion. 
Therefore, although the double-lobed morphology of the narrow component is suggestive of an outflow, the velocity field does not support this interpretation. The morphology and kinematics can be reconciled if we interpret the bright lobes as due to emission from photoionized gas in the regions of the galaxy disk that are intersected by the biconical AGN radiation field that is predicted by the AGN unified model \citep[e.g.][]{Antonucci93}. Our interpretation is based on the model proposed for NGC 4151 by \citet{RobinsonVVA94}: if the edge of the radiation cone emerges from the disk at a shallow angle, then the intersection with the disk surface traces an ellipse, which would result in an elongated surface brightness distribution similar to what is observed (see Section \ref{sec: NLR} and Fig.\ref{fig: cartoon}). 

Fig.\ref{fig: oxigen_overplots} shows that the [OIII]$\lambda$5007 emission matches very well the flux distribution observed for the [NII] emission and the other low-ionization species included in our observation. It therefore seems reasonable to assume that the extended [OIII] emission has the same origin as the low-ionization lines.

% ===========================
\subsection{The broad component - nuclear outflow}

The second kinematic component -- the broad component -- is associated with the strong nuclear line emission responsible for the central ``blob" in the flux distribution maps. It is mostly confined within 1$^{\prime\prime}$ from the nucleus, but shows some extension out to approximately 2$^{\prime\prime}$. This component is actually double, and appears in the emission-line profiles as two components, one blueshifted and the other redshifted relative to the systemic velocity. They are co-spatial and have a relatively large velocity dispersion ($\sigma \approx$ 220 km s$^{-1}$). A reasonable interpretation is that they represent the approaching and receding sides of an outflow.

Spectra presented in Fig.\ref{fig: fits} and the flux maps in the right column of Fig.\ref{fig: [NII]maps} show that the red broad component is brighter than the blue. 
This type of asymmetry (redshifted gas brighter than blueshifted gas) is not what would usually be expected in an outflow. Indeed, emission line blueshifts or blue asymmetries are usually considered evidence for outflows \citep[e.g.][]{ZMSC02}. Here we suggest that the origin of the asymmetry in the brightness of the two components is due to the combination of the dust distribution and morphology of the outflow: we propose that the broad components originate from outflowing discrete clouds containing dust. These clouds are photoionized by radiation from the AGN around which they are distributed, which implies that the ionized face of each cloud is always directed toward the AGN. Therefore blueshifted clouds preferentially present to the observer their non-ionized side; on the other hand, the redshifted clouds present their ionized faces to the observer, thus appearing brighter. 

\begin{figure}
\centering
\includegraphics[trim = 10cm 8cm 1cm 1.5cm, clip = true, scale=0.5]{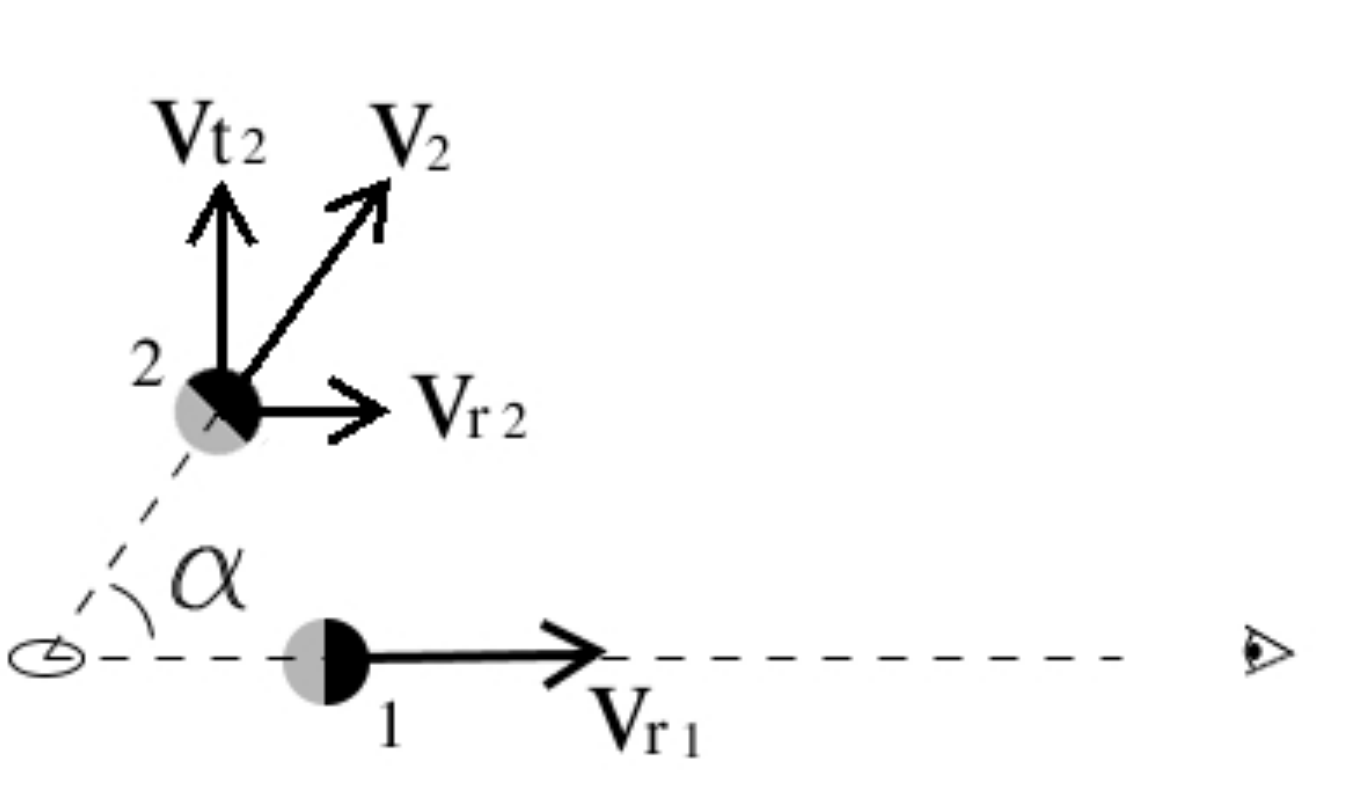}
\caption{Outflowing clouds (circle 1 and 2) are ionized by the AGN accretion disk (ellipse on the left). Blueshifted clouds preferentially show to the observer their non ionized face (in black). Their line emission is attenuated by dust embedded in the cloud. Clouds at large projected distances from the nucleus (large values of $\alpha$, corresponding to small line-of-sight radial velocities, $v_{r}$) are brighter as their ionized face is partially visible to the observer. As a result, the average velocity measured for the blueshifted clouds is skewed toward small values.} 
\label{fig: model_asymmetry}
\end{figure}

The picture outlined above also explains the asymmetry in the velocity of the broad components ($\bar v_\mathrm{blue} \approx -140$ km s$^{-1}$, $\bar v_\mathrm{red} \approx $ 250 km s$^{-1}$): blueshifted clouds with maximal line-of-sight velocity are those at small projected distances from the AGN, e.g. cloud 1 in Fig.\ref{fig: model_asymmetry}. According to our proposed scenario, radiation from these clouds is strongly attenuated as they show the observer their non-ionized face. Blueshifted clouds located at larger projected distances, and therefore characterized by smaller line-of-sight velocities, contribute a greater fraction of the blueshifted emission as their ionized face is partially visible, e.g. cloud 2 in Fig. \ref{fig: model_asymmetry}. On the other side, redshifted clouds at small projected distances from the AGN (and therefore characterized by maximal line-of-sight velocities) always show the observer their ionized faces. As a result, the average velocity measured for the blueshifted clouds is smaller than the velocity measured for the redshifted counterparts.

It seems reasonable to identify the blue- and red-shifted broad components of the line profile with the two components (northern and southern) of the bright central feature in the HST emission line images of \cite{ferruit2000}. If this association is correct, it suggests that the kinematic components are the approaching and receding sides of a bipolar outflow that is aligned with the radiation cone axis.

Maps for the [FeVII] emission lines are consistent with those derived for the [NII] broad component, however there is marginal evidence that the emission is extended along the radiation cones. This is consistent with the idea that the blue- and red-shifted broad components come from a bipolar outflow aligned with the AGN radiation cones.

% ==========================
\subsubsection{Mass outflow rate}

We estimate the mass outflow rate as the ratio between the mass of the outflowing gas and the dynamical time at the nucleus.
The gas mass associated with the broad component is estimated as:

\begin{align}
M_\mathrm{g} &= m_\mathrm{p}n_\mathrm{e}Vf
\label{eq: mgas}
\end{align}

\noindent where m$_{p}$ is the proton mass, $n_{e}$ $\sim$ 1200 cm$^{-3}$ and $V$ are the electron density and the volume of the region where broad components are detected, $f$ is the filling factor.
The filling factor and volume are eliminated by combining eq.\ref{eq: mgas} with the following expression for the H$\alpha$ luminosity:

\begin{align}
L_\mathrm{H_{\alpha}} &\sim n_\mathrm{e}^{2} j_\mathrm{H_{\alpha}}(T)Vf 
\end{align}

\noindent with j$_\mathrm{H_{\alpha}}$(T) = 3.534 $\times$ 10$^{-25}$ cm$^{3}$ erg s$^{-1}$ at T=10000 K \citep{Osterbrock_book89}. Therefore the gas mass is expressed as:

\begin{align}
M_\mathrm{g} &= \frac{m_\mathrm{p}L_\mathrm{H_{\alpha}}}{n_\mathrm{e}j_\mathrm{H_{\alpha}}(T)}.
\end{align}

In the previous section we argued that emission from the blueshifted broad component is attenuated by dust embedded within the emitting clouds, whilst clouds responsible for the redshifted broad component show the observer their ionized side. Therefore, we assume that the red broad component gives a better representation of the intrinsic emission produced by the outflowing gas and we estimate the total H$\alpha$ luminosity due to the outflow as twice the H$\alpha$ luminosity from the red broad component, yielding L$_{H_{\alpha}}$ = 2.5 $\times$ 10$^{40}$ erg s$^{-1}$. The mass of the ionized gas is, therefore, $M_{g} \sim 5 \times 10^{4}$ $M_{\odot}$. This represents the amount of gas responsible for the H$\alpha$ emission from the two ``broad" components.

We estimate the dynamical time as the ratio of the radius of the region where broad components are observed ($\approx$ 2$^{\prime\prime}$ or 152 pc) to the mean velocity of the outflow (250 km s$^{-1}$). This gives $T_{d}\sim 6 \times 10^{5}$ yr.
The corresponding mass outflow rate is $\dot M \sim 0.1$ $M_{\odot}$ yr$^{-1}$ (note that this is a lower limit as it represents the outflowing mass associated with only the ionized side of the clouds). For comparison, the mass accretion rate implied by the bolometric luminosity is $\dot M_{acc} = L_{bol}/\eta c^{2} \sim 3 \times 10^{-4} M_{\sun}$ yr$^{-1}$, where we assume a mass-to-energy conversion efficiency $\eta = 0.1$, $L_{bol} \approx 90 L_{[OIII]}$ \citep{dumasMEN07} and $L_{[OIII]} \approx 2 \times 10^{40}$ erg s$^{-1}$ from the flux measurement in \citet{SBP1989}.

% ======================================
\subsection{Equatorial rotation or expansion?}

In general, the narrow component velocity
field is dominated by circular rotation consistent with the rotation of the large-scale galactic
disk. However, there is strong evidence for a distinct kinematic component extending 2$^{\prime\prime}$ - 3$^{\prime\prime}$ either side of the nucleus, and approximately oriented along the minor axis of the large scale disk. This component appears to be associated with a low surface brightness bar-like structure that is seen crossing the nucleus east to west in the HST WFPC2 image obtained by \citet{ferruit2000}, Sec.\ref{subsec: line_flux}. It is characterized by  increased velocity dispersion and blue- and red-shifted velocity residuals to the west and east of the nucleus, respectively. Notably, the orientation of this structure is roughly perpendicular to the axis of the AGN radiation bicone.

\begin{figure*}[t]
\centering $
\begin{array}{ccc}
\includegraphics[trim = 0cm 0cm 3cm 0cm, clip = true, scale=0.27]{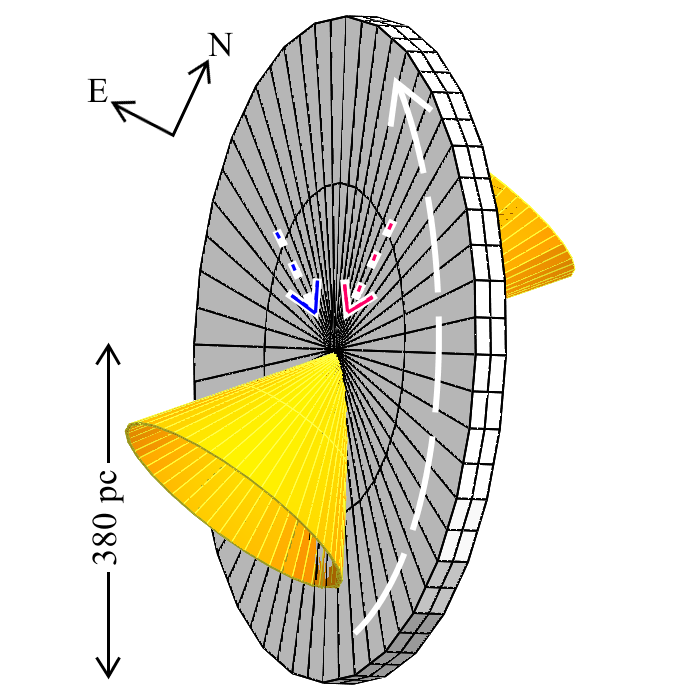} & \includegraphics[trim = 0.5cm 0cm 1cm 0cm, clip = true, scale=0.272]{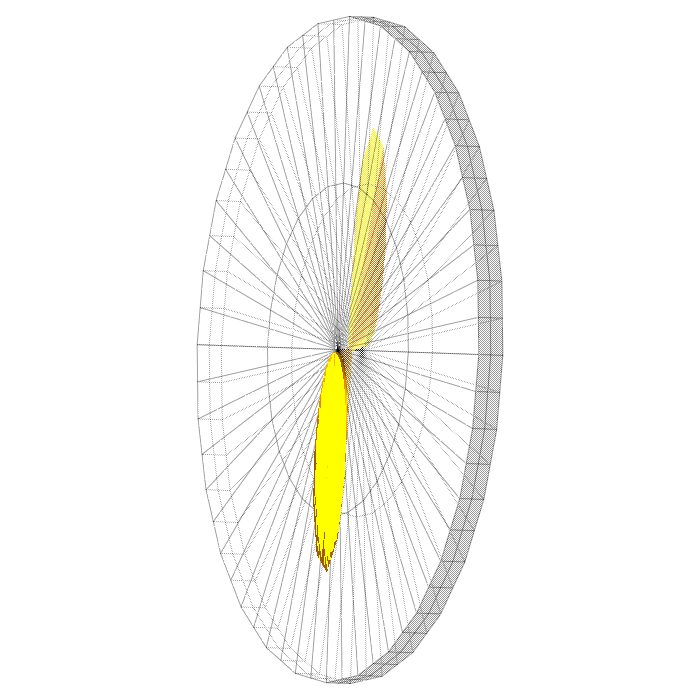} & \includegraphics[trim = 1cm 0cm 3cm 0cm, clip = true, scale=0.27]{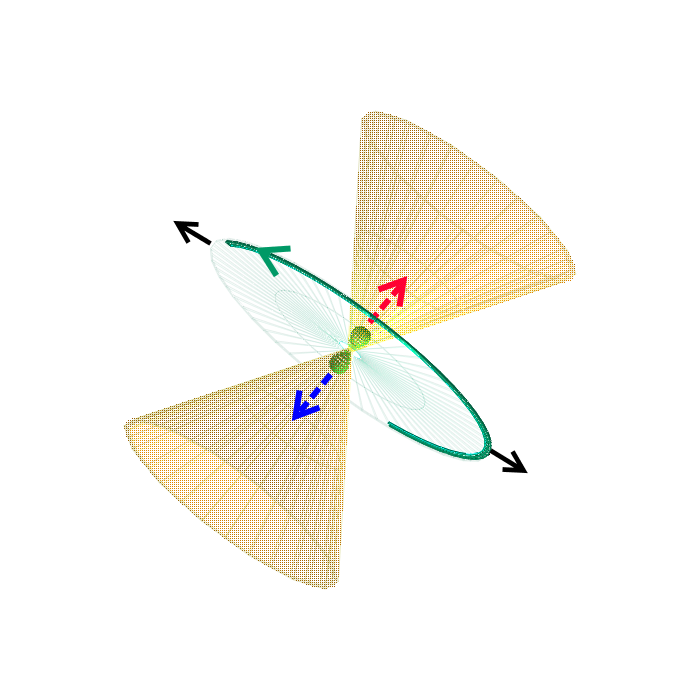} \\
\end{array}$
\caption{The proposed configuration for the nuclear region of NGC 1386 (see Section \ref{sec: NLR}). 
\textit{Left}: the narrow
component of the emission lines originates from ionized gas rotating (long-dashed arrow) and inflowing (short-dashed blue and red arrows) in the plane of the galaxy (large grey disk). The yellow
cones schematically represent the AGN radiation field which escapes the torus in bicones
aligned with the torus axis. The radiation cone axis is inclined at $38^{\circ}$ to the
disk plane and the cones have a half-opening angle of $34^{\circ}$.
\textit{Center}: the bright lobes visible in the flux maps 
are due to photoionization of the disk gas where the radiation cones (omitted for clarity) intersect the disk. 
\textit{Right}: within the central $\approx$ 1$^{\prime\prime}$, the broader emission line components arise mainly from dusty clouds (green blobs) participating in a bipolar
radial outflow (blue and red dashed arrows) aligned with the radiation cone axis. 
Blue- and red-shifted velocity residuals, which
are associated with increased velocity dispersion, suggest that rotation (solid green arrow) 
and/or outflow (short black arrows) is occurring on larger scales (2 - 3$^{\prime\prime}$) in a plane approximately 
perpendicular to the radiation cone axis (thin green disk). The large-scale disk associated
with the narrow emission line component is omitted for clarity. Cartoon created using the \textit{Shape} program \citep{Steffen2010}.} 
\label{fig: cartoon}
\end{figure*}

The velocity residuals can be interpreted as either radial outflow or rotation. In the first case, outflow is favored over inflow on the basis of geometrical arguments: the blue residual is associated with the near side of the galaxy disk. In the second case, the rotation is about an axis approximately coincident with the axis of the AGN radiation cone.
In either case, the outflow/rotation appears to be occurring in a plane inclined at large angle with respect to the large-scale disk and roughly perpendicular to the AGN radiation bicone. We propose that this is the equatorial plane of the torus. It is possible that the residuals result from a combination of rotation and outflow. Several models \citep[e.g.][]{Krolik86,ElitzurShl06,Dorodnitsyn08,Keating12,Wada12,Dorodnitsyn12} propose that the torus itself is formed in an approximately equatorial magneto-hydrodynamical wind or that a wind is blown off the torus. The observed residuals, the enhanced velocity dispersion, and the circum-nuclear surface brightness structure, could conceivably originate from an extension to larger scales (or from a remnant) of such a wind. A cartoon illustrating the proposed scenario is shown in Fig.\ref{fig: cartoon}.

% ======================================
\subsection{Velocity residuals in the large-scale disk}
\label{subsec: vres}

In the outer regions of the FOV, at distances of 2$^{\prime\prime}$ - 5$^{\prime\prime}$ from the nucleus, we find 
blueshifted residuals in the E-NE arc, while redshifted residuals are found to the N-NW and W-SW.
This system is most prominent in the residual maps obtained from 
model 1GK, where the line of nodes and the kinematical center are fixed at the values
determined from stellar kinematics, but can also be discerned at much lower amplitude
in the residual maps obtained from the other two models (2GK and 3GK).
The blueshifted velocities appear in the far side of the galaxy (the south-eastern side, see Fig.\ref{fig: NGC1386}) 
while the redshifted velocities appear in the near side, and are spatially coincident 
with the arc-like dust lanes highlighted in the structure map presented in Fig.\ref{fig: NGC1386}.
The dust lanes appear to trace part of a nuclear spiral, and we suggest that these 
velocity residuals may trace gas streaming inwards along the spiral.

% =============================================================================
% =============================================================================
\subsection{The geometry and kinematics of the nuclear region} 
\label{sec: NLR}

Here we attempt to place the various components of the circum-nuclear
ionized gas in NGC 1386, as characterized by distinct kinematics and spatial distributions, in the
context of an overall structure for the nuclear region. 
We also consider the relationship of these components to the classical emission line regions 
of AGN, namely the narrow line region (NLR) and the extended narrow line region (ENLR).
Our proposed configuration for the nuclear region of NGC 1386 is shown schematically in Fig.\ref{fig: cartoon}.

The gas emitting the narrow kinematic component resides in a disk, which rotates
approximately in the plane of the galaxy. The line emission from this component extends over
the entire FOV (690 $\times$ 530 pc), but is brightest in a double-lobed structure that 
results from photoionization of the disk gas where the AGN radiation cones intersect the disk
(Fig.\ref{fig: cartoon}, left and center). The geometry of the radiation cones can be constrained by
requiring (1) that the disk surface area bounded by the intersection with the cones approximately matches 
the extension and orientation of the bright lobes, and (2) that the line of sight to the nucleus does not pass within the 
radiation cone (i.e., the line of sight must intercept the torus, since NGC 1386 is a Seyfert 2). 
With these constraints, we find that the observed morphology
is approximately reproduced when the radiation cone axis is inclined at $38^{\circ}$ to the
disk plane and the cones have a half-opening angle of $34^{\circ}$ (however, these 
values should be considered illustrative rather than tightly constrained).

The extended emission-line lobes, therefore, do not represent a physical (matter bounded) structure, 
such as an outflow or, as suggested by \citet{SchHnk03}, an edge-on disk; they are merely 
those parts of a disk rotating in the galactic plane that are illuminated (and photoionized) by the AGN radiation emerging from the torus opening. In this respect, they are similar to the $\sim$ kpc scale
``extended narrow line regions'' that are present in many Seyfert galaxies (i.e., interstellar gas clouds photoionized by the anisotropic AGN radiation field, \citealt{UngerPA87}). 
The overall similarity between the morphologies of the extended ($>$ 1$^{\prime\prime}$ north
and south of the nucleus) [OIII]$\lambda$5007 and H$\alpha$ + [NII] emission in HST images \citep{ferruit2000} and the flux distributions in the narrow component of the [NII] (and H$\alpha$) emission lines, 
derived from our IFS data, implies that the extended [OIII] emission has the same origin. 
The sub-structure in the ``lobes'' revealed by the HST images presumably reflects the
non-uniform distribution of gas clouds in the disk.

It is worth noting that Chandra observations of NGC 1386 presented by \citet{LWang2013}
show the presence of elongated X-ray emission, extending over few kpc, with an orientation
consistent with that of the radiation cones inferred here.

As noted in Section \ref{subsec: vres}, the pattern of velocity residuals suggests that inward spiral 
streaming motions may also be present in the disk, although the amplitudes are dependent on
whether or not the gas disk is assumed to be co-planar with the stellar rotation.
\vskip10pt

We consider the gas emitting the broad double component to be participating in a nuclear
outflow, the blue- and red-shifted components representing the approaching and receding sides of
the flow, respectively. Our data do not have sufficient spatial resolution to strongly constrain
the  geometry of the outflow. However, the brightest emission is concentrated within the central 1$^{\prime\prime}$ and may be associated with the elongated double structure which is visible in 
both the structure map presented in Fig.\ref{fig: NGC1386}, and in the HST WFPC2 [OIII] and H$\alpha$ + [NII] images of \citet{ferruit2000}. If this is the case, the morphology suggests a bipolar outflow
approximately aligned with the radiation cone axis. The southern HST component
would then be identified with the approaching, blueshifted, side and the northern component with the receding, redshifted side.

The characteristic properties of the broad component gas, in particular 
its proximity to the nucleus, electron density ($\approx$1000 cm$^{-3}$), velocity dispersion ($\approx$200 km s$^{-1}$) and its complex 
kinematics, lead us to identify it as the NLR in NGC 1386.
In our illustrative cartoon, this is represented by the green blobs in the right panel of Fig.\ref{fig: cartoon}, each one representing a distribution of outflowing clouds.

In addition to the two main kinematic components, that is, rotation in the plane of the
galaxy and a nuclear outflow, there is strong evidence for another component that involves
rotation and/or outflow in the equatorial plane of the torus, and extends to approximately 2 - 3$^{\prime\prime}$ ($\approx$200 pc) either side of the nucleus. We speculate that this structure (represented by the green
disk in the right panel of Fig.\ref{fig: cartoon}) may be evidence for the accretion-disk winds that have been
proposed to explain the formation of the torus. The excitation
mechanism for the relatively faint line emission associated with this kinematic
component is unclear. The gas is (of necessity) located outside the AGN radiation cones.
However, the line ratios in subregions C and D, which sample the emission west and east,
respectively, of the nucleus, are generally consistent with AGN photoionization, although
subregion C in particular, exhibits lower excitation characteristic of LINERs. Thus,
it is possible that this gas is photoionized by an attenuated AGN radiation field ``leaking''
through a clumpy torus. Alternatively, the enhanced velocity dispersion in this region
suggests that shock ionization could also play a role.

% ==============================
% ==============================
\section{Summary and Conclusions} \label{sec: conclusion}

Using the GMOS integral field unit we observed the inner 690 pc of the nearby Seyfert 2 spiral galaxy NGC 1386 in the spectral range 5600 - 7000 \AA\ which includes a number of prominent emission lines ([OI]$\lambda$6302, [FeVII]$\lambda$6087, [NII]$\lambda\lambda$6548,6583, H$\alpha$, [SII]$\lambda\lambda$6716,6731). We modeled the profiles of these lines to produce velocity, velocity dispersion and flux maps with a spatial resolution of 68 pc and a spectral resolution of 66 km s$^{-1}$.

The emission-line flux maps, as well as flux distributions in channel maps, are characterized by two elongated structures to the north and south of the nucleus, a bright nuclear component with FWHM $\approx$ 0\farcs9, and low-level emission extending over the whole FOV. We argue that the elongated structures are due to emission from gas in the galaxy disk photoionized by AGN radiation via a geometry in which the AGN radiation bicone intercepts the galactic disk at a shallow angle. Fainter emission consistent with AGN photoionization is also observed beyond the elongated structures suggesting that some radiation from the AGN escapes through the obscuring torus.

The gas has three distinct kinematic components.
\begin{enumerate}
\item A low velocity dispersion component ($\sigma\approx 90$ km s$^{-1}$), which is consistent with gas rotating in the galaxy disk and includes the bright 
elongated structures (the ``lobes'') observed in the flux map. There is no kinematic
evidence that these structures are distinct from the galaxy disk gas; they are participating in 
the general rotation and, as noted above, can be explained by partial illumination of the disk by the AGN radiation cones. 
However, subtraction of rotation models reveals velocity residuals in the outer regions of the FOV 
that suggest inflow, perhaps in the form of streaming along dusty nuclear spirals.  
We estimate the electron density from the [SII] doublet, finding that the narrow
component has $n_{e} \approx 1000$ cm$^{-3}$ at the nucleus and $\approx 200$ cm$^{-3}$ in the bright extended lobes.

\item A higher velocity dispersion component ($\sigma\approx 220$ km s$^{-1}$) 
characterized by blueshifted and redshifted sub-components at average velocities of $-140$ and $+250$ km s$^{-1}$, respectively. This component corresponds to the bright nuclear emission observed in our flux maps
and is attributed to a compact ($\lesssim 1^{\prime\prime}$) bipolar outflow aligned with the radiation cone axis.
It appears to be resolved into two bright knots by HST observations, which can be
identified with the approaching and receding sides of the outflow. 
We identify it as the narrow line region in NGC 1386. The broad components have electron densities of 
$n_{e} \approx 800$ cm$^{-3}$ for the blueshifted and $\approx 1100$ cm$^{-3}$ for the redshifted components, respectively. 
Using the H$\alpha$ luminosity we estimate a lower limit for the mass outflow rate of $\dot M_{out} \sim 0.1$ M$_{\odot}$ yr$^{-1}$, which is much larger than the accretion rate necessary to sustain the AGN bolometric luminosity, that is $\dot M_{acc} \sim 3 \times 10^{-4}$ M$_{\odot}$ yr$^{-1}$.

\item There is strong evidence for a third kinematical component in the form of velocity residuals that are spatially coincident with a region of enhanced velocity dispersion and a faint, bar-like
emission structure revealed by HST imaging. This component extends 2$^{\prime\prime}$-3$^{\prime\prime}$ either side of the nucleus,
approximately perpendicular to the major axis of the galaxy disk. 
The presence of blueshifted residuals on the near (west) side and redshifted
residuals on the far (east) side of the galaxy is consistent with 
outflow and/or rotation in a plane that is approximately perpendicular to the AGN radiation cone axis. We speculate that this is a wind which is both outflowing in a plane roughly coincident with the equatorial plane of the torus, and rotating about the axis of the radiation cones.
\end{enumerate}

In some respects, the structure and kinematics of the circum-nuclear region of NGC 1386
are consistent with the ``classical'' AGN unification model in which a hydrostatic, optically thick torus prevents isotropic escape of the UV-optical radiation from the AGN.
In particular, the bright extended lobes of line emission can be explained by AGN photoionization where biconical beams of ionizing radiation intersect the galaxy disk. There is also good evidence for a compact bipolar outflow along the axis of the AGN radiation cones; such outflows appear to be a common feature of Seyfert galaxies \citep[e.g.][]{FischerCK13}.

However, we also find features that are not easily explained in the context of 
the ``classical'' unification model. The emission line ratios 
are generally consistent with AGN photoionization over the whole FOV, even for the faint
emission outside the inferred boundaries for the AGN radiation cones. This suggests that
attenuated AGN ionizing radiation is able to escape at large angles from the radiation cone axis, favoring models in which the torus consists of an ensemble of discrete, optically thick
clumps \citep[e.g.][]{NenkovaEtAl08}.

Perhaps more intriguing, however, is the evidence
that rotation and/or outflow is also occurring in the equatorial plane of the torus.
In recent years, our group has found evidence for the presence of similar compact ($\lesssim 100$ pc)
equatorial outflows, i.e. perpendicular to the axis of the radio jet or to the axis of the
ionization cones, in several other active galaxies \citep[][]{CoutoSBA13,SMSBN14,riffelSBR14}; an equatorial outflow has also been suggested by \citet{Su96} to explain the features observed in the radio emission of NGC 5929.
Perhaps we are finding evidence of torus winds similar to those predicted by recent
models \citep[e.g.][and references therein]{Keating12,Dorodnitsyn12}.
These observations hint that equatorial outflows on $\sim$ 100 pc scales may be a common 
feature of AGNs. However, to investigate their nature and ubiquity among the AGN
population, high spatial resolution spectroscopy of a larger sample and detailed modeling
to determine the kinematic and physical properties of the gas will be required.

% ==================================
% ==================================

\section*{Acknowledgments}
We thank the anonymous referee for insightful comments, which helped improve the manuscript.
D.L. thanks M. Richmond and D. Merritt for helpful discussions, W. Steffen and N. Koning for their assistance in the use of the program Shape, and acknowledges support from the 
National Science Foundation under grant no. AST - 1108786. R.A.R. acknowledges support from FAPERGS (project N0. 2366-2551/14-0) and CNPq (project N0. 470090/2013-8 and 302683/2013-5). 

This work is based on observations obtained
at the Gemini Observatory, which is operated by the Association
of Universities for Research in Astronomy, Inc., under a cooperative
agreement with the NSF on behalf of the Gemini partnership:
the National Science Foundation (United States), the Science
and Technology Facilities Council (United Kingdom), the National
Research Council (Canada), CONICYT (Chile), the Australian Research
Council (Australia), Minist«erio da Ciöencia e Tecnologia
(Brazil) and south-eastCYT (Argentina).

We acknowledge the usage of the HyperLeda
database (http://leda.univ-lyon1.fr) and the NASA/IPAC Extragalactic Database (NED) which is operated by the
Jet Propulsion Laboratory, California Institute of Technology, under contract with the
National Aeronautics and Space Administration.

% =============================================================================
% =============================================================================
\bibliography{biblio}
\end{document}